\documentclass[%
 reprint,
superscriptaddress,
nofootinbib,
 amsmath,amssymb,
 aps,
]{revtex4-2}

\usepackage{graphicx}
\usepackage{dcolumn}
\usepackage{bm}

\usepackage{subfigure}

\begin{document}

\preprint{APS/123-QED}

\title{Hidden Markov model tracking of continuous gravitational waves 
 from a neutron star with wandering spin. III. Rotational phase tracking}

\author{A. Melatos}
\email{amelatos@unimelb.edu.au}
\affiliation{School of Physics, University of Melbourne, 
 Parkville, Victoria 3010, Australia}
\affiliation{Australian Research Council Centre of Excellence 
 for Gravitational Wave Discovery (OzGrav), University of Melbourne,
 Parkville, Victoria 3010, Australia}
\author{P. Clearwater}
\affiliation{School of Physics, University of Melbourne, 
 Parkville, Victoria 3010, Australia}
\affiliation{Australian Research Council Centre of Excellence 
 for Gravitational Wave Discovery (OzGrav), University of Melbourne,
 Parkville, Victoria 3010, Australia}
\affiliation{Data61, Commonwealth Scientific and Industrial
 Research Organisation, Corner Vimiera \& Pembroke Roads, 
 Marsfield, NSW 2122, Australia}
\author{S. Suvorova}
\affiliation{School of Physics, University of Melbourne, 
 Parkville, Victoria 3010, Australia}
\affiliation{Australian Research Council Centre of Excellence 
 for Gravitational Wave Discovery (OzGrav), University of Melbourne,
 Parkville, Victoria 3010, Australia}
\affiliation{Department of Electrical and Electronic Engineering, 
 University of Melbourne, Parkville, Victoria 3010, Australia}
\affiliation{School of Electrical and Computer Engineering, 
 RMIT University, Melbourne, Victoria 3000, Australia}
\author{L. Sun}
\affiliation{School of Physics, University of Melbourne, 
 Parkville, Victoria 3010, Australia}
\affiliation{Australian Research Council Centre of Excellence 
 for Gravitational Wave Discovery (OzGrav), University of Melbourne,
 Parkville, Victoria 3010, Australia}
\affiliation{LIGO Laboratory, California Institute of Technology,
 Pasadena, California 91125, USA}
\affiliation{OzGrav-ANU, Centre for Gravitational Astrophysics,
 College of Science, Australian National University,
 Australian Capital Territory 2601, Australia}
\author{W. Moran}
\affiliation{Department of Electrical and Electronic Engineering, 
 University of Melbourne, Parkville, Victoria 3010, Australia}
\affiliation{School of Electrical and Computer Engineering, 
 RMIT University, Melbourne, Victoria 3000, Australia}
\author{R. J. Evans}
\affiliation{Australian Research Council Centre of Excellence 
 for Gravitational Wave Discovery (OzGrav), University of Melbourne,
 Parkville, Victoria 3010, Australia}
\affiliation{Department of Electrical and Electronic Engineering, 
 University of Melbourne, Parkville, Victoria 3010, Australia}

\date{\today}

\begin{abstract}
\noindent
A hidden Markov model (HMM) solved recursively by the Viterbi algorithm
can be configured to search for persistent, quasimonochromatic 
gravitational radiation from an isolated or accreting neutron star,
whose rotational frequency is unknown and wanders stochastically.
Here an existing HMM analysis pipeline is generalized to track 
rotational phase and frequency simultaneously,
by modeling the intra-step rotational evolution according to a 
phase-wrapped Ornstein-Uhlenbeck process,
and by calculating the emission probability using a phase-sensitive
version of the Bayesian matched filter 
known as the $\mathcal{B}$-statistic,
which is more sensitive than its predecessors.
The generalized algorithm tracks signals from isolated and binary sources
with characteristic wave strain $h_0 \geq 1.3\times 10^{-26}$
in Gaussian noise with amplitude spectral density 
$4\times 10^{-24}\,{\rm Hz^{-1/2}}$,
for a simulated observation composed of $N_T=37$ data segments, 
each $T_{\rm drift}=10\,{\rm days}$ long,
the typical duration of a search for the low-mass X-ray binary (LMXB) Sco X$-$1
with the Laser Interferometer Gravitational Wave Observatory (LIGO).
It is equally sensitive to isolated and binary sources
and $\approx 1.5$ times more sensitive than the previous pipeline,
which achieves $h_0 \geq 2.0\times 10^{-26}$ for a comparable search.
Receiver operating characteristic curves
(to demonstrate a recipe for setting detection thresholds)
and errors in the recovered parameters
are presented for a range of practical $h_0$ and $N_T$ values.
The generalized algorithm successfully detects every available synthetic signal
in Stage I of the Sco X$-$1 Mock Data Challenge
convened by the LIGO Scientific Collaboration,
recovering the frequency and orbital semimajor axis 
with accuracies of better than $9.5\times 10^{-7}\,{\rm Hz}$ 
(one part in $\sim 10^8$)
and $1.6\times 10^{-3}\,{\rm lt\,s}$
(one part in $\sim 10^3$)
respectively.
The Viterbi solver runs in $\approx 2\times 10^3$ CPU-hr for an isolated source
and $\sim 10^5$ CPU-hr for a LMXB source
in a typical, broadband ($0.5$-${\rm kHz}$) search,
i.e.\ $\lesssim 10$ times slower than the previous pipeline.
\end{abstract}

\maketitle


\section{\label{sec:level1}Introduction}
Rapidly rotating neutron stars with time-varying mass and current quadrupole
moments are promising targets of searches for continuous-wave
gravitational radiation by long-baseline interferometers such as the
Laser Interferometer Gravitational Wave Observatory (LIGO)
and Virgo
\cite{Riles2013}.
Several classes of isolated and accreting neutron stars are 
predicted to be approaching detection,
if they emit at or near indirect amplitude limits
derived from energy or angular momentum conservation arguments
based on electromagnetic observations.
\cite{Abbott2017PulsarsO1,Abbott2017CrossCorrO1,Woan2018}

Among the challenges faced by such experiments is the fact that
the signal frequency is often unknown or highly uncertain
and wanders stochastically due to irregularities in the star's rotation,
known as {\em spin wandering} or timing noise.
\cite{Cordes1985,Bildsten1997,Mukherjee2018}
For some isolated targets,
such as nonpulsating neutron stars in supernova remnants,
the spin frequency $f_\ast$ of the crust
and corotating magnetosphere cannot be observed, 
e.g.\ central compact objects like Cassiopeia A
or the putative neutron star in SNR 1987A.
\cite{Aasi2015YoungSNR,Sun2016,Sun2018}
In radio pulsars like the Crab, on the other hand,
$f_\ast(t)$ is measured accurately as a function of time $t$
by timing the radio pulsations,
but there is no guarantee that the crust corotates exactly with the
gravitational-wave-emitting quadrupole.
\cite{Abbott2008Crab}
For accreting targets,
such as low-mass X-ray binaries (LMXBs),
\footnote{
In this paper, we follow the usual shorthand of using the term LMXB
interchangeably to refer to either the binary system or the neutron star
therein.
}
the accretion can drive electromagnetic signatures ---
thermal X-ray pulsations or type I X-ray burst oscillations ---
which allow $f_\ast(t)$ to be measured.
However, $f_\ast(t)$ is unknown in some of the brightest sources,
like Scorpius X$-$1 (Sco X$-$1),
which exhibit neither signature.
\cite{Watts2008}
Indirect upper limits on the characteristic gravitational wave strain $h_0$,
\cite{Jaranowski1998}
based on energy conservation in isolated sources
(i.e.\ the star spins down entirely due to gravitational radiation)
and angular momentum conservation in binary sources
(i.e.\ accretion torque balance),
imply
$h_0 \propto \tau^{-1/2}$ and $h_0 \propto F_{\rm X}^{1/2}$ respectively,
where 
$\tau = f_\ast (2 | \dot{f}_\ast | )^{-1}$ denotes the spin-down age,
and $F_{\rm X}$ denotes the X-ray flux.
\cite{Riles2013}
Hence the most promising targets ---
young, isolated objects and X-ray-luminous accretors ---
can be those for which the least is known about $f_\ast(t)$.

One powerful strategy for overcoming the challenge of spin wandering ---
especially in LMXB searches ---
is to track $f_\ast(t)$ with a hidden Markov model (HMM).
\cite{Quinn2001}
Given a time-ordered sequence of observations,
a HMM relates each observation to the system's underlying, hidden state
[e.g.\ $f_\ast(t)$] by an emission probability
(e.g.\ a detection statistic of some type).
The hidden state evolves through a concurrent sequence,
whose step-wise transitions are modelled probabilistically as well
(e.g.\ as a random walk).

In the gravitational wave context,
a HMM solved by the fast, recursive, Viterbi algorithm
\cite{Viterbi1967}
has been implemented as a general-purpose search pipeline and applied
to look for the LMXB Sco X$-$1 
in Advanced LIGO data.
\cite{Abbott2017ViterbiO1,Abbott2019ViterbiO2}
The pipeline exists in two versions.
\begin{itemize}
\item
Version I calculates the emission probability
by summing the maximum likelihood $\mathcal{F}$-statistic
\cite{Jaranowski1998}
at orbital sidebands incoherently without reference to the orbital phase.
\cite{Suvorova2016,Suvorova2017}
Given Gaussian noise with one-sided amplitude spectral density 
$S_h(2f_\ast)^{1/2} = 4\times 10^{-24} \, {\rm Hz^{-1/2}}$,
representative of Advanced LIGO's design sensitivity,
Version I detects isolated sources with 
$h_0 \geq 2\times 10^{-26}$
and binary sources with
$h_0 \geq 8\times 10^{-26}$
and finds 41 out of 50 injected signals in Stage I of the
Sco X$-$1 Mock Data Challenge (MDC).
\cite{Messenger2015,Suvorova2016}
It was applied to data from Advanced LIGO's first observing run (O1)
and returned the upper limit
$h_0 \leq h_0^{95\%} = 5\times 10^{-25}$ ($95\%$ confidence)
at $106\,{\rm Hz}$ for Sco X$-$1,
noting that O1 did not reach full design sensitivity.
\cite{Abbott2017ViterbiO1}
\item
Version II tracks orbital phase as well as $f_\ast(t)$
and sums the sideband power coherently using a Jacobi-Anger
decomposition of the $\mathcal{F}$-statistic.
\cite{Suvorova2017}
Given $S_h(2f_\ast) = 4\times 10^{-24} \, {\rm Hz^{-1/2}}$,
it detects isolated and binary sources with
$h_0 \geq 2\times 10^{-26}$
and finds all 50 injections in Stage I of the Sco X$-$1 MDC.
It is being applied to data from Advanced LIGO's second 
\cite{Abbott2019ViterbiO2}
and third observing runs.
\end{itemize}

In this paper, we extend Version II of the HMM
to track the rotational phase (i.e.\ the phase of the carrier wave)
as well as the orbital phase.
The result is an algorithm (Version III) which performs nearly
as well as a fully coherent matched filter like the $\mathcal{F}$-statistic,
when the phase evolution is known electromagnetically.
It maintains the same level of performance,
when the phase evolution is unknown,
as long as the HMM time-step is chosen to be shorter than the
spin wandering time-scale.
\cite{Mukherjee2018}
Ensuring that the latter condition is satisfied involves
trial and error but is not taxing computationally
for most realistic searches.
Version III of the HMM is built on a phase-dependent version
of the Bayesian matched filter called the ${\cal B}$-statistic 
used in loosely coherent and related continuous-wave searches
\cite{Prix2009,Dergachev2010,Dergachev2012,Whelan2014,Dhurandhar2017,Bero2019}.
It outperforms Versions I and II because
(i) the ${\cal B}$-statistic is more sensitive than the ${\cal F}$-statistic,
and (ii) the in-built requirement of phase continuity reduces
false alarms, as discussed in Section \ref{sec:vit2}.
It leverages the existing, easy-to-use, thoroughly tested
software infrastructure housed in the LIGO Scientific Collaboration
Algorithm Library (LAL).
Several of its subroutines and intermediate data products
are shared by the ${\cal F}$-statistic and Versions I and II of the HMM.
\footnote{
A Viterbi-based algorithm has also been developed to perform
nonparametric, all-sky searches.
\cite{Bayley2019}
Generalizing it to track phase as well as frequency
lies outside the scope of this paper.
}

The paper is structured as follows.
In Sections \ref{sec:vit2}--\ref{sec:vit4} we describe how to modify the
emission and transition probabilities of the HMM to track the
rotational phase.
The performance of the extended HMM is then tested by performing
Monte-Carlo simulations with Gaussian noise
for isolated and binary sources in Sections \ref{sec:vit5}
and \ref{sec:vit6} respectively.
Specifically, the sensitivity is calculated as a function of 
the user-selected false alarm and false dismissal probabilities
and compared for Versions I, II, and III of the HMM.
The accuracy of frequency and phase recovery
as part of a successful detection is also quantified.
Finally we run the extended HMM on data from Stage I of the Sco X$-$1 MDC
in Section \ref{sec:vit7} 
and confirm that it detects every injection easily.
Implications for future gravitational wave searches and their
astrophysical impact are discussed briefly in Section \ref{sec:vit8}.
Among them is the tantalizing possibility that a gravitational wave
detection of spin wandering
(possibly in conjunction with radio/X-ray timing data)
may clarify its physical origin,
which remains a subject of debate in both isolated
\cite{Cordes1985,Alpar1986,Cheng1987,Jones1990,Price2012,Melatos2015}
and accreting
\cite{Taam1988,Baykal1991,Baykal1993,DeKool1993,Bildsten1997,Romanova2004}
systems.

\section{HMM tracking \label{sec:vit2}}
HMM frequency tracking is exploited widely in engineering applications 
ranging from radar and sonar analysis \cite{Paris2003} 
to mobile telephony \cite{White2002,Williams2002}
and has been extended to handle amplitude and phase information
and multiple targets.
\cite{Barrett1993,Xie1991,Xie1993}
It delivers accurate estimation, when the signal-to-noise ratio (SNR) is low,
but the sample size is large, \cite{Quinn2001} 
as in continuous-wave gravitational wave data analysis.
In this section we describe how to generalize a HMM that tracks $f_\ast(t)$
to one that tracks the rotational phase $\Phi_\ast(t)$ (and hence the carrier
phase of the signal) as well as $f_\ast(t)$.
Section \ref{sec:vit2a} sets out the tracking framework
in its general form.
\cite{Streit1990,Quinn2001}
Section \ref{sec:vit2b} explains the central role played by step-wise 
phase continuity in reducing the HMM's false alarm rate.
Section \ref{sec:vit2c} discusses how to discretize the HMM's state space
and the related challenges involved in enforcing phase continuity,
when the emission probability is calculated from the output of a 
frequency-domain matched filter like the $\mathcal{F}$-statistic.
Modified transition and emission probabilities are presented
in Sections \ref{sec:vit3} and \ref{sec:vit4}.

\subsection{General formulation and drift time-scale
 \label{sec:vit2a}}
A HMM is a probabilistic finite state automaton defined by
a hidden (unobservable) state variable, $q(t)$,
and an observable state variable, $o(t)$.
The automaton jumps through a time-ordered sequence of observations,
$O=\{ o(t_0), \dots, o(t_{N_T}) \}$,
at discrete times
$t_0 \leq \dots \leq t_{N_T}$.
In general there exist $N_Q^{N_T+1}$ possible hidden-state paths,
$Q = \{ q(t_0), \dots, q(t_{N_T}) \}$,
which are consistent with $O$.
Here $N_Q$ counts the finite number of discrete values,
that $q(t)$ can take at time $t$.

Given $O$, some paths are more likely than others.
If we assume that the automaton is Markovian,
such that the transition probability from $q(t_n)$ to $q(t_{n+1})$
depends only on $q(t_n)$,
then the probability that $Q$ gives rise to $O$ equals
\begin{eqnarray}
 \Pr(Q|O)
 & = &
 L_{o(t_{N_T}) q(t_{N_T})} A_{q(t_{N_T}) q(t_{N_T-1})}
 \times \dots
 \nonumber \\
 & &
 \times
 L_{o(t_1) q(t_1)} A_{q(t_1) q(t_0)} \Pi_{q(t_0)}~.
\label{eq:vit1}
\end{eqnarray}
In (\ref{eq:vit1}),
\begin{equation}
 A_{q_j q_i}
 = 
 {\rm Pr} [ q(t_{n+1})=q_j | q(t_n) = q_i ]
\label{eq:vit2}
\end{equation}
is the transition probability matrix;
\begin{equation}
 L_{o_j q_i}
 = 
 {\rm Pr} [ o(t_n)=o_j | q(t_n) = q_i ]
\label{eq:vit3}
\end{equation}
is the emission probability matrix,
namely the probability that the system is observed in state $o(t_n)$
while occupying the hidden state $q(t_n)$;
and
\begin{equation}
 \Pi_{q_i}
 =
 {\rm Pr} [ q(t_0) = q_i ]
\label{eq:vit4}
\end{equation}
is the prior vector,
namely the probability that the system occupies the hidden state $q(t_0)$ initially.

To solve the HMM, one seeks the most probable path $Q^\ast(O)$,
which maximizes $\Pr(Q|O)$ given $O$, viz.
\begin{equation}
 Q^\ast(O) = {\rm arg \, max} \, \Pr(Q|O)~.
\label{eq:vit5}
\end{equation}
The maximization can be done in many ways.
In previous gravitational wave applications as well as in this paper,
we employ the Viterbi algorithm,
\cite{Viterbi1967,Quinn2001}
whose logic and pseudocode are summarized briefly in Appendix \ref{sec:vitappa}.
The Viterbi algorithm is a dynamic programming algorithm.
It is computationally efficient, 
executing of order $(N_T+1) N_Q \ln N_Q$ floating point operations.

Table \ref{tab:vit0} summarizes how the general framework above maps onto 
Versions I, II, and III of the HMM.
For each version,
it specifies the intended astrophysical target,
the hidden astrophysical variables being tracked,
the intermediate data inputs distilled from the raw observations
(which go into calculating $L_{o_j q_i}$),
as well as the forms of $A_{q_j q_i}$, $L_{o_j q_i}$, and $\Pi_{q_i}$,
which define the probabilistic structure of the HMM.
The entries in each column are discussed in detail when introduced
in Sections \ref{sec:vit2}--\ref{sec:vit4},
together with full mathematical definitions of the various terms
and symbols, e.g.\ ${\cal F}$, ${\cal J}$, and ${\cal B}$.
In this paper, we take $q(t)=[f_\ast(t),\Phi_\ast(t)]$.
We adopt a flat prior, as in previous work,
\cite{Suvorova2016,Suvorova2017}
and track the phase difference 
$\Phi_\ast(t_{n+1})-\Phi_\ast(t_n)$
across each HMM step;
$\Phi_\ast(0)$ is the result of a historical accident,
which obviates the need to track the absolute phase.

\begin{table*}[!tbh]
	\centering
	\setlength{\tabcolsep}{8pt}
	\begin{tabular}{cccccccc}
		\hline
		\hline
		Version & Target & $q(t)$ & $o(t)$ & $A_{q_j q_i}$ & $L_{o_j q_i}$ & $\Pi_{q_i}$ & Ref. \\
		\hline
		I & isolated & $f_\ast(t)$ & Fourier & 
                  random walk & ${\cal F}$ (max.\ likelihood) & uniform &
                  \cite{Suvorova2016,Abbott2017ViterbiO1} \\
                II & binary & $f_\ast(t)$ & Bessel &
                  random walk & ${\cal J}$ (max.\ likelihood) & uniform &
                  \cite{Suvorova2017,Abbott2019ViterbiO2} \\
                III & isolated & $\Phi_\ast(t)$, $f_\ast(t)$ & Fourier &
                  Ornstein-Uhlenbeck & ${\cal B}$ (Bayesian) & uniform &
                  this paper \\
                    & binary & $\Phi_\ast(t)$, $f_\ast(t)$ & Bessel &
                  Ornstein-Uhlenbeck & ${\cal B}$ (Bayesian) & uniform &
                  this paper \\
		\hline
		\hline
	\end{tabular}
	\caption[version]
{Comparison of HMM Versions I, II, and III:
intended targets (column 2), hidden variables (column 3), 
intermediate data inputs (column 4), and probabilistic structure (columns 5--7).
The entries in each column are discussed in detail 
in Sections \ref{sec:vit2}--\ref{sec:vit4}.
In column 4, the terms Fourier and Bessel refer to ordinary and Bessel-weighted
Fourier transforms of the raw interferometer data respectively,
the latter to account for binary orbital phase,
which go into calculating $L_{o_j q_i}$ as described in Section \ref{sec:vit4}.
In column 5, which defines $A_{q_j q_i}$,
random walk refers to a discrete-time, simple random walk,
and Ornstein-Uhlenbeck refers to continuous-time, damped Brownian motion,
as described in Section \ref{sec:vit3}.
The detection statistics ${\cal F}$, ${\cal J}$, and ${\cal B}$ 
in column 6 are defined mathematically
when first introduced in Sections \ref{sec:vit2}--\ref{sec:vit4}. }
	\label{tab:vit0}
\end{table*}

In gravitational wave applications, 
the underlying, stochastic evolution of $q(t)$ is continuous.
Nonetheless the discrete-time HMM defined by 
(\ref{eq:vit1})--(\ref{eq:vit5}) provides an appropriate analysis framework,
as long as the duration $T_{\rm drift} = t_{n+1}-t_n$
of each HMM step is chosen wisely.
A recipe for choosing $T_{\rm drift}$ in Versions I and II of the HMM
is given in previous papers.
\cite{Suvorova2016,Suvorova2017}
The generalized recipe for Version III
is set out in Appendix \ref{sec:vitappaa},
where the key condition on $T_{\rm drift}$ is given by equation (\ref{eq:vit6}).
One always has $T_{\rm SFT} \leq T_{\rm drift}  \leq T_{\rm obs}$,
where $T_{\rm SFT}$ denotes the duration of the short-time Fourier transforms (SFTs) 
\cite{Mendell2002}
used to compute $L_{o_j q_i}$
(see Section \ref{sec:vit2c} and Appendix \ref{sec:vitappaa}),
and $T_{\rm obs} = N_T T_{\rm drift} \sim 1\,{\rm yr}$ is the total observation time.
The SFTs are a data management device to assist with storage and input-output.
They divide the observing run into short stretches,
typically $T_{\rm SFT}=1800\,{\rm s}$ in length,
during which one assumes that the antenna beam pattern is approximately constant 
(neglecting rotation of the Earth),
and the detector noise is approximately stationary.
They are knitted together to compute a detection statistic such as the $\mathcal{F}$-statistic
coherently over an interval $T_{\rm drift}$.
By contrast,
$T_{\rm drift}$ is a user-selected time interval which contains an integer number of SFTs,
during which one assumes that the system stays within a single HMM state,
if condition (\ref{eq:vit6}) is satisfied.
Detailed implementation instructions,
explaining how the SFTs are converted into `data atoms'
and hence values of the emission probability $L_{o_j q_i}$,
are provided in Ref.\ \cite{Prix2011}.

\subsection{Phase continuity
 \label{sec:vit2b}}
In previous implementations of HMM-based gravitational wave searches,
\cite{Suvorova2016,Abbott2017ViterbiO1,Suvorova2017}
$L_{o(t_n)q_i}$ is computed from the maximum-likelihood,
frequency-domain matched filter called the $\mathcal{F}$-statistic
\cite{Jaranowski1998}
or a close variant,
evaluated over the time interval $t_{n-1} \leq t \leq t_n$.
For an isolated source,
the $\mathcal{F}$-statistic concentrates all the signal power
into a single frequency bin, of width 
$\Delta f_{\rm drift} = (2T_{\rm drift})^{-1}$,
provided that the $T_{\rm drift}$ condition (\ref{eq:vit6}) holds.
For a binary source,
the $\mathcal{F}$-statistic disperses the signal power
into approximately $2M'+1=2{\rm ceil}(2\pi f_\ast a_0)+1$ orbital sidebands,
separated by $P^{-1}$ in frequency,
where $a_0$ is the projected semimajor axis of the binary orbit,
$P$ is the orbital period,
and ${\rm ceil}(\dots)$ returns the lowest integer greater than 
or equal to its argument.
However, it is possible to redirect most of the signal power
into a small subset ($\ll 2M'+1$) of frequency bins by summing the
$\mathcal{F}$-statistic values at the orbital sidebands with
an appropriate weighting,
namely Bessel coefficients arising from the Jacobi-Anger expansion
of the waveform.
If the coefficients are squared Bessel functions,
the sum is incoherent,
and $L_{o(t_n)q_i}$ exhibits a narrow, cuspy peak as a function
of frequency, as in Version I of the HMM
(Bessel-weighted $\mathcal{F}$-statistic).
\cite{Suvorova2017}
If the coefficients include powers of $e^{i\phi_{\rm a}}$,
where $\phi_{\rm a}$ is a reference phase 
(usually defined by the orbit's ascending node),
and the $\mathcal{F}$-statistic is factorized into a product of complex numbers
before summation,
the sum is coherent with respect to orbital phase,
and $L_{o(t_n)q_i}$ contains all the signal power in a single frequency bin,
of width $\Delta f_{\rm drift} = (2T_{\rm drift})^{-1}$,
as in Version II of the HMM ($\mathcal{J}$-statistic).
\cite{Suvorova2017}
In summary, it is always possible to concentrate
all the signal power into a single frequency bin,
by calculating $L_{o(t_n)q_i}$ from the $\mathcal{F}$-statistic
(isolated source) or $\mathcal{J}$-statistic (binary source).
This result is confirmed by numerous Monte Carlo simulations 
in Ref.\ \cite{Suvorova2017}.

There is only one ``correct'' frequency bin at each HMM step,
and $Q^\ast(O)$ either finds it or not.
It is therefore natural to ask 
what extra advantage rotational phase tracking confers,
when the optimal path $Q^\ast(O)$ in Versions I and II of the HMM
already captures the maximum signal power available to {\em any} HMM,
for the reason set out in the previous paragraph.
The answer is that phase tracking increases the detection probability
by sharpening the HMM's ability to discriminate against spurious sequences.
For example, if a strong noise event occurs in the $i$-th frequency bin
at the $n$-th step,
then $Q^\ast(O)$ is likely to contain $q(t_n)=q_i$,
if frequency is the only hidden state variable.
Yet if phase is tracked as well,
the HMM is more likely to reject the spurious path containing $q(t_n)=q_i$
in favor of another path with lower $L_{o(t_n)q_j}$ ($j\neq i$)
but higher $A_{q(t_{n+1})q_j}$ and $A_{q_j q(t_{n-1})}$,
i.e.\ a path
whose transition probabilities into and out of the $n$-th step
are more consistent with phase continuity.
This is equivalent to the distinction between
a semi-coherent and a coherent search.
The latter is $\approx N_T^{1/4}$ times more sensitive than the former
because it effectively reduces the denominator in the SNR
by excluding false alarms that violate phase continuity.

We implement rotational phase tracking by enlarging the state vector
to two dimensions for an isolated source,
with $q(t)=[f_\ast(t),\Phi_\ast(t)]$,
and four dimensions for a binary source,
with $q(t)=[f_\ast(t),a_0(t),\phi_{\rm a}(t),\Phi_\ast(t)]$.
Under normal astrophysical circumstances,
$a_0$ and $\phi_{\rm a}$ are constant throughout a full search
($T_{\rm obs} \lesssim 1\,{\rm yr}$),
so there is no need to track them.
Hence, for both target classes, the HMM reduces to two dimensions,
with $q(t)=[f_\ast(t),\Phi_\ast(t)]$,
except that it is computed on a grid of $(a_0,\phi_{\rm a})$ pairs
for a binary source;
see Section IIA in Ref.\ \cite{Suvorova2017}.
This approach is readily parallelizable
across $(a_0,\phi_{\rm a})$ pairs and sources.

\subsection{Grid resolution
 \label{sec:vit2c}}
How do we select the number of hidden states,
$N_Q=N_{f_\ast} N_{\Phi_\ast}$,
with $N_{f_\ast} = B/\Delta f_{\rm drift}$ and
$N_{\Phi_\ast} = 2\pi/\Delta\Phi_{\rm drift}$,
where $B=\max f_{\ast}- \min f_{\ast}$ 
is the bandwidth,
and $\Delta \Phi_{\rm drift}$ is the width of a phase bin?
There are many valid ways to do this, 
as discussed in Appendix \ref{sec:vitappaa}, noting that
$\Delta f_{\rm drift}$ and $\Delta\Phi_{\rm drift}$
are related through
$\Phi_\ast(t) = 2\pi \int_0^t dt' \, f_\ast(t')$.
The choice comes down to how the HMM emission probability
is calculated from the data, as foreshadowed in Section \ref{sec:vit2a}.
In this paper, we seek to leverage the existing, easy-to-use,
thoroughly tested software infrastructure for frequency-domain
continuous-wave searches maintained in the LAL suite,
including the $\mathcal{F}$-statistic
\cite{Jaranowski1998,Prix2011},
$\mathcal{B}$-statistic
\cite{Prix2009,Dergachev2012,Whelan2014,Dhurandhar2017},
and intermediate data products generated by the $\mathcal{F}$-statistic;
see Section \ref{sec:vit4} in this paper and Section IIIA
in Ref.\ \cite{Suvorova2017}.
These software tools are built around Fourier transforms.
We are therefore obliged to take $\Delta f_{\rm drift}$
to be the half-Nyquist bin width of the $\mathcal{F}$-statistic
evaluated over a time interval of duration $T_{\rm drift}$, viz.\
$\Delta f_{\rm drift} = (2T_{\rm drift})^{-1}$.

The half-Nyquist criterion creates a problem:
small uncertainties in $f_\ast$ of $\pm \Delta f_{\rm drift}$ 
due to binning lead to large uncertainties in $\Phi_\ast$ of 
$\pm 2\pi T_{\rm drift} \Delta f_{\rm drift} = \pm \pi$
when propagated forward over one HMM time-step,
degrading the HMM's ability to track $\Phi_\ast(t)$.
One can circumvent this obstacle by abandoning the frequency domain, 
thereby surrendering its practical advantages.
Alternatively, one can achieve sub-Nyquist frequency resolution
($\ll \Delta f_{\rm drift}$ and hence $N_{\Phi_\ast} \gg 1$)
by modelling the underlying evolution of 
$q(t')=[f_\ast(t'),\Phi_\ast(t')]$
within a HMM time-step
($t_n \leq t' \leq t_n+T_{\rm drift}$).
We adopt the latter approach.
A simple, linear ramp does not improve the situation much,
e.g.\ 
$f_\ast(t')=f_\ast(t_n) \pm (t'-t_n)\Delta f_{\rm drift}/T_{\rm drift}$
implies
$\Phi_\ast(t_{n+1})-\Phi_\ast(t_n) = 2\pi T_{\rm drift} f_\ast(t_n) \pm \pi/2$,
which is still a large fractional uncertainty.
We find instead that evolving $q(t')$ stochastically
according to a phase-wrapped, Ornstein-Uhlenbeck process
(i.e.\ Brownian motion that is $2\pi$-periodic in phase)
yields good practical results.
The approach is described in Section \ref{sec:vit3} and Appendix \ref{sec:vitappb}
and tested against Monte Carlo simulations 
in Sections \ref{sec:vit5} and \ref{sec:vit6}.
It is analogous to a vernier scale,
in which the frequency bins yield a coarse first approximation to the frequency, 
and the phase bins yield a refined approximation.
We find empirically that $N_{\Phi_\ast}=32$ is adequate 
for the transition probabilities assumed in this paper
(see Appendix \ref{sec:vitappaa} and footnote \ref{foot:vit3}).
Sub-Nyquist frequency resolution is routinely achieved in
signal processing problems, where phase tracking is involved,
using a variety of techniques.
\cite{Barrett1993}

\section{Transition probabilities
 \label{sec:vit3}}
In this section we introduce an Ornstein-Uhlenbeck (Brownian) model 
of the stochastic, intra-step evolution of the star's rotation and
hence the signal's frequency and phase.
Transition probabilities $A_{q_jq_i}$
for frequency-phase tracking 
are presented in Section \ref{sec:vit3b}.
The Ornstein-Uhlenbeck model is controlled by two auxiliary parameters.
We explain how to set these parameters given $T_{\rm drift}$
in Section \ref{sec:vit3c}.

\subsection{Stepping forward in frequency and phase
 \label{sec:vit3b}}
In Versions I and II of the HMM,
it is assumed that $f_\ast(t)$ jumps by $-1$, 0, or $+1$ frequency bins
at every step with equal probability $1/3$.
\footnote{
As in previous papers,
we exclude the possibility of impulsive rotational glitches
with $f_\ast(t_{n+1}) - f_\ast(t_n) > \Delta f_{\rm drift}$;
\cite{Melatos2008,Espinoza2011}
see footnote 3 in Ref.\ \cite{Suvorova2016}
and compare Ref.\ \cite{Melatos2020}.
}
In Version III of the HMM, 
we again assume that $f_\ast(t')$ executes an unbiased
random walk for
$t_n \leq t' \leq t_n+T_{\rm drift}$
and choose $T_{\rm drift}$ according to condition (\ref{eq:vit6}),
as discussed in Appendix \ref{sec:vitappaa}.
However we model the intra-step random walk
explicitly as an Ornstein-Uhlenbeck process
that is $2\pi$-periodic in phase.
The aim is to derive $A_{q_j q_i}$ in a way that self-consistently relates
the jumps in $f_\ast(t)$ and $\Phi_\ast(t)$ and allows adequate phase resolution
($N_{\Phi_\ast} \gg 1$), as discussed in Section \ref{sec:vit2c}.

The Ornstein-Uhlenbeck process is described by a pair of stochastic
differential equations,
\begin{eqnarray}
 \frac{df_\ast}{dt}
 & = &
 - \gamma f_\ast + \sigma \xi(t)~,
\label{eq:vit9}
 \\
 \frac{d\Phi_\ast}{dt}
 & = &
 f_\ast~.
\label{eq:vit10}
\end{eqnarray}
It is controlled by two parameters:
$\gamma$, a damping rate, and $\sigma$, a fluctuation amplitude.
The fluctuating torque $\xi(t)$ has white noise statistics, viz.\
\begin{eqnarray}
 \langle \xi(t) \rangle
 & = &
 0~,
\label{eq:vit11}
 \\
 \langle \xi(t) \xi(t') \rangle
 & = &
 \delta(t-t')~,
\label{eq:vit12}
\end{eqnarray}
where $\langle \dots \rangle$ denotes an ensemble average.
We assume that there is no white noise forcing term in (\ref{eq:vit10}),
i.e.\ the principal axes of the gravitational-wave-emitting quadrupole
are fixed in the body frame rotating instantaneously at the frequency $f_\ast(t)$.
In Brownian motion in thermal equilibrium,
$\gamma$ and $\sigma$ are related by the fluctuation-dissipation theorem,
with $\sigma^2/\gamma$ proportional to the system temperature.
Here, in contrast, $\gamma$ and $\sigma$ are independent.
We explain how to choose them in practice in Section \ref{sec:vit3c}.

The stochastic differential equations (\ref{eq:vit9}) and (\ref{eq:vit10})
are equivalent to the forward Fokker-Planck equation
\cite{Gardiner1994}
\begin{equation}
 \frac{\partial p}{\partial t}
 =
 \frac{\partial(\gamma f_\ast p)}{\partial f_\ast}
 -
 \frac{\partial(f_\ast p)}{\partial \Phi_\ast}
 + 
 \frac{\sigma^2}{2}
 \frac{\partial^2 p}{\partial f_\ast^2}~,
\label{eq:vit13}
\end{equation}
whose solution 
$p(t,f_\ast,\Phi_\ast)$
equals the probability density that the hidden state lies in the infinitesimal domain
$(f_\ast,f_\ast+df_\ast) \cup (\Phi_\ast,\Phi_\ast+d\Phi_\ast)$
at time $t$ if it started at $q(0)=[f_\ast(0),\Phi_\ast(0)]$ at $t=0$,
i.e.\
$p(0,f_\ast,\Phi_\ast) =
 \delta[f_\ast -f_\ast(0)] \delta[\Phi_\ast - \Phi_\ast(0) ]$.
Hence evolving
$p(t,f_\ast,\Phi_\ast)$ from $t=t_n$ to $t=t_{n+1}$
is exactly what one needs to calculate the transition probabilities
$A_{q_j q_i}$, as defined by (\ref{eq:vit2}).
Specifically we write
\begin{equation}
 A_{(f_{\ast j},\Phi_{\ast k}) (f_{\ast l},\Phi_{\ast m})}
 =
 p(t_{n+1},f_{\ast j},\Phi_{\ast k} ) 
 \Delta f_{\rm drift} \Delta\Phi_{\rm drift}
\label{eq:vit14}
\end{equation}
with
$f_\ast(t_n)=f_{\ast l}$ and $\Phi_\ast(t_n)=\Phi_{\ast m}$,
where the integers $j$, $l$ and $k$, $m$ index discrete frequency 
and phase bins respectively.
Analytic formulas are derived for $p(t,f_\ast, \Phi_\ast)$
and its characteristic function in Appendix \ref{sec:vitappb}.
\cite{Suvorova2018}

In the Viterbi algorithm, it is sometimes more convenient to calculate
the backward transition probabilities,
$A_{q_j q_i}^{\rm back}
 = \Pr[q(t_{n})=q_j | q(t_{n+1}) = q_i]$.
This can be done by solving the backward Fokker-Planck equation,
which is adjoint to (\ref{eq:vit13}).
Details and formulas are given in Appendix \ref{sec:vitappb}.
The resulting PDF is a $2\pi$-wrapped Gaussian; 
see equations (\ref{eq:vitappb10})--(\ref{eq:vitappb15}).

\subsection{Control parameters
 \label{sec:vit3c}}
How should the control parameters $\gamma$ and $\sigma$ be chosen?
Two conditions must be satisfied during every HMM step:
$\gamma$ must be small enough, such that $\langle f_\ast \rangle$
does not drift by more than one frequency bin, $\Delta f_{\rm drift}$;
and $\sigma$ must be large enough,
so that we have
$\langle f_\ast^2 \rangle - \langle f_\ast \rangle^2
 \approx (\Delta f_{\rm drift})^2$,
i.e.\ probability leaks significantly into the frequency bins on either side
of the starting bin but not much further.
From the moment formulas in Appendix \ref{sec:vitappb},
typical of a diffusion process,
the above conditions reduce to
\begin{equation}
 f_\ast [ 1 -\exp(-\gamma T_{\rm drift}) ] < \Delta f_{\rm drift}
\label{eq:vit17}
\end{equation}
and
\begin{equation}
 \frac{\sigma^2}{2\gamma}
 [ 1 -\exp(-2 \gamma T_{\rm drift}) ] \approx (\Delta f_{\rm drift})^2
\label{eq:vit18}
\end{equation}
respectively for all $f_\ast$ in the observation band.
For a typical LMXB search with $T_{\rm drift} = 10\,{\rm d}$
and $f_\ast \gtrsim 50\,{\rm Hz}$,
we have 
$\Delta f_{\rm drift} / f_\ast \lesssim 1\times 10^{-8}$,
$\gamma T_{\rm drift} \ll 1$, and hence
$\gamma < (2f_\ast T_{\rm drift}^2 )^{-1}$
and
$\sigma \approx (4 T_{\rm drift}^3)^{-1/2}$.

\begin{figure*}
\centering
\subfigure[]
{
\label{fig:vit1a}
\scalebox{0.65}{\includegraphics{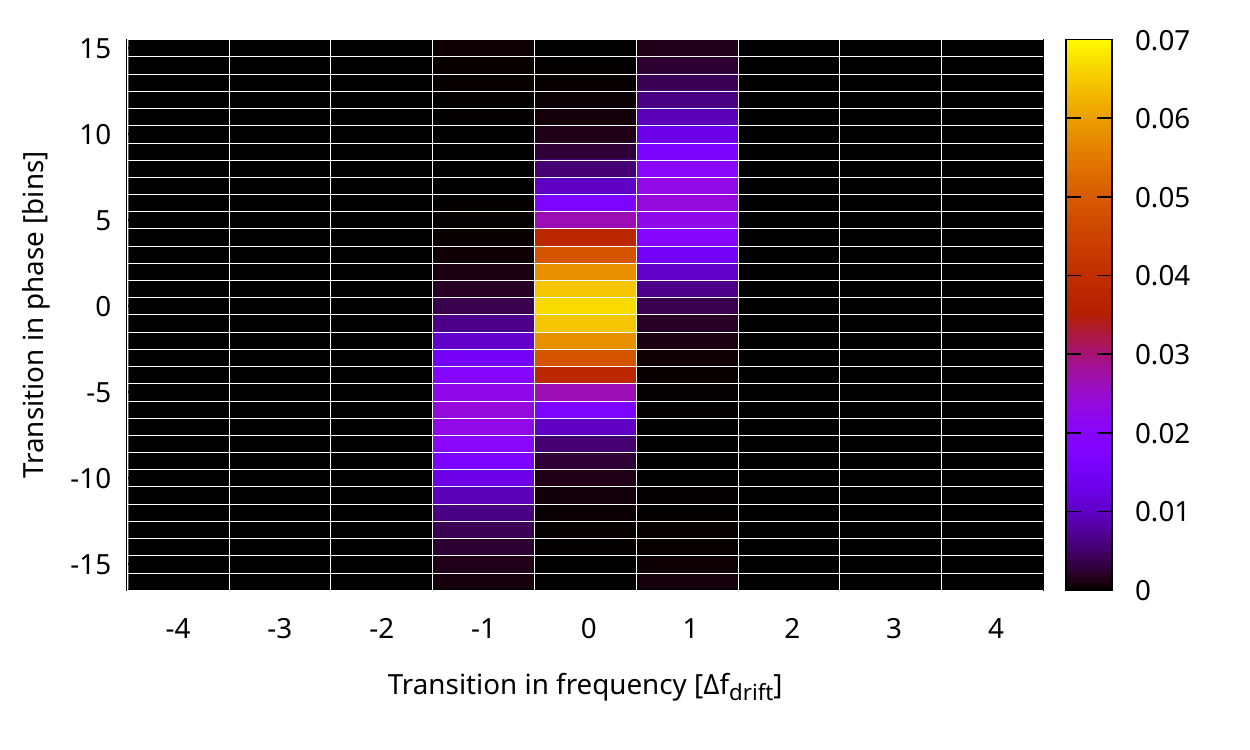}}
}
\subfigure[]
{
\label{fig:vit1b}
\scalebox{0.65}{\includegraphics{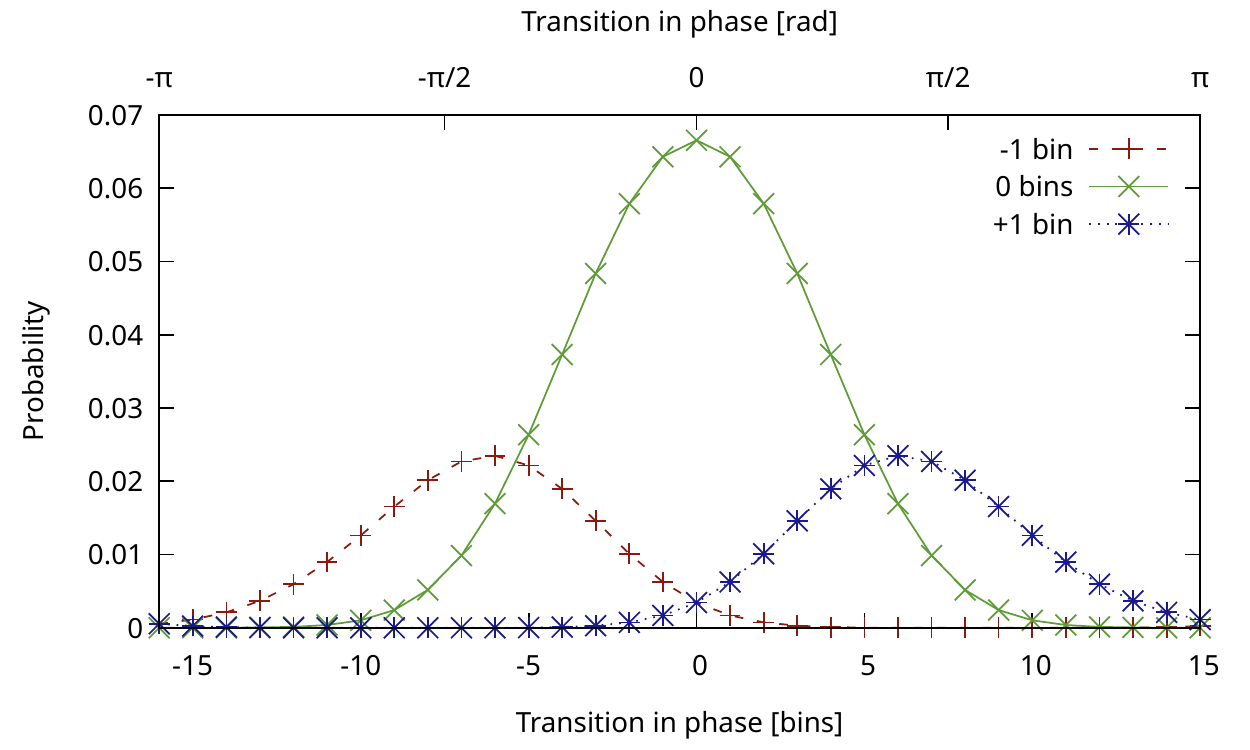}}
}
\caption{
Forward transition probabilities 
$A_{q_j q_i} = \Pr[q(t_{n+1})=q_j | q(t_{n}) = q_i]$
(not normalized)
for $f_\ast(t_{n})=111.0\,{\rm Hz}$, $\Phi_\ast(t_{n})=0\,{\rm rad}$,
$\gamma=1.0\times 10^{-16}\,{\rm s^{-1}}$, 
and $\sigma=3.7\times 10^{-10}\,{\rm s^{-3/2}}$.
(a) Contour plot versus $f_\ast(t_{n+1})-f_\ast(t_n)$ 
and $\Phi_\ast(t_{n+1})-\Phi_\ast(t_n)$.
The color scale is arbitrary; hot colors are high, cool colors are low.
The white grid delineates frequency-phase bins.
The horizontal and vertical axes are labeled by number of bins.
A subset of the hidden state space is plotted for clarity.
(b) Cross-sections at fixed 
$f_\ast(t_{n+1}) - f_\ast(t_n) = 0, \pm \Delta f_{\rm drift}$.
The crosses, plus signs, and asterisks mark phase bins.
The horizontal axes indicate $\Phi_\ast(t_{n+1})-\Phi_\ast(t_n)$
in units of radians (top) and number of bins (bottom).
The backward transition probabilities are identical
but centered on $q(t_{n+1})$ instead of $q(t_{n})$.
}
\label{fig:vit1}
\end{figure*}

Figure \ref{fig:vit1} presents an example of the transition probabilities
for an illustrative choice of $\gamma$ and $\sigma$ satisfying the
constraints in the previous paragraph and used subsequently
in the validation experiments in Sections \ref{sec:vit5} and \ref{sec:vit6}.
Contours of the PDF $A_{q_j q_i}$ in the $f_\ast$-$\Phi_\ast$ plane
are plotted in Figure \ref{fig:vit1}(a).
Three constant-$f_\ast$ cross-sections are plotted versus $\Phi_\ast$
in Figure \ref{fig:vit1}(b).
We find that $p(t_{n+1},f_\ast,\Phi_\ast)$ leaks significantly into the
frequency bins on either side of the starting bin,
with $A_{f_{\ast i\pm1},f_{\ast i}}=0.196$
and $A_{f_{\ast i},f_{\ast i}}=0.608$ (normalized).
In this implementation, the PDF is truncated to give
$A_{f_{\ast i\pm2,3,\dots,},f_{\ast i}}= 0$
to achieve computational savings,
but if one does not truncate one finds
$A_{f_{\ast i\pm2},f_{\ast i}}=3.82\times 10^{-4}$.
The probabilities of jumping up or down in frequency are equal,
as in Version I of the HMM,
while the probability of staying in the same bin is higher ($0.608$)
than in Version I ($0.333$).

In contrast, the PDF extends over many bins in phase,
as is clear from Figure \ref{fig:vit1}(b),
with full-width half-maximum $\approx 1.78 \, {\rm rad}$
(nine bins).
Phase wrapping ensures periodicity in $\Phi_\ast$, 
but for the plotted parameters the PDF is tiny at the edges of the plot,
and it is hard to verify the periodicity by eye.
The initial state $q(t_n)$ determines whether the phase wraps or not.
Figure \ref{fig:vit1}(b) confirms that
phase wrapping alternates between even and odd frequency bins
(and depends on whether $f_\ast$ jumps by zero or $\pm\Delta f_{\rm drift}$),
as discussed in Section \ref{sec:vit2c};
the phase jumps by $\pi$, when the frequency bin at $t_n$ is odd,
and by zero when the frequency bin at $t_n$ is even.
The contours slope diagonally,
because $f_\ast$ and $\Phi_\ast$ are correlated, with 
$\langle f_\ast \Phi_\ast \rangle 
 - \langle f_\ast \rangle \langle \Phi_\ast \rangle \neq 0$;
see equation (\ref{eq:vitappb14})
in Appendix \ref{sec:vitappb}.
The shape of the contours is the same for the forward and backward 
transition probabilities, but the centroid shifts with
$q(t_n)$ and $q(t_{n+1})$ respectively.

The above recipe for setting $\gamma$ and $\sigma$ is sensible but not unique.
The optimal values of the control parameters (and $T_{\rm drift}$)
depend on the waveform of the true signal, 
which is unknown in advance in an astronomical setting.
Altering $\gamma$ and $\sigma$ does not introduce a systematic bias,
because the Ornstein-Uhlenbeck process is symmetric with respect to
positive and negative frequency jumps, 
but in general it increases or decreases the sensitivity modestly.
It is found empirically that HMMs are robust to the exact form of $A_{q_j q_i}$,
which is why the naive choice of $A_{q_j q_i}$ in Version I of the HMM works well.
\cite{Quinn2001}
The extra sensitivity in Version III comes from phase tracking,
which depends weakly on $\gamma$ and $\sigma$,
because the PDF in Figure \ref{fig:vit1} is broad in phase.
When publishing searches with real data,
it is important to emphasize that any upper limits are conditional on
the signal model, which includes $\gamma$, $\sigma$, and $T_{\rm drift}$.

\section{Emission probabilities
 \label{sec:vit4}}
For the class of frequency-domain, continuous-wave searches considered
in this paper, $L_{o_j q_i}$ in (\ref{eq:vit3}) can be expressed in terms of
a suitable frequency-phase detection statistic $G(f_\ast,\Phi_\ast)$ as
\begin{equation}
 L_{o(t_n) q_i}
 \propto 
 \exp[ G(f_{\ast i'},\Phi_{\ast i''}) ]~.
\label{eq:vit19}
\end{equation}
Here
$G(f_{\ast i'},\Phi_{\ast i''})$ is the log likelihood that
$f_\ast(t_{n-1})$ lies in the $i'$-th frequency bin
$[f_{\ast i'}, f_{\ast i'}+\Delta f_{\rm drift} ]$,
and
$\Phi_\ast(t_{n-1})$ lies in the $i''$-th frequency bin
$[\Phi_{\ast i''}, \Phi_{\ast i''}+\Delta\Phi_{\rm drift} ]$,
with $i=i' N_{\Phi_\ast} + i''$,
given the data $o(t_n)$.
\footnote{
Equally one can use some other reference time, e.g. $t_n$.
}
Concretely $o(t_n)$ comprises a set of strain measurements,
numbering $T_{\rm drift}$ multiplied by the interferometer sampling rate,
or their Fourier-transformed counterparts,
sampled during the interval
$t_{n-1} \leq t' \leq t_n$.
There exist many valid ways to construct $G(f_\ast,\Phi_\ast)$,
depending on computational constraints, the data format,
and the assumed model for the evolution of $q(t)=[f_\ast(t),\Phi_\ast(t)]$.

In this paper, we strive to exploit the easy-to-use, 
thoroughly tested software infrastructure in the LAL suite
associated with the $\mathcal{F}$-statistic.
\cite{Jaranowski1998}
We are therefore led to build $G(f_\ast,\Phi_\ast)$ 
as a frequency-domain matched filter,
using as many existing LAL components as possible.
In Versions I and II of the HMM,
$G(f_\ast,\Phi_\ast)$ is constructed as a maximum likelihood estimator
from the ${\cal F}$-statistic (isolated source) 
or a Bessel-weighted sum of ${\cal F}$-statistic values (binary source).
\cite{Suvorova2016,Suvorova2017}
In Version III,
we press into service the phase-dependent generalization
of the Bayesian ${\cal B}$-statistic used in loosely coherent searches.
\cite{Prix2009,Dergachev2010,Dergachev2012,Whelan2014,Dhurandhar2017,Bero2019}
The latter choice is justified against
maximum likelhood alternatives in Appendix \ref{sec:vitappc}.
We review briefly the signal model and its definitions in Section \ref{sec:vit4a},
define the frequency domain intermediate data products that we need
(e.g.\ complex Fourier amplitudes generated by the LAL) in Section \ref{sec:vit4b},
and present a formula for $G(f_\ast,\Phi_\ast)$ 
in terms of the ${\cal B}$-statistic in Section \ref{sec:vit4c}.

\subsection{Signal model and likelihood
 \label{sec:vit4a}}
The gravitational wave signal measured at the Earth from a biaxial rotor
can be written as a linear combination of eight independent components,
\cite{Jaranowski1998}
\begin{equation}
 h(t)
 =
 \sum_{i=1}^4 A_{1i} h_{1i}(t) + A_{2i} h_{2i}(t)~.
\label{eq:vit20}
\end{equation}
In (\ref{eq:vit20}), $A_{1i}$ and $A_{2i}$ are arbitrary amplitudes
set by the source,
and $h_{1i}(t)$ and $h_{2i}(t)$ are defined in Ref.\ \cite{Jaranowski1998}
as sinusoidal functions of $\Phi(t)$ and $2\Phi(t)$ respectively,
where $\Phi(t)$ is the signal phase at the detector
[note: $\Phi(t) \neq \Phi_\ast(t)$ in general].
The amplitudes of $h_{1i}(t)$ and $h_{2i}(t)$ are modulated diurnally
by the antenna beam-pattern functions $a(t)$ and $b(t)$, defined by
equations (12) and (13) respectively in Ref.\ \cite{Jaranowski1998}.

Following equations (18) and (96) in Ref.\ \cite{Jaranowski1998},
we split the signal phase into five terms,
\begin{eqnarray}
 \Phi(t) 
 & = &
 2\pi f_0 [t+\Phi_{\rm m}(t;\alpha,\delta)]
 + \Phi_{\rm s}[t;f_0^{(k)},\alpha,\delta]
 \nonumber \\
 & &
 - 2\pi f_0 a_0 \sin (2\pi t / P - \phi_{\rm a})
 + \Phi_{\rm w}(t)~.
\label{eq:vit25}
\end{eqnarray}
In (\ref{eq:vit25}), $f_0$ is the signal frequency at the detector,
\footnote{
One has $f_0 \neq f_\ast(t)$ in general.
$f_\ast(t)$ is the true, underlying spin frequency of the star,
which we cannot measure directly and which forms one component 
of the hidden state.
$f_0$ is any arbitrary frequency, where the emission probability 
and associated phase model (\ref{eq:vit25}) are evaluated,
which may or may not coincide with $f_\ast(t)$,
depending on where in the parameter space we look.
}
$\Phi_{\rm m}$ is a time shift produced by the diurnal and annual
motions of the detector and source relative to the Solar System barycentre,
$\Phi_{\rm s}$ is a phase shift combining the latter two effects
with the intrinsic, deterministic, {\em secular} evolution
of the source through the frequency derivatives $f_0^{(k)}=d^kf_0/dt^k$
($k\geq 1$)
(see equation (14) in Ref.\ \cite{Jaranowski1998}),
the fourth term ($\propto a_0$) is the Doppler modulation
produced by the source's orbital motion in a binary system,
and $\Phi_{\rm w}(t)$ is the phase accumulated from {\em stochastic}
spin wandering.
The sky position of the source (right ascension $\alpha$, declination $\delta$)
enters $\Phi_{\rm m}$ and $\Phi_{\rm s}$.
Naturally it is possible to absorb the binary orbit and stochastic
spin wandering into $f_0^{(k)}$,
and hence absorb the fourth and fifth terms in (\ref{eq:vit25})
into $\Phi_{\rm s}$,
but it is clearer to keep the contributions separate in what follows.

The output from a single interferometer is given by $x(t)=h(t)+n(t)$,
where $n(t)$ denotes additive noise.
The normalized log likelihood after measuring the time series $x(t)$
over the interval $0\leq t \leq T_{\rm obs}$ is proportional to
\begin{equation}
 \ln \Lambda' 
 = 
 (x \| h) - \frac{1}{2} (h \| h)~,
\label{eq:vit26}
\end{equation}
where we define the inner product
\begin{equation}
 (x \| y )
 =
 \frac{2}{T_{\rm obs}}
 \int_0^{T_{\rm obs}} dt \, x(t) y(t)~.
\label{eq:vit27}
\end{equation}
In Versions I and II of the HMM,
the emission probability is computed by maximizing $\ln \Lambda'$
in (\ref{eq:vit26}) with respect to the amplitudes $A_{1i}$ and $A_{2i}$
and evaluating the result on a grid of $f_0$ values to find the peak.
(This procedure is not exactly the same as maximizing over $A_{1i}$,
$A_{2i}$, and $f_0$ simultaneously.)
The result is a sum of terms quadratic in $(x \| h_{1i} )$ or
$(x \| h_{2i} )$,
which can be computed from Fourier-transformed interferometer data
as discussed in Section IIID in Ref.\ \cite{Jaranowski1998}
and Appendix \ref{sec:vitappc} below;
see also Ref.\ \cite{Prix2011}
and Section IIA in Ref.\ \cite{Suvorova2017}.
In Version III of the HMM,
the emission probability is computed via the Bayesian ${\cal B}$-statistic,
defined in Section \ref{sec:vit4c}.
The ${\cal B}$-statistic can be computed efficiently
from the same, Fourier-transformed interferometer data used by
Versions I and II.
We define the relevant Fourier integrals in Section \ref{sec:vit4b}.

\subsection{Fourier integrals
 \label{sec:vit4b}}
The waveforms $h_{1i}(t)$ and $h_{2i}(t)$ in (\ref{eq:vit20})
are amplitude modulated by the antenna beam pattern functions $a(t)$ and $b(t)$.
The log likelihood $\ln \Lambda'$ in (\ref{eq:vit26}) is a function of $(x \| h )$,
which reduces to calculating the Fourier transforms of $x(t) a(t)$ and $x(t) b(t)$, 
because one has $(x \| h_{11}) = (x \| a(t) \cos\Phi(t) )$ for example.
For an isolated source ($a_0=0$),
let us define the Fourier integrals
\cite{Jaranowski1998}
\begin{eqnarray}
 F_{1a}(f_0)
 & = &
 \int_0^{T_{\rm obs}} dt_{\rm b} \,
 x[t(t_{\rm b})] a[t(t_{\rm b})] 
 e^{-i\Phi_{\rm s}[t(t_{\rm b})] -2\pi i f_0 t_{\rm b}}~,
\label{eq:vit29}
 \\
 F_{1b}(f_0)
 & = &
 \int_0^{T_{\rm obs}} dt_{\rm b} \,
 x[t(t_{\rm b})] b[t(t_{\rm b})] 
 e^{-i\Phi_{\rm s}[t(t_{\rm b})] -2\pi i f_0 t_{\rm b}}~,
\label{eq:vit30}
\end{eqnarray}
where $t_{\rm b} = t+\Phi_{\rm m}(t)$ defines a barycentered time coordinate
$t_{\rm b}$ related implicitly to $t$ through the time shift arising from
the Earth's motion.
In this paper, we neglect the secular frequency evolution of the source,
e.g.\ due to electromagnetic braking,
and set $f_0^{(k)}=0$ for $k\geq 1$ 
and hence $\Phi_{\rm s}[t(t_{\rm b})]=0$.
It is easy to keep $f_0^{(k)}\neq 0$ 
in (\ref{eq:vit29}) and (\ref{eq:vit30}) if desired.
We also specialize without loss of generality to the case $A_{1i}=0$,
corresponding to a search for one signal frequency (as opposed to two
simultaneously).

For a binary source ($a_0\neq 0$),
the integrands in (\ref{eq:vit29}) and (\ref{eq:vit30})
feature an extra, Doppler-modulated phase factor 
$\exp[2\pi i f_0 a_0 \sin(2\pi t/P - \phi_{\rm a})]$,
derived from (\ref{eq:vit25}).
Upon expanding this factor using the Jacobi-Anger identity,
we find that $F_{1a}$ and $F_{1b}$ should be replaced in $L_{o(t_n) q_i}$ by
\begin{eqnarray}
 J_{1a}(f_0)
 =
 \sum_{s=-M'}^{M'}
 J_s(2\pi f_0 a_0) e^{-is\phi_{\rm a}}
 F_{1a}(f_0+s/P)~,
\label{eq:vit31}
 \\
 J_{1b}(f_0)
 =
 \sum_{s=-M'}^{M'}
 J_s(2\pi f_0 a_0) e^{-is\phi_{\rm a}}
 F_{1b}(f_0+s/P)~,
\label{eq:vit32}
\end{eqnarray}
where $J_s$ denotes a Bessel function of order $s$ of the first kind.
Equations (\ref{eq:vit31}) and (\ref{eq:vit32}) add together 
the Fourier amplitudes in orbital sidebands coherently,
by taking into account the relative orbital phases of the sidebands.
\cite{Suvorova2017}
The infinite sums are truncated,
because one has $J_s(x) \ll 1$ for $s \gg x$
to a good approximation.
\cite{Sammut2014,SCO-X1-2015}

It turns out that the emission probability $L_{o(t_n) q_i}$ in (\ref{eq:vit19})
can be calculated easily from $F_{1a}$ and $F_{1b}$ (isolated source)
or $J_{1a}$ and $J_{1b}$ (binary source)
in every HMM implementation we consider.
The maximum likelihood formulas for $L_{o(t_n) q_i}$
in Versions I and II of the HMM are quoted in Appendix \ref{sec:vitappc},
where it is shown that they (and their phase-dependent generalizations)
are poorly suited to rotational phase tracking.
The ${\cal B}$-statistic adopted in this paper
for Version III of the HMM is presented next
in Section \ref{sec:vit4c}.

The Fourier integrals (\ref{eq:vit29}) and (\ref{eq:vit30})
are taken formally over $0\leq t \leq T_{\rm obs}$.
In practice, to facilitate data management, 
the integral is subdivided into `atoms'.
\cite{Prix2011}
Each atom corresponds to one SFT, which is convolved with a
sliding-window sinc function to increase the frequency resolution
from $(2T_{\rm SFT})^{-1}$ to $(2T_{\rm obs})^{-1}$,
as required by (\ref{eq:vit29}) and (\ref{eq:vit30}).
The reader is referred to Section 4.2 in Ref.\ \cite{Prix2011}
for full details; see also Section IIIA in Ref.\ \cite{Suvorova2017}.
In this paper,
following Ref.\ \cite{Prix2011},
we approximate $a(t)$ and $b(t)$ as piecewise-constant during each SFT.

\subsection{Phase-dependent ${\cal B}$-statistic
 \label{sec:vit4c}}
The ${\cal B}$-statistic \cite{Prix2009}
is a Bayesian alternative to the maximum likelihood ${\cal F}$-statistic
\cite{Jaranowski1998}.
It is derived from the likelihood function $\Lambda'$ in (\ref{eq:vit26})
combined with an isotropic prior on the source orientation (i.e.\ spin axis).
Its detection efficiency is $\approx 5$ per cent greater than that
of the ${\cal F}$-statistic,
and it is arguably motivated better astrophysically;
the ${\cal F}$-statistic implicitly assumes a uniform prior on the amplitude, 
whereas the ${\cal B}$-statistic favors lower amplitudes,
which is more realistic.
\cite{Prix2009}
In practice, however, the ${\cal F}$-statistic has proved more popular
than the ${\cal B}$-statistic,
having been preferred in various published LIGO searches,
e.g.\ \cite{Aasi2014PulsarsS6,Abbott2017PulsarsO1} (targeted),
\cite{Aasi2015YoungSNR,Abbott2017NGC6544} (directed),
and \cite{Abbott2017TimeDomainFstat,Abbott2017EatH} (all-sky),
as well as forming the basis of Versions I and II of the HMM.
\cite{Suvorova2016,Suvorova2017,Abbott2017ViterbiO1,Sun2018}
This is because:
(i) the advantage held by the ${\cal B}$-statistic in terms of
detection efficiency is small;
\cite{Prix2009,Dergachev2012,Whelan2014}
(ii) the ${\cal F}$-statistic software in the LAL was developed first
and is now thoroughly tested;
and (iii) the ${\cal B}$-statistic involves numerical integrals,
which are relatively expensive computationally,
although fast approximations do exist.
\cite{Dergachev2012,Whelan2014,Bero2019}

In this section, 
we present a phase-dependent version of the ${\cal B}$-statistic
formulated for loosely coherent searches.
\cite{Dergachev2012}.
The associated emission probability is calculated from (\ref{eq:vit29})
and (\ref{eq:vit30}) (isolated source)
or (\ref{eq:vit31}) and (\ref{eq:vit32}) (binary source),
i.e.\ the same intermediate data products as Versions I and II of the HMM.
We settle on this choice after testing several phase-dependent generalizations
of the maximum likelihood ${\cal F}$-statistic, 
as discussed in Appendix \ref{sec:vitappc}.
Empirically we find that:
(i) none perform as well as the ${\cal B}$-statistic nor offer
any discernible improvement over Versions I and II of the HMM;
(ii) a HMM based on the ${\cal B}$-statistic approaches the theoretical
sensitivity of a fully coherent search;
and (iii) the sensitivity improvement exceeds the $\approx 5$ per cent
advantage of the ${\cal B}$-statistic over the ${\cal F}$-statistic
without any phase dependence, \cite{Prix2009}
so phase tracking is clearly playing a role.
Of course these empirical findings do not constitute a formal proof,
that the phase-dependent ${\cal B}$-statistic always outperforms any
phase-dependent maximum likelihood estimator, cf.\ Ref.\ \cite{Prix2009}.
However such a formal proof lies outside the scope of this paper
and is unnecessary at this stage given the excellent performance achieved 
in tests with synthetic data in Sections \ref{sec:vit5} and \ref{sec:vit6}.
Other competing estimators will be tested in future work,
e.g.\ the phase-relaxed ${\cal F}$-statistic.
\cite{Cutler2012}

Instead of maximizing $\Lambda'$ in (\ref{eq:vit26})
with respect to $A_{1i}$ and $A_{2i}$,
we marginalize it (by Bayes's theorem) over uniform priors 
in three source-dependent variables:
(i) the polarization angle, $\psi$;
(ii) the cosine of the inclination angle, $\cos\iota$;
and (iii) the characteristic wave strain, $h_0$.
\cite{Dergachev2012,Whelan2014,Dhurandhar2017}
Let us define 
\begin{equation}
 A_+ = \frac{h_0}{2} (1 + \cos^2\iota)
\label{eq:vit33}
\end{equation}
and
\begin{equation}
 A_\times = h_0 \cos\iota
\label{eq:vit34}
\end{equation}
to be the real amplitudes of the plus and cross polarizations respectively,
which can be related to $A_{2i}$
as explained in Ref.\ \cite{PrixWhelan2007}.
(For simplicity we consider the popular case $A_{1i}=0$ here.)
Following Ref.\ \cite{Dergachev2012},
let us also define the auxiliary complex variables
\begin{equation}
 w_1'
 = 
 (2 h_0)^{-1} (A_+ \cos 2\psi + i A_\times \sin 2\psi)
\label{eq:vit35}
\end{equation}
and
\begin{equation}
 w_2'
 = 
 (2 h_0)^{-1} (A_+ \sin 2\psi - i A_\times \cos 2\psi)~,
\label{eq:vit36}
\end{equation}
which satisfy the identities
\begin{equation}
 1 = 
 | w_1' + i w_2' |^{1/2}
 + | w_1' - i w_2' |^{1/2}~,
\label{eq:vit37}
\end{equation}
$2 h_0 w_1' = A_{21} - i A_{23}$, and
$2 h_0 w_2' = A_{22} - i A_{24}$.
In terms of the above definitions,
we obtain the following expression for the marginalized likelihood:
\cite{Dergachev2012}
\begin{eqnarray}
 {\cal B} 
 & = &
 \int_0^\pi d\psi \int_{-1}^1 d(\cos\iota) \int_0^{h_0^{\rm max}} dh_0 \,
\nonumber \\
 & & 
 \times 
 \exp\left( h_0 U - \frac{h_0^2 V}{2} \right),
\label{eq:vit38}
\end{eqnarray}
with
\begin{equation}
 U = 
 w_1'^\ast R_{1a}(f_0,\Phi_0) 
 + w_2'^\ast R_{1b}(f_0,\Phi_0)~,
\label{eq:vit39}
\end{equation}
\begin{equation}
 V = A |w_1'|^2 + 2 C {\rm Re} (w_1' w_2'^\ast) + B |w_2'|^2~,
\label{eq:vit40}
\end{equation}
\begin{equation}
 R_{1a}(f_0,\Phi_0)={\rm Re}[\exp(-i\Phi_0) F_{1a}(f_0) ]~,
\label{eq:vit41}
\end{equation}
and
\begin{equation}
 R_{1b}(f_0,\Phi_0)={\rm Re}[\exp(-i\Phi_0) F_{1b}(f_0) ]~.
\label{eq:vit42}
\end{equation}
In (\ref{eq:vit40}), we have
$A=(a||a)$, $B=(b||b)$, $C=(a||b)$,
and $C \ll {\rm min}(A,B)$ for most sky positions.
\cite{Bero2019}
The ${\cal B}$-statistic peaks,
when the trial phase $\Phi_0$ in (\ref{eq:vit39}), (\ref{eq:vit41}) and (\ref{eq:vit42})
matches the true signal phase at the detector,
viz.\ $\Phi(t)$ in (\ref{eq:vit25}).

The $h_0$ integral in (\ref{eq:vit38}) is not normalized as it stands;
the HMM disregards multiplicative constants.
Hence we can take the limit $h_0^{\rm max} \rightarrow \infty$
without loss of generality
and express the $h_0$ integral in closed form as an error function.
In loosely coherent searches, ${\cal B}$ is maximized with respect to $\Phi_0$.
\cite{Dergachev2012}
We cannot do the same here, 
because we track the rotational phase and therefore
need ${\cal B}$ to depend on $\Phi_0$.
The final result, expressed again in the notation of Ref.\ \cite{Dergachev2012},
is given by
\begin{eqnarray}
 {\cal B}(f_0,\Phi_0) 
 & = &
 \int_0^\pi d\psi \int_{-1}^1 d(\cos\iota)
 \, \left( \frac{\pi}{2V} \right)^{1/2}
\nonumber \\
 & & 
 \times
 \exp\left( \frac{U^2}{2V} \right)
 \left[
  1 + {\rm erf}\left( \frac{U}{\sqrt{2V}} \right)
 \right],
\label{eq:vit43}
\end{eqnarray}
with
\begin{equation}
 {\rm erf}(x) 
 =
 \frac{2}{\sqrt{\pi}}
 \int_{0}^x dy \, \exp(-y^2)~.
\label{eq:vit44}
\end{equation}
The double integral in (\ref{eq:vit43}) is evaluated numerically 
by Simpson's rule in what follows.

\begin{figure*}
\centering
\subfigure[]
{
\label{fig:vit2a}
\scalebox{0.65}{\includegraphics{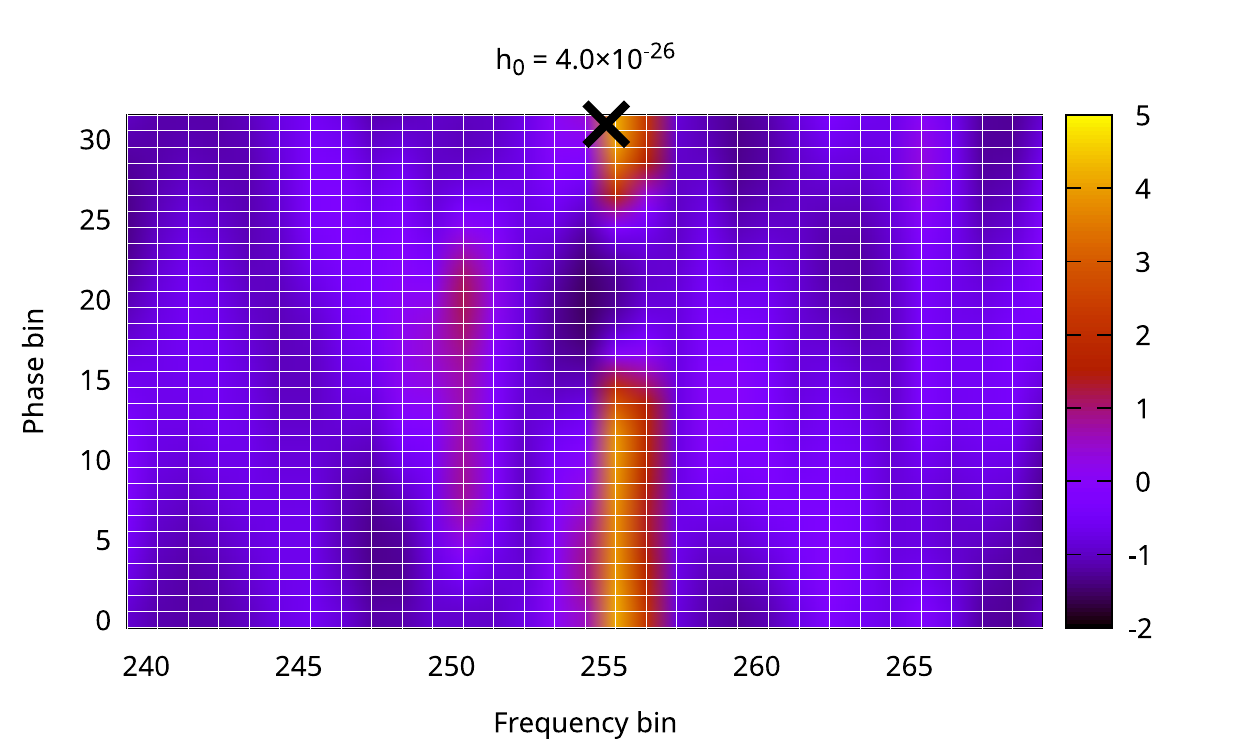}}
}
\subfigure[]
{
\label{fig:vit2b}
\scalebox{0.65}{\includegraphics{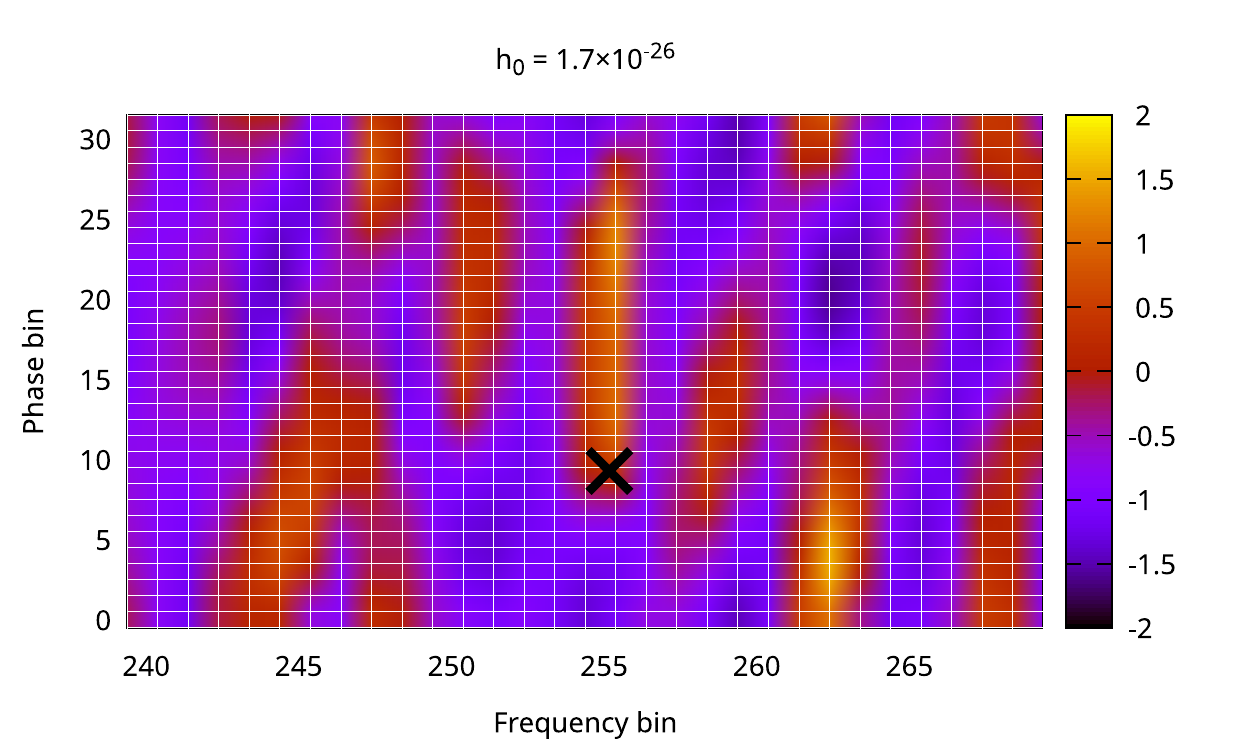}}
}
\caption{
Logarithm of the emission probability,
$G(f_0,\Phi_0)=\ln {\cal B}(f_0,\Phi_0)$
(not normalized),
represented by its contours in the $f_0$-$\Phi_0$ plane
(color scale arbitrary).
(a)
Stronger injection; $h_0=4.0\times 10^{-26}$.
(b)
Weaker injection; $h_0=1.7\times 10^{-26}$.
The injections are marked by crosses;
their parameters are listed in Table \ref{tab:vit1} (isolated source).
The observation $o(t_n)$ consists of $T_{\rm obs}=10\,{\rm days}$
of data ($N_T=1$).
The white grid delineates frequency-phase bins with
$\Delta f_{\rm drift} = 5.8\times 10^{-7} \, {\rm Hz}$ and 
$\Delta \Phi_{\rm drift} = \pi/16$.
A subset of the hidden states is plotted for clarity.
The noise is Gaussian,
with $S_h(f_0)^{1/2} = 4\times 10^{-24}\,{\rm Hz^{-1/2}}$
as in Table \ref{tab:vit1}.
}
\label{fig:vit2}
\end{figure*}

Figure \ref{fig:vit2} presents examples of the emission probability
for two signals from an isolated source injected into Gaussian noise 
with $S_h(f_0)=4\times 10^{-24}$.
The contour plots in the figure are generated from 
$T_{\rm obs}= 10\,{\rm days}$ of synthetic data.
In Figure \ref{fig:vit2}(a) the stronger injection is clearly detectable,
with $h_0 = 4.0\times 10^{-26}$,
and the emission probability peaks near the correct $(f_0,\Phi_0)$ bin.
Figure \ref{fig:vit2}(b) displays the weaker injection,
with $h_0 = 1.7\times 10^{-26}$.
There is a hot spot near the correct bin,
but it does not stand out visually from the other, noise-generated hot spots.
The constant-$f_0$ cross-section does not peak at the correct value of $\Phi_0$, 
although it still has roughly the same functional form
as in Figure \ref{fig:vit2}(a).
Note that the emission probability is not always unimodal
in the vicinity of the injection as it is in Figure \ref{fig:vit2}(a).
When the complex arguments of $F_{1a}$ and $F_{1b}$ differ sufficiently,
${\cal B}$ develops two peaks as a function of $\Phi_0$
(at fixed $f_0$),
only one of which corresponds to the signal. 
The tests in Section \ref{sec:vit5} show that the HMM is effective 
at resolving this ambiguity and identifying the true peak
for $N_T>1$.

\section{Isolated neutron star
 \label{sec:vit5}}
We begin by testing Version III of the HMM on synthetic data generated by
injecting the signal from an isolated neutron star
into additive, Gaussian noise.
Section \ref{sec:vit5a} describes the injection procedure.
Tracking results are presented in Section \ref{sec:vit5b}
for a representative sample of synthetic data.
A systematic, threshold-based strategy for identifying signal candidates
during an astrophysical search is described in Section \ref{sec:vit5c}
and is applied to characterize the performance of the HMM
in Section \ref{sec:vit5d}.
The accuracy with which the HMM reconstructs the true hidden state sequence
given a successful detection is quantified in Section \ref{sec:vit5e}.
Versions I (isolated source) and III of the HMM are compared at each stage.
Versions II (binary source) and III are compared in Section \ref{sec:vit6}. 

\subsection{Synthetic data
 \label{sec:vit5a}}
The signal phase corresponding to an isolated neutron star
is given by (\ref{eq:vit25}) with $a_0=0$.
In the tests below,
the stochastic component of the injected phase evolves during the interval
$t_n \leq t' \leq t_{n+1}$ according to
$\Phi_{\rm w}(t')
 = 
 2\pi [ (t'-t_n)^3 \ddot{f}_\ast(t_n) / 6  
  + (t'-t_n)^2 \dot{f}_\ast(t_n)/2
  + (t'-t_n) f_\ast(t_n) ]
 + \Phi_{\rm w}(t_n)$,
where
$\ddot{f}_\ast(t') = \ddot{f}_\ast(t_n)$
is drawn randomly from a uniform PDF while ensuring that 
$| f_\ast(t_{n+1}) - f_\ast(t_n) | \leq \Delta f_{\rm drift}$
is satisfied,
and $\dot{f}_\ast(t')$, $f_\ast(t')$, and $\Phi_{\rm w}(t')$ are continuous
from one HMM step to the next.
This prescription is neither unique nor necessarily optimal;
it is one of many, equally valid approaches.
We assume for simplicity that there is no secular frequency drift,
i.e.\ $\langle \dot{f}_\ast(t) \rangle = 0$.
Incorporating $\langle \dot{f}_\ast(t) \rangle \neq 0$ is straightforward;
it is already part of LAL implementations of the ${\cal F}$-statistic, for example.
However it is unnecessary in many astrophysical settings,
because the HMM with the transition probabilities defined in Section \ref{sec:vit3}
and Figure \ref{fig:vit1}
automatically handles secular spin evolution with
$\langle | \dot{f}_\ast(t) | \rangle \lesssim \Delta f_{\rm drift} / T_{\rm drift}$ 
as a matter of course.

The step-wise evolution of $\ddot{f}_\ast$ differs deliberately 
from the step-wise evolution of $\dot{f}_\ast$ modeled by the
fluctuating torque $\xi(t)$ in (\ref{eq:vit9}),
which underpins the transition probabilities in Section \ref{sec:vit3}
and Appendix \ref{sec:vitappb}.
In general we do not know the functional form of the spin wandering 
in astrophysical sources.
\cite{Mukherjee2018}
Hence it is prudent to assume different forms of wandering in the
test injections and transition probabilities,
to double-check the robustness of the algorithm.
The injection parameters are quoted in Table \ref{tab:vit1}
and are the same as those in Ref.\ \cite{Suvorova2016} 
to facilitate comparison,
except that in this paper 
$f_\ast(t_0)$ and $\Phi_\ast(t_0)$ are chosen randomly
(from uniform PDFs covering the ranges in Table \ref{tab:vit1})
as a self-blinding precaution.
The synthetic data are generated using 
{\em Makefakedata{\textunderscore}v4} in the LAL.

\begin{table}
	\centering
	\setlength{\tabcolsep}{8pt}
	\begin{tabular}{lll}
		\hline
		\hline
		Parameter & Value & Units \\
		\hline
		$\Phi_\ast(t_0)$ & $[0,2\pi]$ & rad\\
		$f_\ast(t_0)$ & $[111.0,111.1]$ & Hz \\
		$\dot{f}_\ast(t_0)$ & 0 & Hz\,s$^{-1}$ \\
		$\psi$ & 4.08407 & rad\\
		$\cos\iota$ & 0.71934 & $-$ \\
		$\alpha$ & 4.27570 & rad\\
		$\delta$ & $-$0.27297 & rad\\
		$S_h (f_0)^{1/2}$ & $4 \times 10^{-24}$ & Hz$^{-1/2}$ \\
		\hline
		\hline
	\end{tabular}
	\caption[parameters]{Injection parameters used to create the synthetic data 
analysed in Sections \ref{sec:vit5} and \ref{sec:vit6}. 
Different tests employ different subsets of the ranges in the first two lines.}
	\label{tab:vit1}
\end{table}

\subsection{Representative example
 \label{sec:vit5b}}
\begin{figure*}
	\centering
	\subfigure[]
	{
		\label{fig:vit3a}
		\scalebox{0.65}{\includegraphics{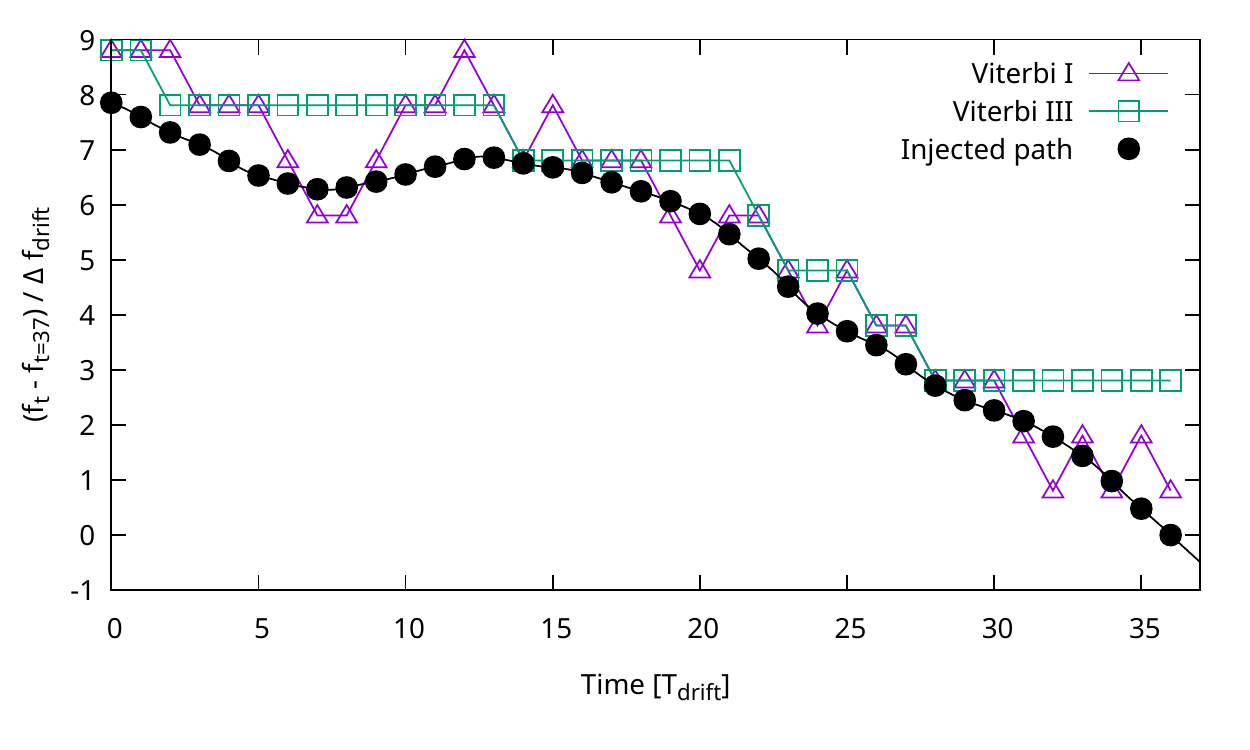}}
	}
	\subfigure[]
	{
		\label{fig:vit3c}
		\scalebox{0.65}{\includegraphics{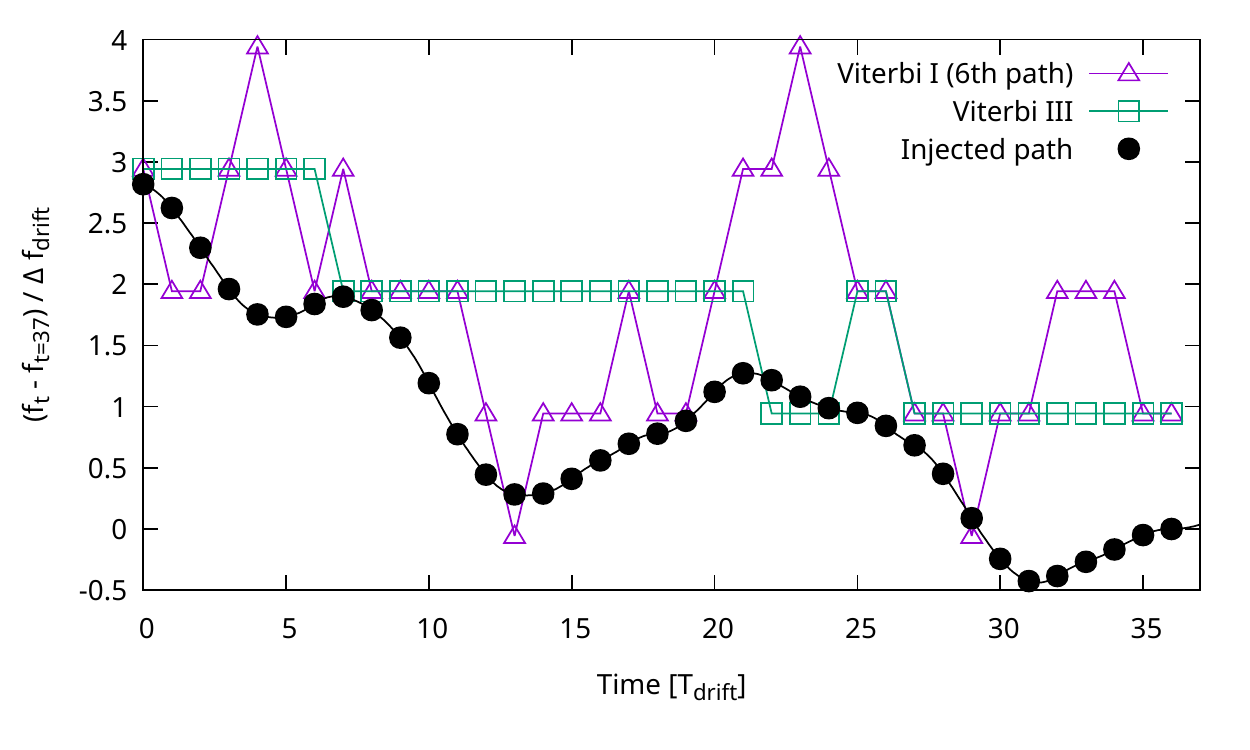}}
	}
	\subfigure[]
	{
		\label{fig:vit3e}
		\scalebox{0.65}{\includegraphics{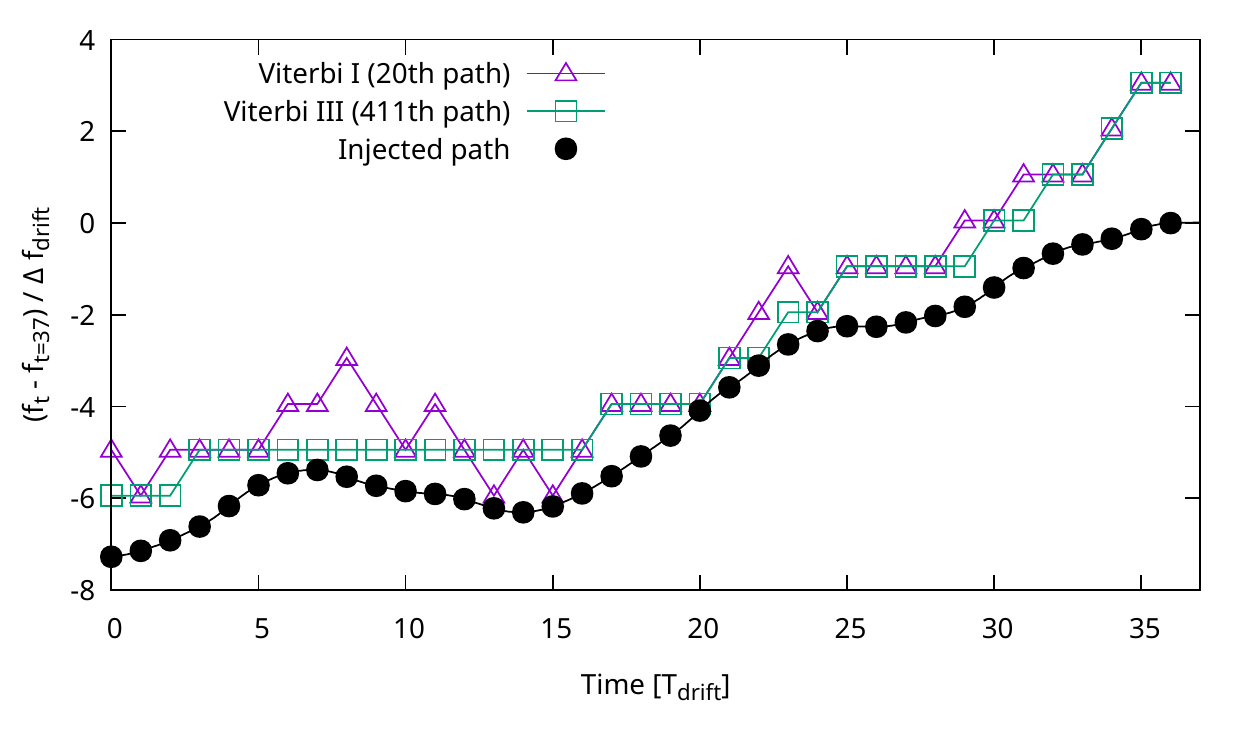}}
	}
	\caption{
Sample tracking output from
Versions I (purple curves) and III (green curves) of the HMM
for three injected signals (black curves) from an isolated neutron star 
with source parameters drawn from Table \ref{tab:vit1} and
$h_0/10^{-26} = 1.7$ [panel (a)],
$1.3$ [panel (b)],
and $1.1$ [panel (c)].
The purple and green curves are the best-matching frequency paths
(with minimum path-integrated, root-mean-square error $\varepsilon_{f_\ast}$;
see Section \ref{sec:vit5e}),
centred on $f_\ast(t_{N_T})$ 
and plotted in units of $\Delta f_{\rm drift}$;
they are not necessarily the optimal path $Q^\ast(O)$.
The optimal path matches well 
[i.e.\ within two frequency bins of $f_\ast(t)$ for all $t$]
for Versions I and III in (a) and Version III in (b).
The optimal path matches poorly
for Version I in (b) and Versions I and III in (c);
indeed it lies outside the border of the plot.
We plot instead the paths with minimum $\varepsilon_{f_\ast}$,
viz.\ the sixth, 20-th, and 411-th Viterbi paths respectively,
which lie within a few frequency bins of $f_\ast(t)$ purely by chance 
but are of no practical use in an astrophysical search.
Control parameters:
$\gamma=1.0\times 10^{-16}\,{\rm s^{-1}}$, 
$\sigma=3.7\times 10^{-10}\,{\rm s^{-3/2}}$.
}
	\label{fig:vit3}
\end{figure*}

Figure \ref{fig:vit3} illustrates the output of Versions I and III of the HMM
for three typical, injected signals with $h_0/10^{-26} = 1.7$, $1.3$, and $1.1$.
The strongest signal is detectable by both versions of the HMM,
the intermediate signal is detectable by Version III only,
and the weakest signal is detectable by neither version.
The figure displays the frequency path
that best matches the injected $f_\ast(t)$.
When the signal is detected, 
the best-matching frequency path is also the optimal HMM path, 
i.e.\ the frequency component of $Q^\ast(O)$.
The frequency is recovered accurately,
with root mean square errors of
$\varepsilon_{f_\ast} = 6.5\times 10^{-7}\,{\rm Hz} = 1.1 \Delta f_{\rm drift}$
and
$\varepsilon_{f_\ast} = 5.9\times 10^{-7}\,{\rm Hz} = 1.0 \Delta f_{\rm drift}$
for Version III in Figures \ref{fig:vit3}(a) and \ref{fig:vit3}(b) respectively.
Note that the injected $f_\ast(t)$ traces a piecewise-parabolic path,
because $\ddot{f}_\ast(t)$ is piecewise-constant (see Section \ref{sec:vit5a}).
In contrast, the frequency path recovered by Version III of the HMM,
which obeys the Ornstein-Uhlenbeck transition probabilities
in Appendix \ref{sec:vitappb},
is piecewise-constant in the figure,
because the HMM jumps between discrete frequency bins 
of width $\Delta f_{\rm drift}$.

\begin{figure*}
	\centering
	\subfigure[]
	{
		\label{fig:vitappd1a}
		\scalebox{0.65}{\includegraphics{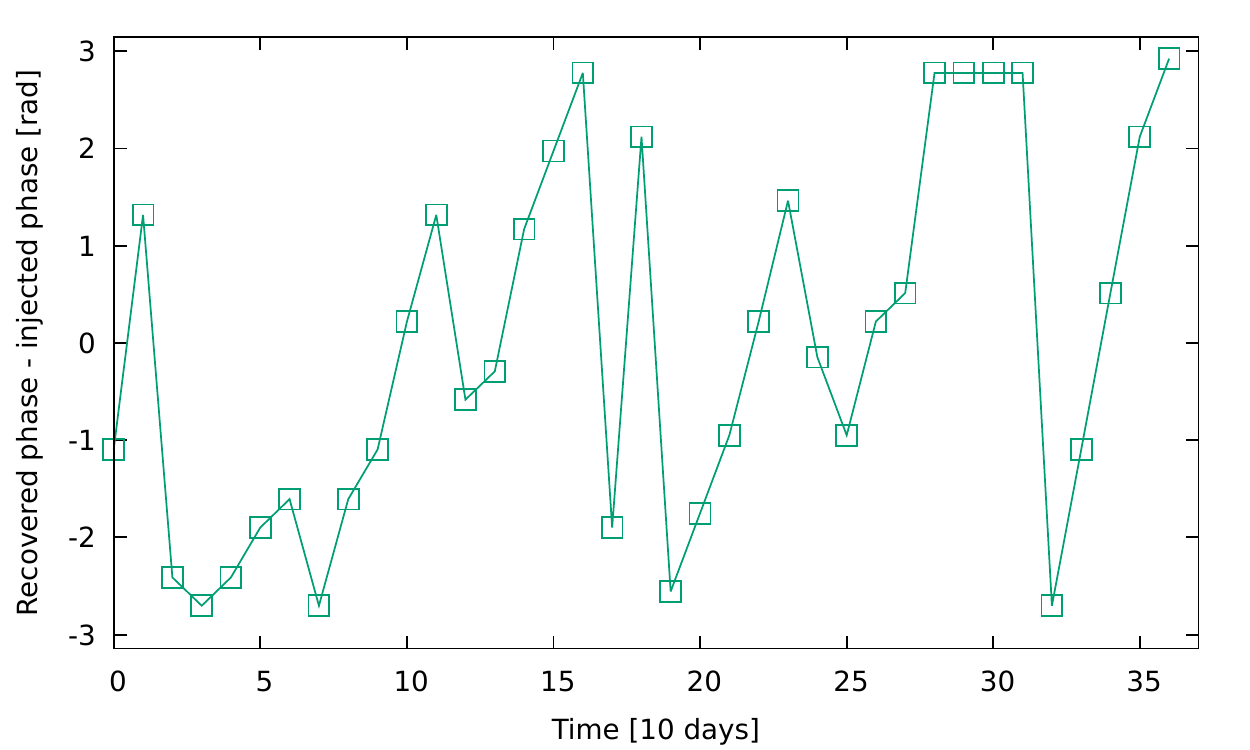}}
	}
	\subfigure[]
	{
		\label{fig:vitappd1b}
		\scalebox{0.65}{\includegraphics{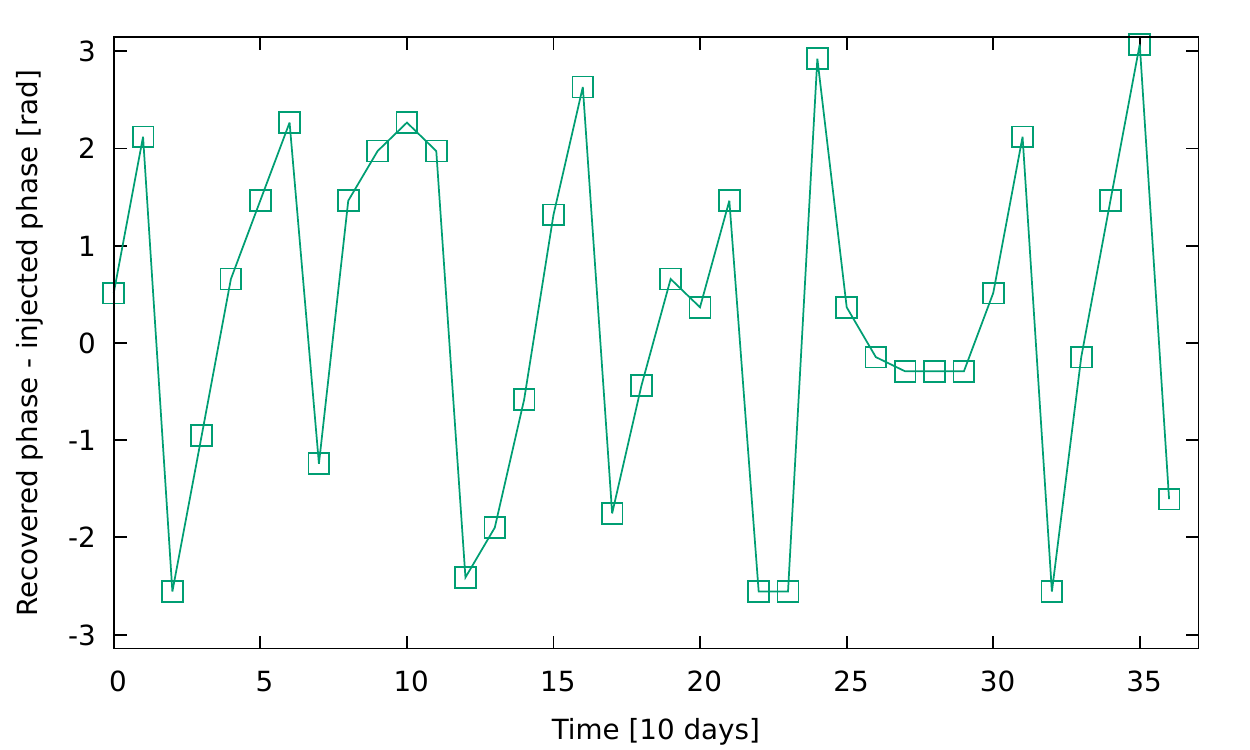}}
	}
	\subfigure[]
	{
		\label{fig:vitappd1c}
		\scalebox{0.65}{\includegraphics{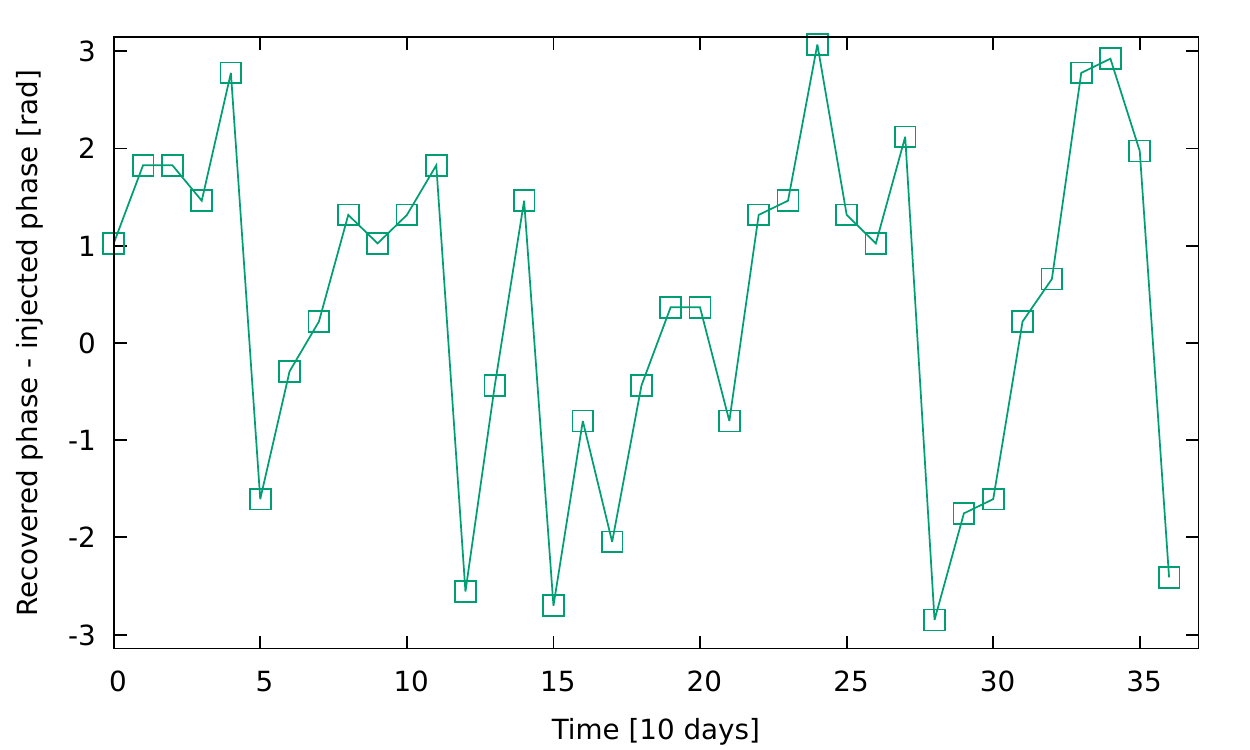}}
	}
	\caption{
Accuracy of HMM phase reconstruction. Absolute pointwise phase error (in rad) between the injected phase 
and the phase recovered by Version III of the HMM
for the three isolated sources in Figure \ref{fig:vit3},
plotted in the range $[-\pi,\pi]$ versus time (in units of $10\,{\rm days}$).
Panels (a), (b), and (c) correspond one-to-one to the panels
in Figure \ref{fig:vit3}.
}
	\label{fig:vitappd1}
\end{figure*}

Figure \ref{fig:vitappd1} displays the absolute error between 
the injected and recovered phase as a function of time for the
three isolated sources studied in Figure \ref{fig:vit3}.
Superficially the phase reconstruction in Figure \ref{fig:vitappd1}
looks worse than the corresponding frequency reconstruction 
in Figure \ref{fig:vit3}.
The ${\cal B}$-statistic concentrates the signal power
into at most two adjacent frequency bins yet spreads it out over multiple
phase bins.
As seen in Figure \ref{fig:vit2}(a),
${\cal B}(f_0,\Phi_0)$ is nearly a delta function in frequency 
(like the ${\cal F}$- and ${\cal J}$-statistics in Versions I and II of the HMM)
but has full-width half-maximum $\approx \pi$ in phase.
Nevertheless, although the phase tracking is imperfect,
it delivers improved sensitivity on balance,
if one compares Figure \ref{fig:vit3}(a)
with Figure \ref{fig:vit3}(b) for example.
This improvement does not occur simply because Version III of the HMM
uses the ${\cal B}$-statistic, which in its phase-maximized form
is $\approx 5$ per cent more sensitive than the ${\cal F}$-statistic
(see Section \ref{sec:vit4c}).
\cite{Prix2009}
To verify this, 
we repeat the tests in Figures \ref{fig:vit3} and \ref{fig:vitappd1}
while artificially scrambling the phase,
i.e.\ randomizing $\Phi_\ast(t_n)$ at every HMM step
while keeping $f_\ast(t)$ continuous as in Section \ref{sec:vit5a}.
Phase randomization converts the Version III detection of the injection
with $h_0 = 1.3\times 10^{-26}$ into a nondetection
while having no effect on the Version I results.

When the signal is not detected, $Q^\ast(O)$ is clearly wrong,
e.g.\ $\varepsilon_{f_\ast} = 1.2\times 10^{-2}\,{\rm Hz}
 = 2.1\times 10^4 \Delta f_{\rm drift}$ 
for Version III in Figure \ref{fig:vit3}(c).
The agreement looks better in the figure but artificially so.
The minimum-$\varepsilon_{f_\ast}$ paths plotted in the figure
turn out to be the sixth, 20-th, and 411-th Viterbi paths 
[i.e.\ not $Q^\ast(O)$]
for the nondetections using
Version I in Figure \ref{fig:vit3}(b) and
Versions I and III in Figure \ref{fig:vit3}(c) respectively.
While these do lie within a few frequency bins of $f_\ast(t)$ by chance,
they are of no practical use in an astrophysical search,
where the true $f_\ast(t)$ is unknown.
The optimal path is plotted whenever possible in Figure \ref{fig:vit3}
but it always lies far outside the border of the plot,
when the signal is not detected.

The PDFs of $\ln {\cal B}$ in pure noise and for a relatively strong injection 
are compared in Appendix \ref{sec:vitappea} for completeness.
They do not follow a chi-squared distribution,
unlike the ${\cal F}$-statistic,
because marginalizing over $\psi$, $\cos\iota$, and $h_0$
in (\ref{eq:vit38}) implicitly enforces constraints between
the amplitudes in (\ref{eq:vit20}),
so that $\ln {\cal B}$ is not a sum of independent squares.

\subsection{Detection strategy
 \label{sec:vit5c}}
We assess the performance of Version III of the HMM
within the Neyman-Pearson framework applied to other continuous wave
search pipelines developed by the LIGO Scientific Collaboration.
\cite{Riles2013}
Specifically, we generate receiver operating characteristic (ROC) curves
for a range of $h_0$ and $N_T$ values,
generalizing the tests in Ref.\ \cite{Suvorova2018} to include the
time-dependent antenna beam-pattern functions $a(t)$ and $b(t)$.
The aims of the exercise are:
(i) to characterize the sensitivity given 
user-selected false alarm and false dismissal probabilities, 
denoted by $P_{\rm a}$ and $P_{\rm d}$ respectively;
and (ii) to develop a practical recipe for how to subdivide
the full data set (duration $T_{\rm obs}$) into $N_T$ segments
of duration $T_{\rm drift}$.

To generate a ROC curve, 
i.e.\ a graph of $1-P_{\rm d}$ versus $P_{\rm a}$,
we must first define precisely what a detection means.
This is not trivial for HMM-based algorithms.
In Versions I and II of the HMM,
the probability that a Viterbi path terminates in a particular
frequency bin is correlated with the termination probability
for the $2N_T$ nearest bins,
because HMM paths terminating in neighboring bins 
share common subpaths in general.
The problem worsens in Version III of the HMM,
where the tails of 
$p(t_{n+1},f_{\ast j},\Phi_{\ast k})$ in (\ref{eq:vit14})
extend outside the range 
$|f_{\ast j}-f_{\ast}(t_n)| \leq \Delta f_{\rm drift}$
and wrap through $2\pi$ in phase,
as calculated in Appendix \ref{sec:vitappb}
(see also Section \ref{sec:vit2c}).
For example, the chance of encountering a false alarm within $\sim N_T$ bins
of another false alarm is higher than encountering it elsewhere.

Several valid ways exist to handle the above correlations.
In this paper, we adopt the following approach.
First, we divide the full search space into disjoint parcels 
of width $2N_T \Delta f_{\rm drift}$ in frequency and $2\pi$ in phase,
which we call `blocks'.
Each block contains $2N_T N_\Phi$ frequency-phase bins.
(We check that the results do not change significantly,
if the frequency width of the blocks is $k N_T \Delta f_{\rm drift}$ 
with $k \gtrsim 2$, in Appendix \ref{sec:vitappe}.)
Starting with multiple realizations of pure noise (i.e.\ $h_0=0$),
we calculate 
\begin{eqnarray}
 S_i 
 & = &
 \max_{|i'- i| \leq N_T} \max_{0\leq \Phi_{\ast''} \leq 2\pi}
 \nonumber \\
 & & 
 \ln \Pr[Q^\ast(O)|O; q^\ast(t_{N_T})=(f_{\ast i'},\Phi_{\ast''})]
\label{eq:vit45}
\end{eqnarray}
in the block centered on the $i$-th frequency bin.
In (\ref{eq:vit45}), $S_i$ is the HMM log likelihood for the optimal path $q^\ast(t)$
terminating at a given frequency-phase bin,
$q^\ast(t_{N_T})=(f_{\ast i'}, \Phi_{\ast ''})$,
maximized over all the frequency-phase bins in the block centered at frequency $f_{\ast i}$,
with
$| f_{\ast i'} - f_{\ast i} | \leq N_T \Delta f_{\rm drift}$
and
$0 \leq \Phi_{\ast ''} \leq 2\pi$.
We call $S_i$ the `block score'
and write it as $S$ henceforth as shorthand.
\footnote{
The block score does not equal the Viterbi score used in previous work,
\cite{Suvorova2017,Abbott2017ViterbiO1}
e.g.\ equations (29)--(31) in Ref.\ \cite{Suvorova2017}.
The latter quantity is defined as the number of standard deviations
that $\ln \Pr[Q^\ast(O)|O; q^\ast(t_{N_T})=q_i]$ in the $i$-th bin
stands away from the mean,
where the mean and standard deviation are computed over
the full search band (width $B$) for one realization.
}
We then define a threshold $S_{\rm th}(f)$,
where $f$ is the central frequency of the block,
such that an analyst-selected fraction $P_{\rm a}$ of the realizations are false alarms,
i.e.\ they return 
$S >S_{\rm th}(f)$.
(The dependence on $f$ is weak.)
We then repeat the exercise after injecting a signal $h_0 > 0$
into multiple noise realizations.
A block with $S > S_{\rm th}(f)$ is flagged as a candidate.
If any subset of the frequency component of the injected path,
$\{ f_\ast(t_1),\dots,f_\ast(t_n) \}$,
overlaps with the block, 
the candidate counts as a successful detection;
otherwise the candidate is a false alarm.
\footnote{
There is no advantage in also testing for phase overlap with
$\{ \Phi_\ast(t_1),\dots,\Phi_\ast(t_n) \}$,
because ${\cal B}(f_0,\Phi_0)$ is a broad function of $\Phi_0$;
see Figure \ref{fig:vit2}.
}
We check below that the results do not change significantly,
if we require a minimum of (say) half the injected path
to overlap with the block.
Conversely, a false dismissal occurs,
when zero candidates overlap even partially with
the one or two blocks containing the injected signal.
\footnote{
It is always possible that the highest $S$ value in a block is a 
false alarm, while the second-highest (say) is a real signal,
because nearby HMM paths are correlated.
In practice it is imprudent to claim a detection in a 
genuine, astrophysical search under such circumstances; 
the pragmatic response is to wait for more data.
}

Sample histograms of the block score $S$ in (\ref{eq:vit45}) are presented
in Appendix \ref{sec:vitappea} as a validation test.
Noise-only and noise-plus-injection histograms are visibly separate,
when the detection threshold is exceeded,
demonstrating the discriminating power of the HMM.
The PDFs of $S$ and $\ln {\cal B}$ have different functional forms,
brought about 
by the maximization steps in the Viterbi algorithm and (\ref{eq:vit45}).
\cite{Suvorova2017}

Continuous wave searches are typically subdivided into sub-bands
of width
$\Delta f_{\rm sub} \sim 1\,{\rm Hz}$ ($0.6\,{\rm Hz}$ in this paper).
Sub-bands are a housekeeping device
to handle the practicalities of data management 
(e.g.\ storage and input-output overhead on a compute cluster).
They are not the same as blocks,
which are logical units in the detection strategy above.
It is therefore necessary to convert the block-based false alarm
probability, $P_{\rm a}$, to a sub-band-based false alarm probability,
$P_{\rm a}'$, using the binomial theorem, 
viz.\ $P_{\rm a}' = 1- (1-P_{\rm a})^{N'}$
with $N' = \Delta f_{\rm sub} / (2 N_T \Delta f_{\rm drift})$.
\cite{Sammut2014}
Note that $S_{\rm th}(f)$ is a slow function of $f$ over $\sim 1\,{\rm Hz}$,
so it is usually good enough to use its midpoint value
across the whole sub-band.
\cite{Sammut2014,SCO-X1-2015}
The above approach mimics the one adopted in previous searches
for Sco X$-$1 with the Sideband algorithm,
where frequency bins are correlated over windows of width
$(2M'+1) \Delta f_{\rm drift}$,
i.e.\ the width of the Bessel comb of orbital sidebands
\cite{Sideband-Messenger2007,Sammut2014,Messenger2015,SCO-X1-2015}.
The threshold $S_{\rm th}(f)$ is also a function of $N_T$,
as discussed in Appendix \ref{sec:vitappeb}.

\subsection{ROC curves
 \label{sec:vit5d}}
\begin{figure*}
	\centering
	{
		\scalebox{0.65}{\includegraphics{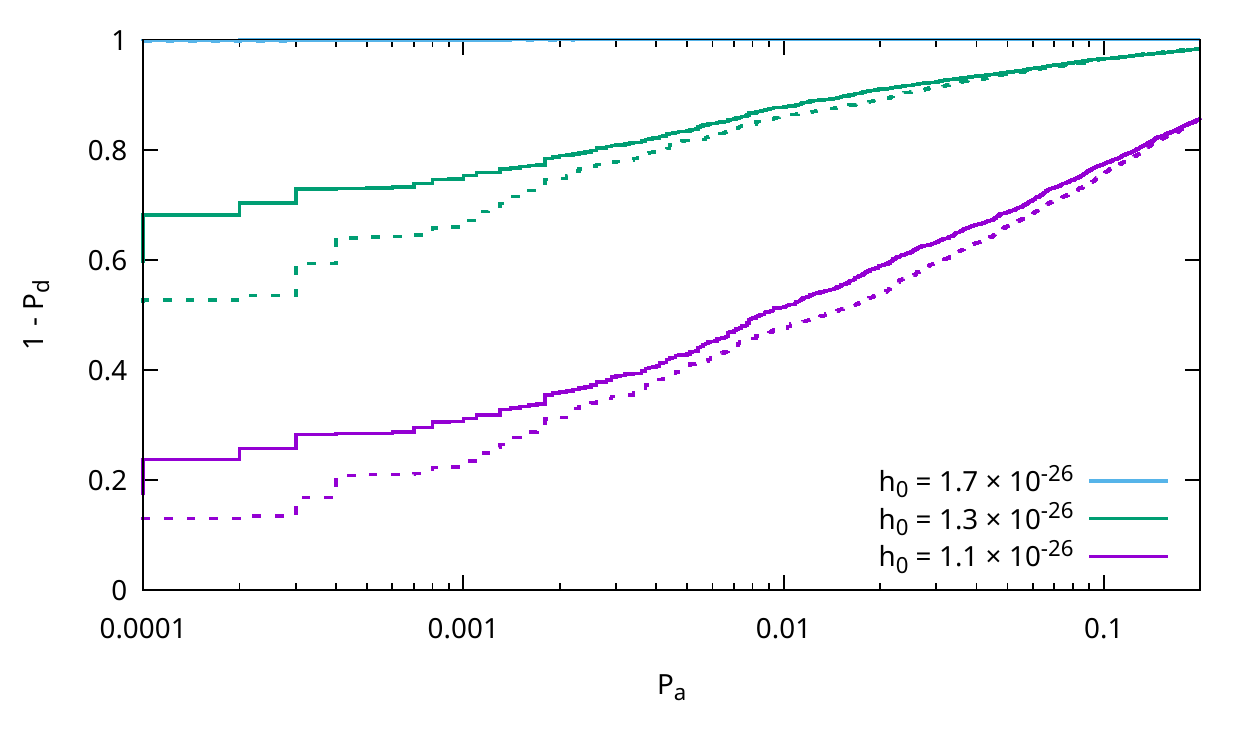}}
	}
	\caption{
Comparative HMM performance for an isolated source.
ROC curves for $h_0/10^{-26} = 1.7$ (blue curve; hidden under top border),
1.3 (green curve), and 1.1 (purple curve)
and the source parameters in Table \ref{tab:vit1}.
The false alarm probability $P_{\rm a}$
and detection probability $1-P_{\rm d}$
are plotted on the horizontal and vertical axes respectively.
Solid and dashed curves correspond to HMM Versions III and I respectively
with $T_{\rm drift}=10\,{\rm d}$ and $N_T=37$.
Control parameters:
$\gamma=1.0\times 10^{-16}\,{\rm s^{-1}}$, 
$\sigma=3.7\times 10^{-10}\,{\rm s^{-3/2}}$.
Realizations: $10^4$ per curve.
}
	\label{fig:vit5}
\end{figure*}

A key question for any detection algorithm is how the trade-off between
$P_{\rm a}$ and $P_{\rm d}$ adjusts,
as the SNR changes.
To this end, we present ROC curves in Figure \ref{fig:vit5} for
$h_0/10^{-26} = 1.7$, $1.3$, and $1.1$,
$S_h(f_0)^{1/2} = 4\times 10^{-24} \, {\rm Hz}^{-1/2}$,
$T_{\rm drift}=10\,{\rm d}$, $N_T=37$,
and the source parameters in Table \ref{tab:vit1},
adhering to the detection strategy in Section \ref{sec:vit5c}.
Results from Versions III and I of the HMM are plotted as
solid and dashed curves respectively.
The Version III curve for $h_0=1.7\times 10^{-26}$ overlaps
with the top border of the figure and is invisible.
The Version III  curve for $h_0=1.3\times 10^{-26}$ gives $P_{\rm d} \approx 0.1$
for $P_{\rm a}=10^{-2}$,
a popular combination in published LIGO searches,
e.g.\ Ref.\ \cite{Abbott2017ViterbiO1}.
In comparison, Version I of the HMM achieves the same $(P_{\rm a},P_{\rm d})$
combination for $h_0 \approx 2\times 10^{-26}$,
i.e.\ its sensitivity is $\approx 1.5$ times lower.
\cite{Suvorova2016}
Version III of the HMM is a fairly reliable detection algorithm
even at low false alarm probabilities,
with $P_{\rm d} < 0.4$ for $P_{\rm a} \geq 10^{-4}$.
The detection probability for $P_{\rm a}=10^{-2}$
drops below $1-P_{\rm d} = 0.5$ for $h_0 \leq 1.1\times 10^{-26}$.

A practical task when applying the HMM is to estimate in advance,
how its performance scales with the volume of data available,
and how the data and parameter space should be subdivided to maximize performance.
Appendix \ref{sec:vitappeb} quantifies how the detection probability
scales with $N_T$ under two  practical scenarions:
(i) $T_{\rm drift}$ is fixed, 
so that the volume of data increases, as $N_T$ increases;
and (ii) $T_{\rm obs}$ is fixed,
so that a fixed volume of data is subdivided into more coherent segments,
as $N_T$ increases.
In scenario (i),
$1-P_{\rm d}$ rises monotonically with $N_T$, as expected.
In scenario (ii), $1-P_{\rm d}$ peaks,
when $T_{\rm obs} / N_T$ matches the characteristic time-scale over which
$f_\ast(t)$ fluctuates intrinsically,
also as expected.
The block score threshold $S_{\rm th}$ is calculated versus $N_T$
for both scenarios.
Appendix \ref{sec:vitappec} checks for completeness, 
that the ROC curves are insensitive to how the blocks are partitioned.
It is found that $P_{\rm d}$ changes by $\leq 3$ per cent at fixed $P_{\rm a}$
(with $10^{-4} \leq P_{\rm a} \leq 1$)
for block bandwidths $2k N_T \Delta f_{\rm drift}$ 
in the range $0.243 \leq k \leq 2.00$,
independent of the absolute position of the leftmost bin in the block.

\subsection{Accuracy
 \label{sec:vit5e}}
Previous numerical experiments with Versions I and II of the HMM
demonstrate that the tracking accuracy is bounded by the Nyquist criterion.
\cite{Suvorova2016,Suvorova2017,Suvorova2018}
When an injected signal is detected successfully, 
the root mean square error integrated along the path satisfies
$\varepsilon_{f_\ast} \lesssim \Delta f_{\rm drift}$,
whereas one typically finds
$\varepsilon_{f_\ast} \gg \Delta f_{\rm drift}$
for false alarms.
The representative examples in Figure \ref{fig:vit3} suggest
that this remains true for Version III of the HMM,
with
$\varepsilon_{f_\ast}/\Delta f_{\rm drift} = 1.0$ (detection), 
1.1 (detection), and $2\times 10^4$ (nondetection)
for $h_0/10^{-26} = 1.7$, 1.3, and 1.1 respectively.
Versions I and III are equally accurate in Figure \ref{fig:vit3}(a),
for example,
with
$\varepsilon_{f_\ast} \lesssim \Delta f_{\rm drift}$.
The tendency for Version III to dwell somewhat longer in certain
frequency bins follows from $A_{q_j q_i}$ in Figure \ref{fig:vit1}.

\begin{figure*}
	\centering
	\subfigure[]
	{
		\label{fig:vit9a}
		\scalebox{0.65}{\includegraphics{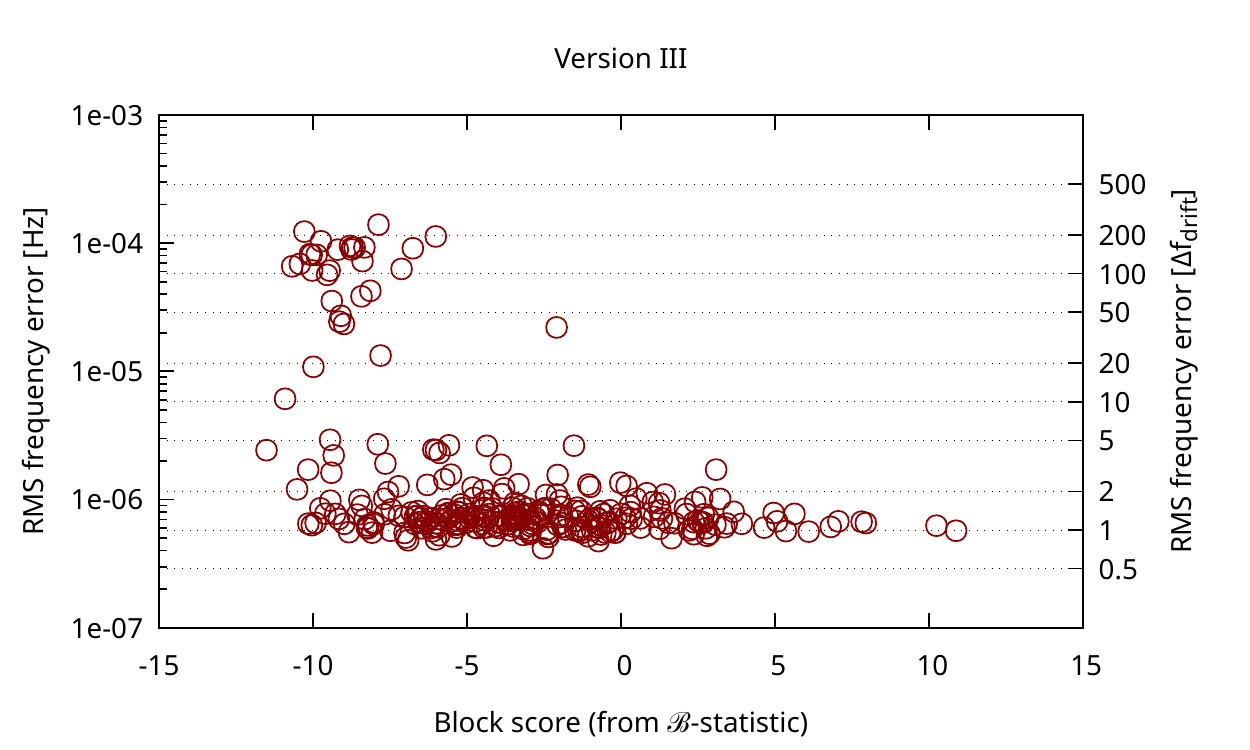}}
	}
	\subfigure[]
	{
		\label{fig:vit9b}
		\scalebox{0.65}{\includegraphics{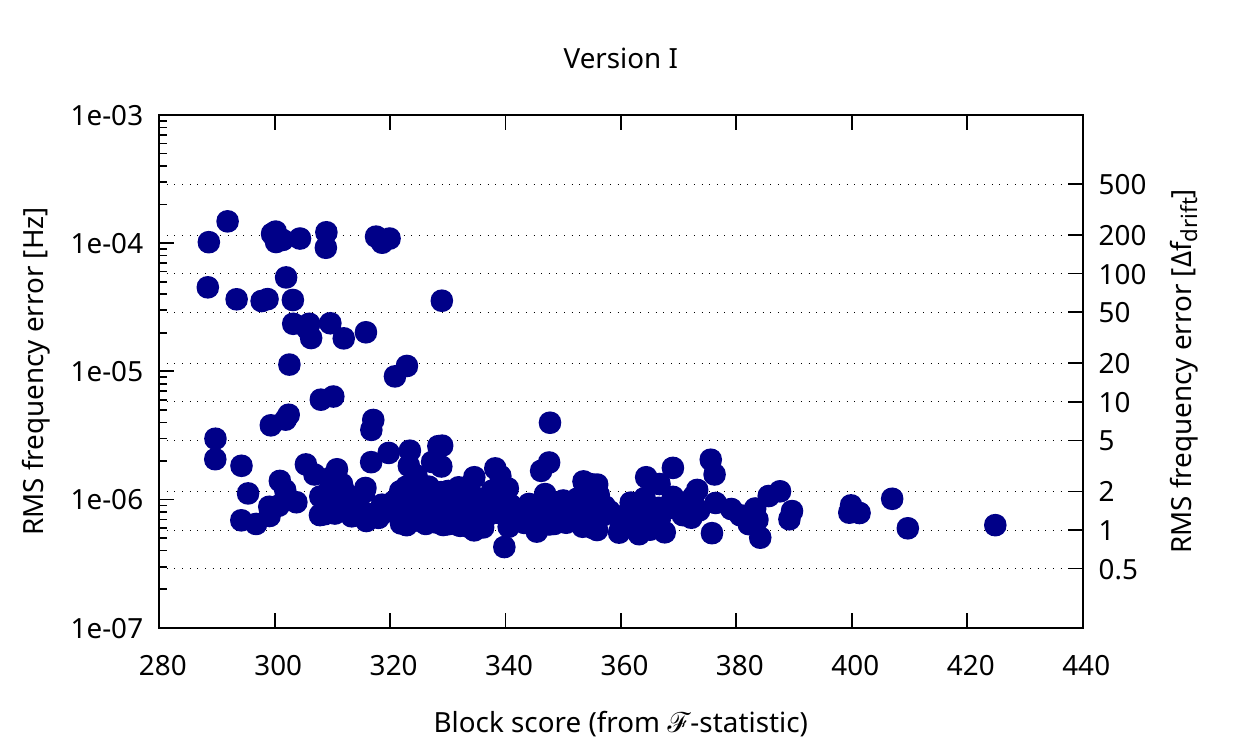}}
	}
	\caption{
Tracking accuracy of the HMM.
Root mean square frequency error $\varepsilon_{f_\ast}$
(left axis in units of Hz,
right axis in units of $\Delta f_{\rm drift}=5.8\times 10^{-7}\,{\rm Hz}$)
integrated along the optimal path in the highest-ranked block
versus the block score $S$.
(a) Version III (red, open circles)
with $N_T=37$ and $T_{\rm drift}=10\,{\rm d}$.
(b) Version I (blue, filled circles)
with $N_T=37$ and $T_{\rm drift}=10\,{\rm d}$.
Realizations: $3\times 10^2$ per panel.
Each realization comprises 20 contiguous, 37-bin blocks,
one of which contains an injected signal with
$h_0=1.3\times 10^{-26}$
and the source parameters in Table \ref{tab:vit1}.
The block scores in (a) and (b)
should not be compared as they arise from different statistics
(${\cal B}$ and ${\cal F}$ respectively).
}
	\label{fig:vit9}
\end{figure*}

We quantify the tracking accuracy systematically through Figure \ref{fig:vit9},
which displays $\varepsilon_{f_\ast}$ for the optimal path in 
the highest-ranked block 
against the block score $S$.
Versions III and I of the HMM are displayed in Figures \ref{fig:vit9}(a)
and \ref{fig:vit9}(b) respectively.
The plotted symbols, each corresponding to one realization,
separate into two clusters:
detections at the bottom right,
with $S \gtrsim S_{\rm th}(f)$ 
and $\varepsilon_{f_\ast}\lesssim \Delta f_{\rm drift}$,
and nondetections at the top left,
with $S \lesssim S_{\rm th}(f)$ 
and $\varepsilon_{f_\ast}\gg \Delta f_{\rm drift}$.
A handful of points form a bridge between the clusters,
because a few realizations
produce false alarms with $S > S_{\rm th}(f)$ but
$\varepsilon_{f_\ast}\gg \Delta f_{\rm drift}$,
e.g.\ the point with $S\approx -2.4$ and 
$\varepsilon_{f_\ast}\approx 2.5\times 10^{-5}\,{\rm Hz}$ 
in Figure \ref{fig:vit9}(a).
These accidents are expected;
phase consistency sometimes happens by chance in the noise
along a path with fortuitously high ${\cal B}$ values.
Occasionally the tracker achieves a good match
with $\varepsilon_{f_\ast}\lesssim \Delta f_{\rm drift}$
even for $S \lesssim S_{\rm th}(f)$,
corresponding to a false dismissal in a real search.
About 5 per cent of the latter events occur accidentally,
when the signal block happens to rank highest
(out of 20 blocks in Figure \ref{fig:vit9})
due to features in the noise (even with $h_0=0$).
Note that no threshold is applied explicitly in constructing Figure \ref{fig:vit9},
although implicitly $S_{\rm th}(f)$ falls near the value of $S$
below which $\varepsilon_{f_\ast}\gg \Delta f_{\rm drift}$ typically occurs.

The significant uncertainty in phase tracking,
exemplified by Figure \ref{fig:vitappd1},
does not impair the accuracy of frequency tracking
reported in Figure \ref{fig:vit9},
as discussed in Sections \ref{sec:vit2c} and \ref{sec:vit5b}.
However, it does circumscribe the astrophysical questions that can be answered.
Knowing the phase evolution more accurately can help distinguish
between astrophysical emission mechanisms,
in situations where the frequency evolution is not informative enough.
A time-domain version of the HMM offers one possible way to achieve better
phase tracking, at the cost of stepping outside the well-tested frequency-domain
software infrastructure in the LAL suite.
Designing a time-domain HMM is a goal of future work.

\section{Neutron star in a binary
 \label{sec:vit6}}
We now repeat the tests in Section \ref{sec:vit5} for a neutron star
in a binary system.
The HMM structure and search procedure remain unchanged,
except that $F_{1a}(f_0)$ and $F_{1b}(f_0)$ are replaced by
$J_{1a}(f_0)$ and $J_{1b}(f_0)$ respectively in the ${\cal B}$-statistic
via (\ref{eq:vit38})--(\ref{eq:vit44}).
Appendix \ref{sec:vitapped} verifies that this replacement leads to
minimal loss of signal power;
the Doppler sidebands collapse into a single frequency bin
without discernible leakage into neighboring bins,
just like for the ${\cal J}$-statistic.
In Section \ref{sec:vit6a} we present tracking results 
for a representative sample of synthetic data.
ROC curves are discussed in Section \ref{sec:vit6b}.
In Section \ref{sec:vit6c} we plot the ${\cal B}$-statistic
as a function of the orbital parameters $a_0$ and $\phi_{\rm a}$,
in order to inform the gridding strategy for future searches,
e.g.\ for LMXBs.
Versions II and III of the HMM are compared at each stage.

\subsection{Representative example
 \label{sec:vit6a}}
\begin{figure*}
	\centering
	\subfigure[]
	{
		\label{fig:vit10a}
		\scalebox{0.65}{\includegraphics{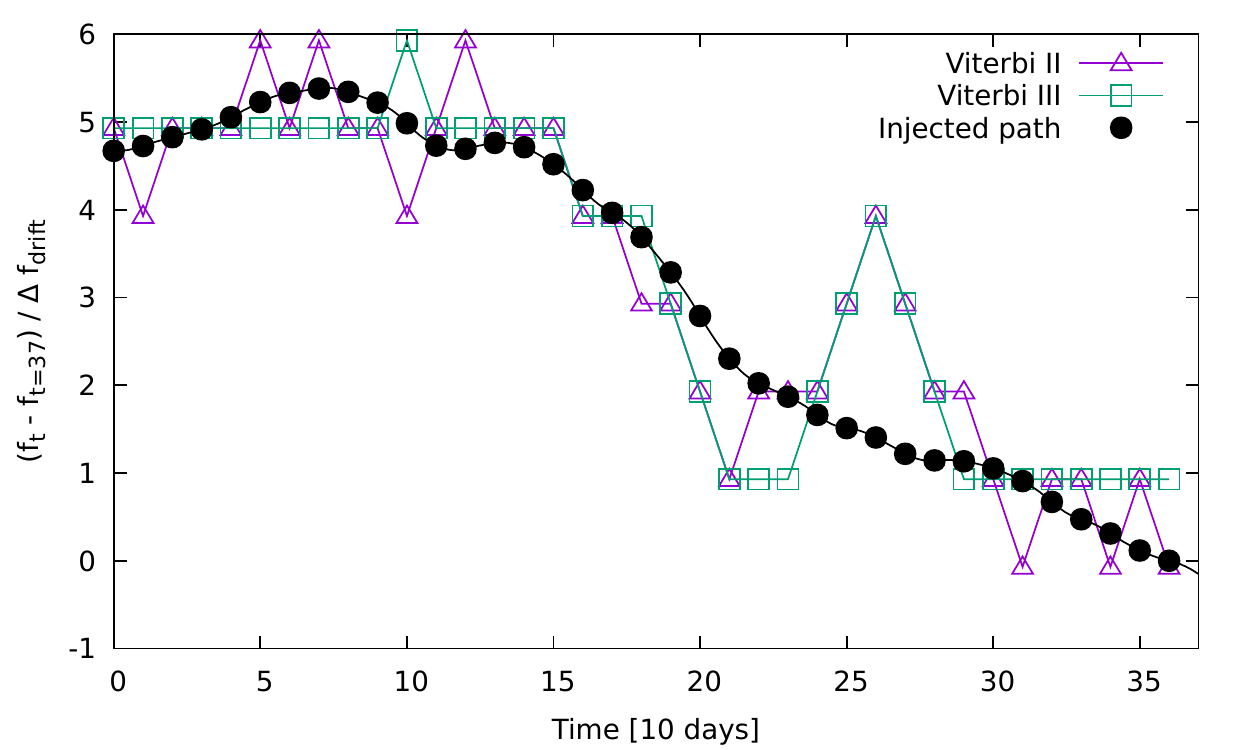}}
	}
	\subfigure[]
	{
		\label{fig:vit10c}
		\scalebox{0.65}{\includegraphics{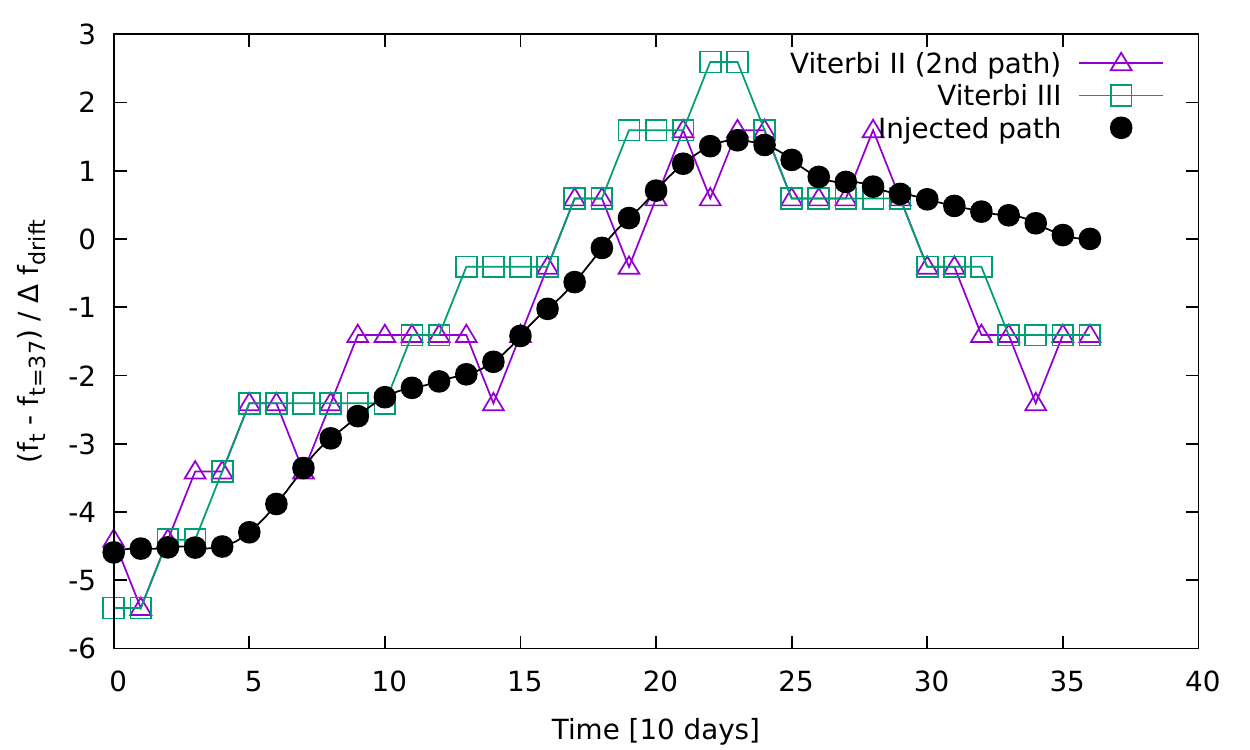}}
	}
	\subfigure[]
	{
		\label{fig:vit10e}
		\scalebox{0.65}{\includegraphics{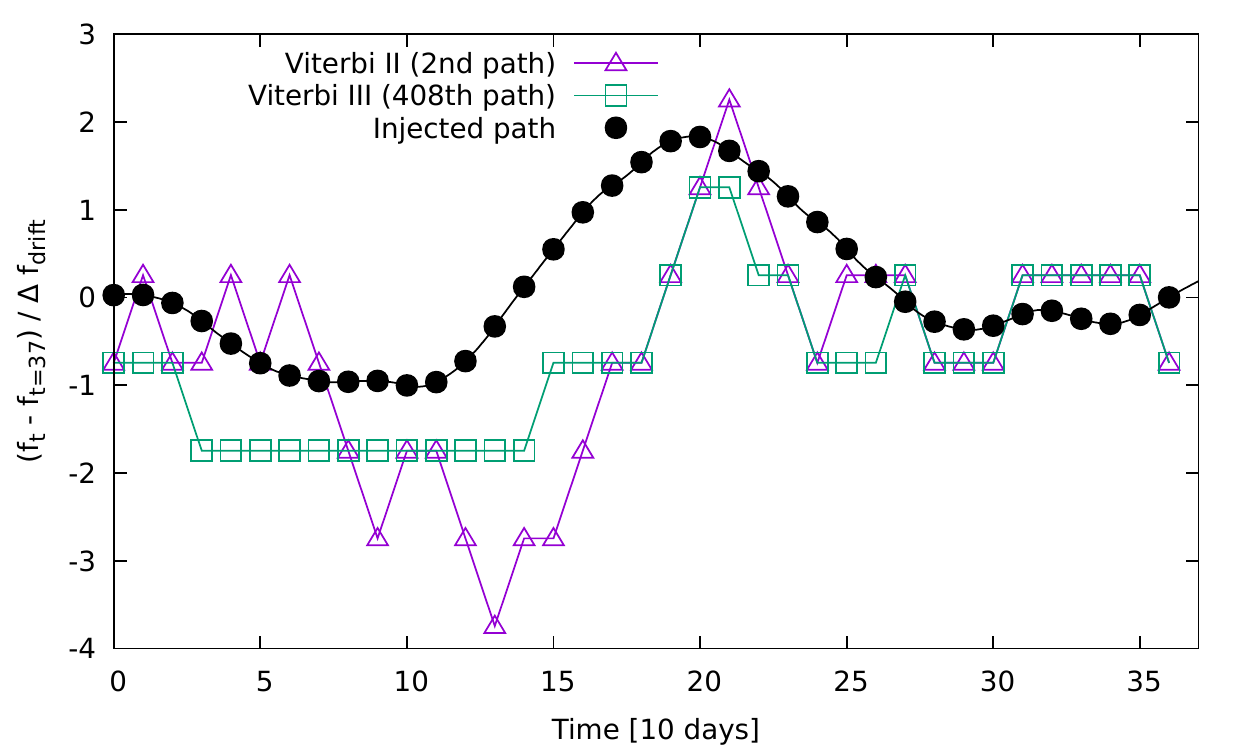}}
	}
	\caption{
Sample tracking output from
Versions II (purple curves) and III (green curves) of the HMM
for three injected signals (black curves) from a source
in a binary with parameters drawn from Tables \ref{tab:vit1} and \ref{tab:vit2} with
$h_0/10^{-26} = 1.7$ [panel (a)], 
$1.3$ [panel (b)], and $1.1$ [panel (c)],
plotted on the same axes as in Figure \ref{fig:vit3}.
The purple and green curves are the best-matching frequency paths
(with minimum $\varepsilon_{f_\ast}$);
they are not necessarily the optimal path $Q^\ast(O)$.
The optimal path matches well 
[i.e.\ within a few frequency bins of $f_\ast(t)$ for all $t$]
for Versions II and III in (a) and Version III in (b).
The optimal path matches poorly
for Version II in (b) and Versions II and III in (c);
indeed it lies outside the border of the plot.
We plot instead the paths with minimum $\varepsilon_{f_\ast}$,
viz.\ the second, second, and 408-th Viterbi paths respectively,
which lie within a few frequency bins of $f_\ast(t)$ purely by chance
but are of no practical use in an astrophysical search.
Control parameters:
$\gamma=1.0\times 10^{-16}\,{\rm s^{-1}}$, 
$\sigma=3.7\times 10^{-10}\,{\rm s^{-3/2}}$.
}
	\label{fig:vit10}
\end{figure*}

The signal phase corresponding to a binary neutron star
is given by (\ref{eq:vit25}) with $a_0\neq 0$.
Figure \ref{fig:vit10} illustrates the output of Versions II and III of the HMM
for three injected signals of the above form
with the same $h_0$ values as in Section \ref{sec:vit5a},
viz.\ $h_0 / 10^{-26} = 1.7$, $1.3$, and $1.1$.
The parameters of the binary orbit are quoted in Table \ref{tab:vit2},
with $a_0$ and $\phi_{\rm a}$ set at the midpoints of their ranges.
The stochastic component of the injected phase, $\Phi_{\rm w}(t)$,
evolves according to the algorithm in Section \ref{sec:vit5a}.

The results in Figure \ref{fig:vit10} resemble those in Figure \ref{fig:vit3}.
Both HMM versions detect the strongest signal, 
but only Version III detects the intermediate signal.
Neither detects the weakest signal.
Version III is $\approx 1.4$ times more sensitive than Version II,
and its sensitivity is approximately the same for isolated and binary sources.
\footnote{
This is consistent with previous work: 
Version II of the HMM is sensitive down to the same $h_0$ value,
$h_0 \approx 2\times 10^{-26}$,
for a binary source as Version I is for an isolated source.
}
Once the HMM fails to detect a signal, 
the optimal Viterbi path stands many bins away from the injected path 
and normally falls outside the plotted region.
The agreement in Figures \ref{fig:vit10c} and \ref{fig:vit10e}
looks better than it actually is,
because we plot the minimum-$\varepsilon_{f_\ast}$ paths, 
which turn out to be the second, second, and 408-th Viterbi paths
for the nondetections using Version II in Figure \ref{fig:vit10c} 
and Versions II and III in Figure \ref{fig:vit10e} respectively.
Such coincidental successes are useless in an astrophysical search,
where the true $f_\ast(t)$ is unknown.
Similarly, it may seem that Version II outperforms Version III 
on the $h_0=1.1\times 10^{-26}$ injection, because the 
minimum-$\varepsilon_{f_\ast}$ paths
are the second (Version II) versus the 408-th (Version III).
Again this is misleading: paths other than the first are not ranked
consistently by the Viterbi algorithm, 
and besides Version III has 32 times more paths than Version II
(and a different bin numbering system) because it tracks both
$f_\ast$ and $\Phi_\ast$.

The phase component of $Q^\ast(O)$ is discussed briefly for completeness
in Appendix \ref{sec:vitappd}.

\begin{table*}
	\centering
	\setlength{\tabcolsep}{6pt}
	\begin{tabular}{llll}
		\hline
		\hline
		Parameter & Value & Units & Description\\
		\hline
		$P$ & 68023.7 & s & Orbital period\\
		$a_0$ & [1.26,1.62] & lt-s & Projected orbital semimajor axis \\
                $\phi_{\rm a}$ & $[0,2\pi]$ & --- & Reference orbital phase \\
		$e$ & 0.0 & --- & Orbital eccentricity \\
		\hline
		\hline
	\end{tabular}
	\caption{Orbital parameters used to create the synthetic data 
for the binary sources analysed in Section \ref{sec:vit6}. 
}
	\label{tab:vit2}
\end{table*}

\subsection{ROC curves
 \label{sec:vit6b}}
\begin{figure*}
	\centering
	{
		\scalebox{0.65}{\includegraphics{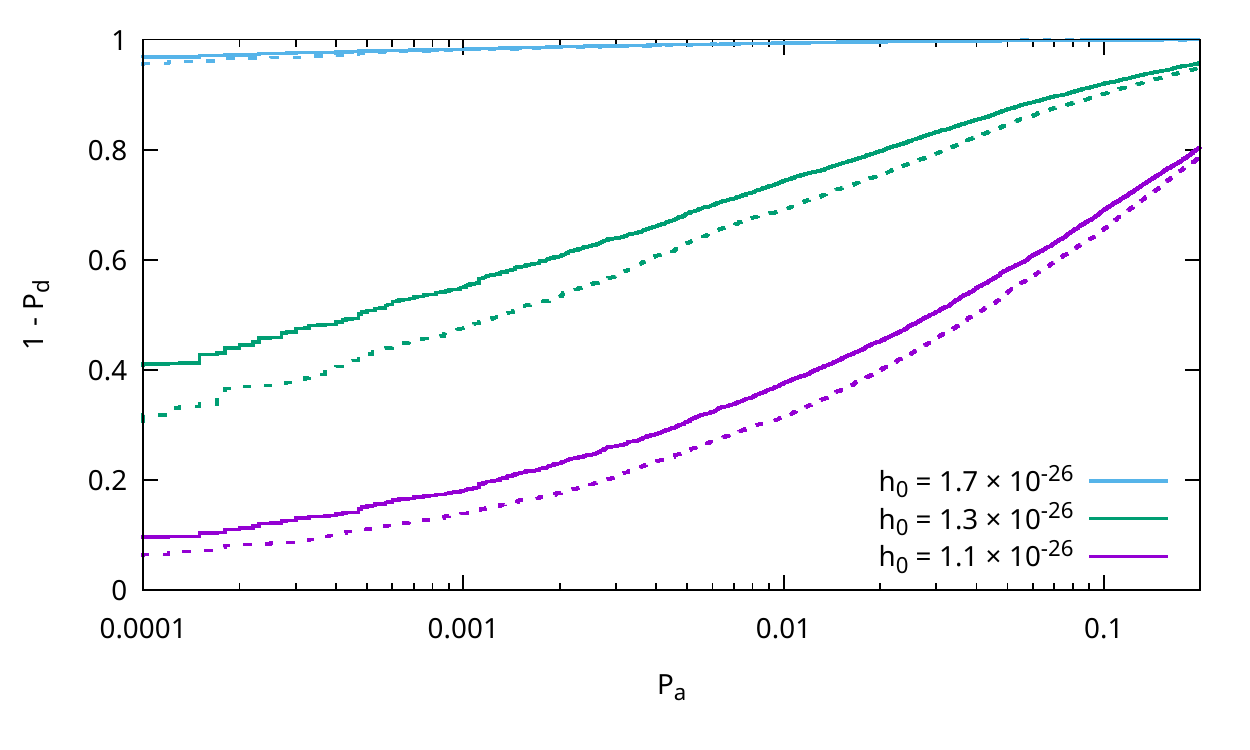}}
	}
	\caption{
Comparative HMM performance for a source in a binary.
ROC curves for sources with $h_0/10^{-26} = 1.7$ (blue curve),
1.3 (green curve), and 1.1 (purple curve)
and the source parameters in Tables \ref{tab:vit1} and \ref{tab:vit2}.
The false alarm probability $P_{\rm a}$
and detection probability $1-P_{\rm d}$
are plotted on the horizontal and vertical axes respectively.
Solid and dashed curves correspond to HMM Versions III and II respectively.
Control parameters:
$\gamma=1.0\times 10^{-16}\,{\rm s^{-1}}$, 
$\sigma=3.7\times 10^{-10}\,{\rm s^{-3/2}}$.
Realizations: $10^4$ per curve.
}
	\label{fig:vit11}
\end{figure*}

In order to characterize the sensitivity of the HMM systematically,
we compute ROC curves for the same three signal amplitudes
in Figure \ref{fig:vit10},
viz.\ $h_0 / 10^{-26} = 1.7$, $1.3$, and $1.1$.
The results are plotted in Figure \ref{fig:vit11},
where solid and dashed curves correspond to Versions III and II of the HMM
respectively.
In the regime of practical interest, viz.\
$5\times 10^{-3} \leq P_{\rm a} \leq 2\times 10^{-1}$,
Version III of the HMM delivers a detection probability
$\approx 0.05$ higher than Version II at the same $P_{\rm a}$,
a significant advantage when operating near the detection limit.
Replacing $F_{1a}(f_0)$ and $F_{1b}(f_0)$
with $J_{1a}(f_0)$ and $J_{1b}(f_0)$ in the ${\cal B}$-statistic
leads to similar tracking performance
for isolated and binary sources,
although there is some modest loss of sensitivity in the latter case.
For example,
a detection probability of $\approx 0.75$ is achieved in Figure \ref{fig:vit11}
for a binary source with $h_0=1.3\times 10^{-26}$, 
given $P_{\rm a}=10^{-2}$,
compared to $\approx 0.90$ for an isolated source with the same $h_0$ 
in Figure \ref{fig:vit5}.
\cite{Abbott2017ViterbiO1}
This is because the Jacobi-Anger decomposition 
(\ref{eq:vit31}) and (\ref{eq:vit32})
accounts for the binary motion imperfectly when combined
with the ${\cal B}$-statistic,
due to some covariance between the orbital and carrier phases
in the orbital sidebands.
For $h_0 \geq 1.7\times 10^{-26}$, the performance is almost identical,
as in Appendix \ref{sec:vitapped}.

Monte Carlo simulations confirm that the performance of the HMM 
as a function of $N_T$ for $T_{\rm drift}$ or $T_{\rm obs}$ fixed
is the same as in the case of isolated sources
(see Figures \ref{fig:vit6} and \ref{fig:vit7} respectively
in Appendix \ref{sec:vitappe}).
The results are not plotted to avoid repetition.

\subsection{Sensitivity to orbital parameters
 \label{sec:vit6c}}
Electromagnetic observations normally supply
prior constraints on LMXB orbital parameters.
\cite{Watts2008,Galloway2014,Premachandra2016,Wang2018}
For many objects, including Sco X$-$1,
the electromagnetic measurement of $P$ through high-resolution
optical spectroscopy is accurate enough,
that a search over $P$ is unnecessary.
In contrast, searches over $a_0$ and $\phi_{\rm a}$ are usually required.
\cite{Leaci2015}

\begin{figure}
	\centering
	\scalebox{0.45}{\includegraphics{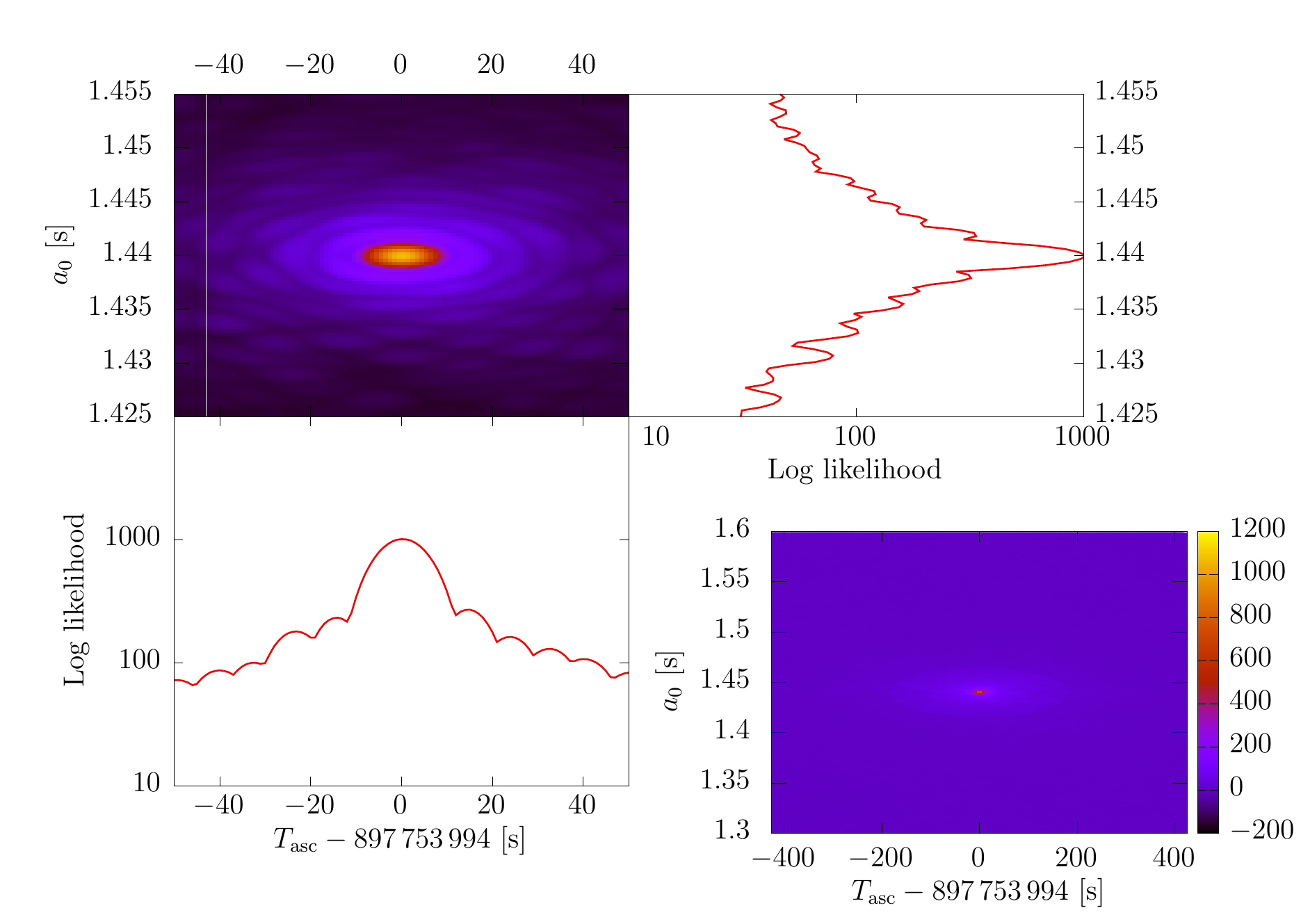}}
	\caption{
HMM performance as a function of binary orbital elements.
Log probability $\ln \Pr[Q^\ast(O)|O]$ versus $a_0$ and $T_{\rm asc}$
for a strong (${\rm SNR} \gg 1$) binary-star signal with constant $f_\star$ 
observed during $N_T=37$ 10-day segments. 
({\em Top left.})
Contours of $\ln \Pr[Q^\ast(O)|O]$ on the $T_{\rm asc}$-$a_0$ plane,
centered on the injected values $T_{\rm asc}^{\rm true}$ and $a_0^{\rm true}$.
Hot colors stand for the highest log probabilities. 
({\em Top right.})
Cross-section through the peak of 
$\ln \Pr[Q^\ast(O)|O]$ versus $a_0$ 
for $T_{\rm asc}=T_{\rm asc}^{\rm true}$.
({\em Bottom left.})
Cross-section through the peak of 
$\ln \Pr[Q^\ast(O)|O]$ versus $T_{\rm asc}$ 
for $a_0 = a_0^{\rm true}$.
({\em Bottom right.})
Zoomed out version of the top left panel.
Injection parameters:
$h_0=8\times 10^{-26}$, $f_\star=111.1\,{\rm Hz}$, 
$a_0^{\rm true}=1.44 \, {\rm lt\,s}$,
and $T_{\rm asc}^{\rm true}=897 753 994\,{\rm s}$ (arbitrary orbital phase), 
characteristic of Scorpius X$-$1;
see also Tables \ref{tab:vit1} and \ref{tab:vit2}.
}
	\label{fig:vit13}
\end{figure}

Figure \ref{fig:vit13} displays $\ln \Pr[Q^\ast(O)|O]$
for Version III of the HMM
as a function of $a_0$ and 
$T_{\rm asc} = \phi_{\rm a} P / (2\pi)+{\rm constant}$,
where $T_{\rm asc}$ is the time of ascending node.
The log probability is evaluated at the true, injected value of $f_\ast$
and maximized with respect to $\Phi_\ast$,
for a strong signal with $h_0=8\times 10^{-26}$
tracked over $N_T=37$ steps.
Starting from the panel at the bottom right of the figure,
we observe that $\ln \Pr[Q^\ast(O)|O]$ peaks strongly around the
true, injected orbital elements
$a_0^{\rm true}$ and $T_{\rm asc}^{\rm true}$.
The top left panel zooms into the peak (note the magnified scale)
and shows that it is encircled
by ``ripples'' reminiscent of a diffraction pattern.
The ripples are visible more clearly in the cross-sections at
$T_{\rm asc} = T_{\rm asc}^{\rm true}$
and
$a_0 = a_0^{\rm true}$,
graphed in the top right and bottom left panels respectively.
Both cross-sections are sinc-like,
except that the nodes do not touch zero;
$\Pr[Q^\ast(O)|O]$ is positive definite.
Qualitatively the features in Figure \ref{fig:vit13} match those
observed in Figure 4 in Ref.\ \cite{Suvorova2017}
for the ${\cal J}$-statistc HMM (Version II),
although the scales are not comparable of course.

In practice, in a search with real data,
the grid spacings in $a_0$ and $T_{\rm asc}$ 
are set according to a parameter space metric
and depend on the search frequency $f_0$.
\cite{Leaci2015}
For example, the LIGO O2 search for Sco X$-$1 with HMM Version II
employs 768 $a_0$ bins of width
$2.3\times 10^{-3}\,{\rm lt \ s}$
[with $1.45 \leq a_0 / (1\,{\rm lt\,s}) \leq 3.25$]
at $f_0 = 60\,{\rm Hz}$,
compared to 8227 $a_0$ bins of width
$2.2\times 10^{-4}\,{\rm lt \ s}$
at $f_0 = 650\,{\rm Hz}$.
\cite{Wang2018,Abbott2019ViterbiO2}
The resolution is chosen to yield a mismatch of $\leq 10\%$
in the squared SNR,
as defined by equation (5) in Ref.\ \cite{Leaci2015},
the worst case being when the signal straddles
the boundary between two bins.
Without being comparable directly,
the above approach is consistent with Figure \ref{fig:vit13}:
the squared SNR is of the same order as $\ln \Pr[Q^\ast(O)|O]$,
and $\ln \Pr[Q^\ast(O)|O]$ drops off by $\leq 10\%$ from its peak
for $|a_0 - a_0^{\rm true}| \lesssim 10^{-3} \, {\rm lt\,s}$
in the top right panel of Figure \ref{fig:vit13}
and for $|T_{\rm asc}-T_{\rm asc}^{\rm true} | \lesssim 5 \, {\rm s}$
in the bottom left panel of Figure \ref{fig:vit13}.
Convenient formulas 
for the number of $a_0$ and $T_{\rm asc}$ templates
in terms of the desired mismatch are given in Section V
of Ref.\ \cite{Leaci2015}.

\section{Sco X-1 MDC: a realistic example
 \label{sec:vit7}}

\subsection{Synthetic data
 \label{sec:vit7a}}
The Sco X$-$1 MDC is a project to compare systematically
the performance of published continuous-wave search pipelines
on a level playing field under simulated Advanced LIGO conditions.
\cite{Messenger2015}
The MDC predates HMM Versions I and II.
It evaluates the relative proficiency of five pipelines against criteria
including sensitivity, computational cost, and accuracy in parameter
estimation.
The pipelines are based on the
CrossCorr, \cite{Dhurandhar2008,Chung2011,Whelan2015}
TwoSpect, \cite{TwoSpectGoetz2011,Meadors2016}
Radiometer, \cite{RadiometerBallmer2006,Radiometer2007,Radiometer2011}
Sideband, \cite{Sideband-Messenger2007,Sammut2014} 
and Polynomial \cite{Polynomial-Putten2010} algorithms.
Method papers describing each algorithm are cited in the previous sentence.
Since the MDC was published,
two of the pipelines have completed searches
using Advanced LIGO data from O1 and O2,
\cite{Abbott2017CrossCorrO1,Abbott2019Radiometer}
as have HMM Versions I and II.
\cite{Abbott2017ViterbiO1,Abbott2019ViterbiO2}
Two other pipelines have completed searches using
Initial LIGO data from Science Run 6 (S6).
\cite{SCO-X1-2015,Meadors2017}
It should be noted that O1 and O2 do not achieve Advanced LIGO's
design sensitivity, approximated in the MDC as 
$S_h(f_0)^{1/2} \approx 4\times 10^{-24}\,{\rm Hz^{-1/2}}$
(Gaussian recolored).

The MDC enables an important check on the results in previous sections
under realistic yet controlled conditions
on a data set generated by an independent party.
Of course, the MDC is no longer closed,
as it was in its original incarnation;
the TwoSpect, Radiometer, Sideband, and Polynomial pipelines
competed blindly in Ref.\ \cite{Messenger2015},
before the injection parameters were revealed,
and the CrossCorr pipeline analysed the data in self-blinded mode,
after the injection parameters were revealed.
In this paper we preserve the etiquette of a self-blinded analysis
but note in fairness that some of the authors participated 
in previous analyses of the same data with HMM Versions I and II.
\cite{Suvorova2016,Suvorova2017}
We also note that $f_\ast(t)$ does not wander for any of the injected signals,
even though we allow for wandering in the HMM transition probabilities.
Extensive testing in previous published work demonstrates,
that the HMM delivers equal sensitivity,
whether $f_\ast(t)$ wanders or not,
as long as $T_{\rm drift}$ satisfies (\ref{eq:vit6}),
\cite{Suvorova2016,Suvorova2017}
in line with theoretical expectations.
\cite{Quinn2001}
Strictly speaking, however, the analysis in this section
checks the sensitivity and accuracy of Version III of the HMM; 
it does not test its robustness to spin wandering.
(Indeed nor did the original MDC study with the five pipelines
in Ref.\ \cite{Messenger2015}.)
A future incarnation of the MDC including spin wandering,
drawing on the analysis in Ref.\ \cite{Mukherjee2018},
is currently being prepared and should be encouraged.

The parameters of the 50 injected signals in Stage I (version 6)
of the MDC are listed in Table III
in Ref.\ \cite{Messenger2015}.
They are designed to resemble Sco X$-$1,
with $0.050 \leq 2f_\ast / (1\,{\rm kHz}) \leq 1.5$
and orbital elements similar to those measured electromagnetically.
\cite{Galloway2014,Premachandra2016,Wang2018}
Since the original MDC release,
the data for three injections,
with indexes 65, 66, and 75 in Ref.\ \cite{Messenger2015},
are no longer accessible due to human error.
They are omitted from the analysis below,
which is restricted to 47 injections.

\subsection{Search procedure
 \label{sec:vit7b}}
The analysis is conducted as follows in order to copy approximately
some of the steps in a search with real LIGO data.
\begin{enumerate}
\item
Starting from $f_0=50\,{\rm Hz}$ and defining sub-bands in increments
of $0.1\,{\rm Hz}$,
we identify the sub-band containing the injected signal.
The partition is similar to the O2 Sco X$-$1 search with Version II of the HMM,
which implemented $0.6$-${\rm Hz}$ sub-bands,
without $f_\ast$ being known of course.
\cite{Abbott2019ViterbiO2}
In effect this step is self-blinded to a good approximation,
because there are $(0.1\,{\rm Hz}) / \Delta f_{\rm drift} = 1.7\times 10^5$
frequency bins in the sub-band, 
any single one of which can contain the injected signal in principle.
\item
An orbital grid is laid out in $a_0$ and $T_{\rm asc}$
as for the HMM O2 Sco X$-$1 search.
The grid spacings in $a_0$ and $T_{\rm asc}$ are given by
$1.2\times 10^{-4} (f_0/0.3\,{\rm kHz})^{-1} \, {\rm lt\,s}$
and
$0.89 (f_0/0.3\,{\rm kHz})^{-1} (a_0/1.44\,{\rm lt\,s})^{-1} \, {\rm s}$
within the electromagnetic priors 
$1.45 \leq a_0 / (1\,{\rm lt\,s}) \leq 3.25$
and
$1164543014 \leq T_{\rm asc} / (1\,{\rm s}) \leq 1164543614$
respectively.
\cite{Wang2018,Abbott2019ViterbiO2}
The grid spacings are one quarter of what is predicted by 
the parameter space metric via
Eqs (70) and (71) in Ref.\ \cite{Leaci2015},
assuming a squared-SNR mismatch of $\leq 10\%$.
The safety factor $1/4$ is discussed further below.
Strictly speaking the grid spacing varies from one $f_0$ bin to the next,
but in practice it is kept uniform within each $0.1$-${\rm Hz}$ sub-band,
substituting the sub-band midpoint into the above formulas 
as a good approximation.
\item
A grid is also laid out in orbital period $P$,
with grid spacing
$1.0 (f_0/0.3\,{\rm kHz})^{-1} (a_0/1.44\,{\rm lt\,s})^{-1} \, {\rm s}$
involving the same safety factor 1/4 from step 2 above,
based on Eqs (70) and (71) in Ref.\ \cite{Leaci2015} 
in the regime $P \ll T_{\rm drift}$.
This is a new step.
Some of the MDC injections are not exactly at $P=68023.7\,{\rm s}$,
the central value returned by electromagnetic observations,
\cite{Galloway2014,Premachandra2016,Wang2018}
although they are close to it.
Previous MDC analyses ignore the slight mismatch,
motivated by the parameter space metric 
which implies that one $P$ template is sufficient,
because the experimental uncertainty ($\pm 0.04\,{\rm s}$)
is less than the metric-based resolution $\approx 0.2\,{\rm s}$.
\cite{Leaci2015,Suvorova2017}
They search $P=68023.7\,{\rm s}$ only and are still successful;
for example, Version II of the HMM finds all 50 injections thus.
\cite{Suvorova2017}
However Version III of the HMM,
which is more sensitive to weaker signals,
also depends more sensitively on $P$.
\item
Version III of the HMM is executed on $4\times 4\times 4$
adjacent triples $(a_0,T_{\rm asc},P)$ centered on the
injection.
(The MDC analysis is executed on a subset of the grid for
computational economy; in an astrophysical search,
we scan the whole grid.)
Each triple $(a_0,T_{\rm asc},P)$ is accompanied by an $f_0$ scan
divided into 
$(0.1\,{\rm Hz}) / \Delta f_{\rm drift} / N_T = 4671$ blocks
as described in Section \ref{sec:vit5c}.
The highest log probability among these $64\times 4671$
$(f_0,a_0,T_{\rm asc},P)$ combinations becomes the block score
according to (\ref{eq:vit45}).
\item
The root mean square frequency error $\varepsilon_{f_\ast}$
is calculated along the optimal, wandering Viterbi track 
as in Section \ref{sec:vit5e}.
Absolute, signed errors 
$\varepsilon_{a_0}$ and $\varepsilon_{T_{\rm asc}}$
are also calculated for $a_0$ and $T_{\rm asc}$ respectively 
as the injected minus recovered values for the optimal Viterbi track.
This approach is adopted deliberately to stay consistent with
previous MDC analyses,
which verify the accuracy of the top candidate in a block 
instead of quantifying the false alarm probability.
\cite{Messenger2015,Suvorova2016}
In a search with real data, one would instead compare the block score 
with a threshold set by $P_{\rm a}$
and follow up any candidates through a veto procedure.
\cite{Abbott2017ViterbiO1,Abbott2019ViterbiO2}
\end{enumerate}

\subsection{Signal detectability
 \label{sec:vit7c}}
The results of analysing the MDC data with Version III of the HMM
are presented in Table \ref{tab:vit3}.
Each line of the table corresponds to one injection,
indexed as in Ref.\ \cite{Messenger2015} (first column).
The injection parameters $f_\ast$, $a_0$, and $T_{\rm asc}$
are quoted along with the respective errors
$\varepsilon_{f_\ast}$, $\varepsilon_{a_0}$, and $\varepsilon_{T_{\rm asc}}$
in the parameter values recovered by the HMM.
Two simulated interferometers (H1 and L1) are employed,
chiefly to preserve consistency with the previous MDC analysis 
involving Version II of the HMM.
\citep{Suvorova2017}
The data start at GPS time 1230338490 and are divided into
$N_T=37$ segments with $T_{\rm drift}=10\,{\rm d}$.

Version III detects 47 out of 47 available injections.
The outcome is reassuring but not surprising.
Version II also detects every signal,
and Version III is $\approx 1.5$ times more sensitive than Version II
according to the results in Sections \ref{sec:vit5} and \ref{sec:vit6}.
The signal amplitudes are quoted in the second and third columns
of Table \ref{tab:vit3} in terms of $h_0$ and 
$h_0^{\rm eff}=h_0 2^{-1/2} 
 \{ [ (1+\cos^2\iota)/2 ]^2 + \cos^2\iota \}^{1/2}$ respectively.
The source inclination influences detectability through
the relative weighting of the plus and cross polarizations,
and $h_0^{\rm eff}$ serves as an amplitude proxy
which normalizes for this effect,
as verified in Ref.\ \cite{Suvorova2017}
(see the tests in Section V A and Figure 5 of the latter reference).
The quietest detected signals from the $h_0$ and $h_0^{\rm eff}$
perspectives are injections 90
($h_0 = 6.8\times 10^{-26}$)
and 64
($h_0^{\rm ref} = 5.7\times 10^{-26}$)
respectively.
Both lie well above the Version III sensitivity limit 
$h_0 \geq 1.3\times 10^{-26}$ 
established in Sections \ref{sec:vit5} and \ref{sec:vit6}.
The conclusions are not affected
by the absence of injections 65, 66, and 75,
which are all relatively strong
[$7.7\leq h_0 / (10^{-25}) \leq 9.3$]
and are detected easily by Versions I and II with two interferometers.

\subsection{Accuracy
 \label{sec:vit7d}}
Although Versions II and III both detect all the injections,
Version III recovers the true signal parameters more accurately.
The fifth column of Table \ref{tab:vit3} indicates that Version III
recovers $f_\ast(t)$ with a root mean square error across
$N_T=37$ segments of
$\varepsilon_{f_\ast}\leq 9.5\times 10^{-7}\,{\rm Hz}
 \leq 2\Delta f_{\rm drift}$.
(Recall that the optimal Viterbi track is free to wander,
whereas the injections are stationary.)
Indeed 33 out of 47 injections are recovered with
$\varepsilon_{f_\ast}\leq \Delta f_{\rm drift}$.
Essentially parameter estimation is limited by the spectral resolution.
In contrast, Version II of the HMM recovers
27 out of 50 injections with 
$\varepsilon_{f_\ast} \approx P^{-1} \gg \Delta f_{\rm drift}$,
much worse than the spectral resolution, viz.\
$1\leq \varepsilon_{f_\ast} / (10^{-5} \,{\rm Hz}) \leq 2$;
see Table IV in Ref.\ \cite{Suvorova2017}.
The step up from $\varepsilon_{f_\ast} \sim \Delta f_{\rm drift}$
to $\varepsilon_{f_\ast} \sim P^{-1}$ occurs,
because Version II sometimes converges on the orbital sidebands
$f_\ast \pm P^{-1}$, 
whereas Version III always converges on the central peak $f_\ast$
for the MDC injections.
Interestingly, no strong correlation is found between
$\varepsilon_{f_\ast}$ and $h_0^{\rm eff}$ with Version III.
Once the HMM detects a signal, 
$\varepsilon_{f_\ast} \lesssim \Delta f_{\rm drift}$
is grid-limited and essentially random.
A similar lack of correlation is observed for Version II.
\cite{Suvorova2017}

Version III is also more accurate than Version II when recovering
the orbital elements.
The seventh column of Table \ref{tab:vit3} indicates that Version III
recovers $a_0$ with an absolute error of 
$| \varepsilon_{a_0} | \leq 1.6\times 10^{-3} \, {\rm lt\,s}$.
This amounts to $\lesssim 5$ times the grid resolution,
which decreases $\propto f_0^{-1}$ from
$6.6\times 10^{-4}\,{\rm lt\,s}$ at $f_0=54.5\,{\rm Hz}$ (injection 1)
to
$2.6\times 10^{-5}\,{\rm lt\,s}$ at $f_0=1.37\,{\rm kHz}$ (injection 98).
Although the maximum value of $| \varepsilon_{a_0} |$ is comparable
for Versions II and III,
Version III recovers 26 out of 47 injections with
$| \varepsilon_{a_0} | \leq 1\times 10^{-4} \, {\rm lt\,s}$,
whereas Version II only recovers eight out of 50 injections with
$| \varepsilon_{a_0} | \leq 1\times 10^{-4} \, {\rm lt\,s}$.
Interestingly Version III underestimates $a_0$ 41 out of 47 times.
It is currently unclear why this happens, 
and more tests are needed to explore the behavior and
check if it is a statistical fluctuation.

The ninth column of Table \ref{tab:vit3} indicates
that Version III recovers $T_{\rm asc}$ with an absolute error of
$| \varepsilon_{T_{\rm asc}} | \leq 11\,{\rm s}$,
i.e.\ $\lesssim 5$ times the grid resolution,
which decreases $\propto f_0^{-1} a_0^{-1}$ from
$\approx 5\,{\rm s}$ at $f_0=54.5\,{\rm Hz}$ (injection 1)
to
$\approx 0.2\,{\rm s}$ at $f_0=1.37\,{\rm kHz}$ (injection 98).
The $T_{\rm asc}$ estimates compare favorably with 
the orbital phase errors
$| \varepsilon_{\phi_{\rm a}} | 
 = 2\pi | \varepsilon_{T_{\rm asc}} | / P
 \leq 1.0\times 10^{-3}$ 
yielded by Version II.
The maximum $\phi_{\rm a}$ error is comparable in Versions II and III,
but Version III recovers 21 out of 47 injections with
$|\varepsilon_{T_{\rm asc}}| \leq 0.5\,{\rm s}$,
whereas Version II recovers only five out of 50 injections with
$|\varepsilon_{T_{\rm asc}}| \leq 0.5\,{\rm s}$.
The $T_{\rm asc}$ results 
parallel the behavior observed in $\varepsilon_{a_0}$.

Accuracy of parameter estimation is a better diagnostic
for illustrating the superiority of Version III
in the MDC context
than (say) the minimum number of segments required to detect a signal.
Version II detects 43 out of 50 injections
with $N_T=1$ and the remaining seven with $N_T \leq 13$
($T_{\rm drift}=10\,{\rm d}$).
\cite{Suvorova2017}
There is not much room for Version III to outperform
against this measure but for the record it does:
it detects every injection except the two weakest 
(indexes 64 and 90) with $N_T=1$.

\begin{table*}[!tbh]
	\centering
	\setlength{\tabcolsep}{2pt}
\begin{tabular}{lcccccccc}\hline \hline
Index & $h_0\ (10^{-25})$ & $h_0^{\mathrm{eff}}\ (10^{-25})$ & $f_\ast$ (Hz) & $\varepsilon_{f_\star}$ (Hz) & $a_0$ (s) & $\varepsilon_{a_0}$ (s) & $T_{\mathrm{asc}}$ (s) & $\varepsilon_{T_{\mathrm{asc}}}$ (s) \\
\hline
1 & 4.160 & 2.706 & 54.498391348174 & $4.342 \times 10^{-7}$ & 1.37952 & $-9.518 \times 10^{-4}$ & 1245967666.02 & -11.12 \\
2 & 4.044 & 2.511 & 64.411966012332 & $4.229 \times 10^{-7}$ & 1.76461 & $4.803 \times 10^{-4}$ & 1245967592.98 & -5.27 \\
3 & 3.565 & 3.463 & 73.795580913582 & $6.836 \times 10^{-7}$ & 1.53460 & $-1.585 \times 10^{-3}$ & 1245967461.35 & -5.66 \\
5 & 1.250 & 1.154 & 93.909518008164 & $5.104 \times 10^{-7}$ & 1.52018 & $-8.158 \times 10^{-5}$ & 1245966927.93 & 2.22 \\
11 & 3.089 & 1.399 & 154.916883586097 & $3.464 \times 10^{-7}$ & 1.39229 & $4.297 \times 10^{-5}$ & 1245967559.97 & 2.67 \\
14 & 2.044 & 1.286 & 183.974917468730 & $3.553 \times 10^{-7}$ & 1.50970 & $-7.066 \times 10^{-4}$ & 1245967551.05 & -3.63 \\
15 & 11.764 & 4.169 & 191.580343388804 & $3.612 \times 10^{-7}$ & 1.51814 & $-4.484 \times 10^{-4}$ & 1245967298.45 & 0.10 \\
17 & 3.473 & 1.253 & 213.232194220000 & $2.244 \times 10^{-7}$ & 1.31021 & $-7.427 \times 10^{-5}$ & 1245967522.54 & 1.74 \\
19 & 6.031 & 2.437 & 233.432565653291 & $3.189 \times 10^{-7}$ & 1.23123 & $-1.060 \times 10^{-4}$ & 1245967331.14 & 1.27 \\
20 & 9.710 & 3.434 & 244.534697522529 & $3.941 \times 10^{-7}$ & 1.28442 & $-4.418 \times 10^{-4}$ & 1245967110.97 & -1.10 \\
21 & 1.815 & 0.792 & 254.415047846878 & $5.561 \times 10^{-7}$ & 1.07219 & $7.354 \times 10^{-5}$ & 1245967346.40 & -1.24 \\
23 & 2.968 & 1.677 & 271.739907539784 & $3.922 \times 10^{-7}$ & 1.44287 & $-2.731 \times 10^{-4}$ & 1245967302.29 & -2.22 \\
26 & 1.419 & 1.172 & 300.590450155009 & $3.342 \times 10^{-7}$ & 1.25869 & $-1.721 \times 10^{-4}$ & 1245967177.47 & -1.87 \\
29 & 4.275 & 3.131 & 330.590357652653 & $4.893 \times 10^{-7}$ & 1.33070 & $-6.673 \times 10^{-5}$ & 1245967520.83 & -0.84 \\
32 & 10.038 & 4.391 & 362.990820993568 & $1.870 \times 10^{-7}$ & 1.61109 & $-2.790 \times 10^{-4}$ & 1245967585.56 & 0.24 \\
35 & 16.402 & 9.183 & 394.685589797695 & $3.466 \times 10^{-7}$ & 1.31376 & $-1.059 \times 10^{-4}$ & 1245967198.05 & 1.75 \\
36 & 3.864 & 1.539 & 402.721233789014 & $5.075 \times 10^{-7}$ & 1.25484 & $-6.642 \times 10^{-5}$ & 1245967251.35 & 0.79 \\
41 & 1.562 & 0.746 & 454.865249156175 & $2.651 \times 10^{-7}$ & 1.46578 & $-1.896 \times 10^{-4}$ & 1245967225.75 & 0.36 \\
44 & 2.237 & 1.996 & 483.519617972096 & $8.346 \times 10^{-8}$ & 1.55221 & $-1.446 \times 10^{-4}$ & 1245967397.86 & 0.13 \\
47 & 4.883 & 1.992 & 514.568399601819 & $2.824 \times 10^{-7}$ & 1.14020 & $-1.637 \times 10^{-4}$ & 1245967686.81 & 0.33 \\
48 & 1.813 & 0.745 & 520.177348201609 & $6.614 \times 10^{-7}$ & 1.33669 & $-3.329 \times 10^{-5}$ & 1245967675.30 & 0.15 \\
50 & 1.093 & 1.027 & 542.952477491471 & $5.178 \times 10^{-7}$ & 1.11915 & $-2.302 \times 10^{-4}$ & 1245967927.48 & -1.47 \\
51 & 9.146 & 3.372 & 552.120598886904 & $6.501 \times 10^{-7}$ & 1.32783 & $6.253 \times 10^{-5}$ & 1245967589.54 & -0.94 \\
52 & 2.786 & 1.550 & 560.755048768919 & $4.209 \times 10^{-7}$ & 1.79214 & $-6.193 \times 10^{-5}$ & 1245967377.20 & 0.61 \\
54 & 1.518 & 1.256 & 593.663030872532 & $5.792 \times 10^{-7}$ & 1.61276 & $-3.115 \times 10^{-5}$ & 1245967624.53 & 0.30 \\
57 & 1.577 & 0.788 & 622.605388362863 & $5.260 \times 10^{-7}$ & 1.51329 & $-5.596 \times 10^{-5}$ & 1245967203.21 & -1.00 \\
58 & 3.416 & 1.287 & 641.491604906276 & $6.158 \times 10^{-7}$ & 1.58443 & $-1.418 \times 10^{-4}$ & 1245967257.74 & 0.16 \\
59 & 8.835 & 4.981 & 650.344230698489 & $7.830 \times 10^{-7}$ & 1.67711 & $-1.422 \times 10^{-4}$ & 1245967829.90 & -0.69 \\
60 & 2.961 & 2.467 & 664.611446618250 & $7.197 \times 10^{-7}$ & 1.58262 & $5.343 \times 10^{-5}$ & 1245967612.31 & -0.41 \\
61 & 6.064 & 2.158 & 674.711567789201 & $4.978 \times 10^{-7}$ & 1.49937 & $-1.037 \times 10^{-4}$ & 1245967003.32 & -0.01 \\
62 & 10.737 & 3.853 & 683.436210983289 & $8.223 \times 10^{-7}$ & 1.26951 & $-4.060 \times 10^{-5}$ & 1245967453.97 & -0.00 \\
63 & 1.119 & 0.745 & 690.534687981171 & $6.762 \times 10^{-7}$ & 1.51824 & $-3.958 \times 10^{-5}$ & 1245967419.39 & -0.18 \\
64 & 1.600 & 0.570 & 700.866836291234 & $5.143 \times 10^{-7}$ & 1.39993 & $-6.909 \times 10^{-5}$ & 1245967596.12 & -0.96 \\
67 & 4.580 & 1.623 & 744.255707971300 & $3.620 \times 10^{-7}$ & 1.67774 & $-1.551 \times 10^{-4}$ & 1245967084.30 & 0.27 \\
68 & 3.696 & 1.844 & 754.435956775916 & $4.000 \times 10^{-7}$ & 1.41389 & $-8.960 \times 10^{-5}$ & 1245967538.70 & 0.38 \\
69 & 2.889 & 1.053 & 761.538797037770 & $3.693 \times 10^{-7}$ & 1.62613 & $-1.239 \times 10^{-4}$ & 1245966821.55 & 0.03 \\
71 & 2.923 & 1.232 & 804.231717847467 & $3.238 \times 10^{-7}$ & 1.65203 & $8.338 \times 10^{-6}$ & 1245967156.55 & 0.30 \\
72 & 1.248 & 0.792 & 812.280741438401 & $4.597 \times 10^{-7}$ & 1.19649 & $-1.325 \times 10^{-4}$ & 1245967159.08 & 0.87 \\
73 & 2.444 & 0.936 & 824.988633484129 & $9.533 \times 10^{-7}$ & 1.41715 & $-6.960 \times 10^{-5}$ & 1245967876.83 & 0.82 \\
76 & 3.260 & 1.725 & 882.747979842807 & $4.813 \times 10^{-7}$ & 1.46249 & $-8.305 \times 10^{-5}$ & 1245966753.24 & -0.17 \\
79 & 4.681 & 1.656 & 931.006000308958 & $2.697 \times 10^{-7}$ & 1.49171 & $-7.243 \times 10^{-5}$ & 1245967290.06 & 0.14 \\
83 & 5.925 & 2.186 & 1081.398956458276 & $7.176 \times 10^{-7}$ & 1.19854 & $-3.862 \times 10^{-5}$ & 1245967313.93 & -1.02 \\
84 & 11.609 & 7.184 & 1100.906018344283 & $7.529 \times 10^{-7}$ & 1.58972 & $-6.257 \times 10^{-6}$ & 1245967204.15 & -0.35 \\
85 & 4.553 & 1.633 & 1111.576831848269 & $8.018 \times 10^{-7}$ & 1.34479 & $-9.497 \times 10^{-5}$ & 1245967049.35 & -0.90 \\
90 & 0.684 & 0.618 & 1193.191890630547 & $4.053 \times 10^{-7}$ & 1.57513 & $-7.212 \times 10^{-5}$ & 1245966914.27 & -0.21 \\
95 & 4.293 & 3.059 & 1324.567365220908 & $5.198 \times 10^{-7}$ & 1.59169 & $-1.443 \times 10^{-5}$ & 1245967424.76 & 0.53 \\
98 & 5.404 & 1.948 & 1372.042154535880 & $7.448 \times 10^{-7}$ & 1.31510 & $-7.340 \times 10^{-5}$ & 1245966869.92 & -0.34 \\
\hline
\hline
\end{tabular}
	\caption{
Results of tracking the 47 available injections in the Sco X$-$1 MDC,
sorted by index from Ref.\ \cite{Messenger2015},
using Version III of the HMM to track phase and frequency.
}
	\label{tab:vit3}
\end{table*}

\section{Conclusions
 \label{sec:vit8}}
A HMM coupled with a step-wise matched filter provides an efficient,
semi-coherent way to detect and track the unknown signal frequency 
of a quasimonochromatic, continuous gravitational wave source
with spin wandering driven by internal processes (isolated source)
or accretion (binary source).
In previous work HMMs have searched for the LMXB Sco X$-$1
in LIGO O1 and O2 data using frequency domain,
maximum likelihood matched filters:
the Bessel-weighted ${\cal F}$-statistic (Version I), 
which does not track orbital phase,
and the Jacobi-Anger ${\cal J}$-statistic (Version II),
which does.
Here we generalize existing HMM pipelines to track
rotational phase as well as orbital phase (Version III).
In the emission probability,
the ${\cal J}$-statistic is replaced by a phase-sensitive version of
the Bayesian ${\cal B}$-statistic introduced for loosely coherent searches.
The data are input as SFTs,
leveraging the well-tested software infrastructure in the LAL.
In the transition probability,
the intra-step spin wandering is modeled according to a
phase-wrapped Ornstein-Uhlenbeck process.
A recipe for choosing the Ornstein-Uhlenbeck control parameters,
$\gamma$ and $\sigma$, is given in Section \ref{sec:vit3c}.
A revised detection strategy based on block scores
is described in Section \ref{sec:vit5c}.

The sensitivity of Version III of the HMM is quantified.
The ROC curves in Sections \ref{sec:vit5d} and \ref{sec:vit6b}
give $P_{\rm d} \geq 0.9$ (isolated source)
and $P_{\rm d} \geq 0.75$ (binary source),
when the characteristic wave strain
satisfies $h_0 \geq 1.3\times 10^{-26}$,
with $P_{\rm a} = 10^{-2}$.
Hence Version III is $\approx 1.5$ times more sensitive than Version II.
The requirement of phase continuity from one HMM step to the next
lowers $P_{\rm a}$ at fixed $h_0$ 
and increases $1-P_{\rm d}$ at fixed $P_{\rm a}$.
Performance is optimized,
when $T_{\rm drift}$ matches the source's spin wandering time-scale.
The results depend weakly on $\gamma$, $\sigma$, 
the block width when calculating the block score,
and the location of the block boundary.

The tracking accuracy is quantified
in Sections \ref{sec:vit5e} and \ref{sec:vit6c}.
It is found that the root mean square frequency error
is bounded spectrally and is therefore near-optimal,
with $\varepsilon_{f_\ast} \lesssim \Delta f_{\rm drift}$
when an injected signal is detected successfully and
$\varepsilon_{f_\ast} \gg \Delta f_{\rm drift}$ otherwise.
The absolute errors in the orbital elements are limited to
$\lesssim 5$ times the grid resolution in $a_0$ and $\phi_{\rm a}$
(or equivalently $T_{\rm asc}$)
set by the parameter space metric.
\cite{Leaci2015}
The HMM log probability peaks unimodally at the correct value
in the $a_0$-$T_{\rm asc}$ plane, with a sinc-like cross-section
(see Figure \ref{fig:vit13}).
The accuracy is confirmed by the performance of Version III
of the HMM in the Sco X$-$1 MDC (in self-blinded mode).
It finds 47 out of 47 injections currently available
(out of 50 originally)
with $N_T=37$, $T_{\rm drift}=10\,{\rm d}$,
and two simulated interferometers,
achieving accuracies of
$| \varepsilon_{f_\ast} | \leq 9.5\times 10^{-7} \, {\rm Hz}$,
$| \varepsilon_{a_0} | \leq 1.6\times 10^{-3} \, {\rm lt \, s}$,
and
$| \varepsilon_{T_{\rm asc}} | \leq 11 \, {\rm s}$.
Version III is less prone to converging on the sidebands
$f_\ast \pm P^{-1}$ and is systematically more accurate,
e.g.\ it recovers 26 out of 47 injections with
$| \varepsilon_{a_0} | \leq 1\times 10^{-4} \, {\rm lt \, s}$,
whereas Version II only achieves such accuracy eight times out of 50.
The gridding strategy adopted here, 
which is to implement conservatively
the parameter space metric in Ref.\ \cite{Leaci2015} 
as described in Section \ref{sec:vit7b},
should be regarded as a first pass.
Optimizing the gridding strategy is postponed to future work,
in the context of a search with real data
(which introduces other relevant constraints).
Stage II of the MDC will test the robustness of the HMM and
other algorithms like CrossCorr \cite{Whelan2015} and TwoSpect 
\cite{Meadors2016} to spin wandering.
Previous studies demonstrate that the HMM handles signals 
with and without spin wandering with equal dexterity,
as long as $T_{\rm drift}$ satisfies condition (\ref{eq:vit6}).
\citep{Suvorova2016,Suvorova2017}

The HMM in this paper is solved by the Viterbi algorithm,
which exploits dynamic programming.
The additional phase tracking step inevitably slows down 
Version III of the HMM compared to Version II,
with the number of operations scaling approximately
$\propto N_Q \ln N_Q$ (see Section \ref{sec:vit2a}),
and $N_Q$ increasing by a factor $\sim 10$.
Overall, however, the implementation remains fast,
processing $\approx 0.3\,{\rm Hz}$ per CPU-hr
for one choice of $(a_0,T_{\rm asc},P)$,
approximately $10$ times slower than Version II.
Viterbi-based continuous wave searches have proved amenable
to being implemented on graphical processing units,
which can shorten the run time $\approx 40$-fold.
\cite{Dunn2020}
The computational savings from an optimized implementation
on graphical processing units
can be re-invested to extend the astrophysical ambition of an analysis,
e.g.\ by targeting LMXBs other than Sco X$-$1 \cite{Watts2008}.
Savings can also be re-invested to expand the scope of Viterbi-based,
nonparametric, all-sky searches
and searches for wandering instrumental lines.
\cite{Bayley2019}

What conclusions can we expect to draw about the astrophysical causes 
of spin wandering, when the HMM ultimately detects a real signal?
At present it is hard to say.
Neutron star models involve a great deal of uncertain physics,
which will blur the interpretation of any HMM detection,
whether it involves Versions I, II, or III,
unless the detection itself reveals some unexpected and informative signature.
Electromagnetic observations may improve the situation.
Consider, for example, an LMXB where one observes simultaneously
the X-ray flux $F_X(t)$ and the wandering spin $f_\ast(t)$.
One might hope to cross-correlate the fluctuations in $F_X$ and $\dot{f}_\ast$ 
and thereby test the accretion physics.
\cite{Mukherjee2018}
However, the traditional assumption $F_X \propto \dot{M} \propto \dot{f}_\ast$,
where $\dot{M}$ denotes the mass accretion rate,
does not always hold for various reasons,
e.g.\ nonconservative mass transfer,
hydromagnetic contributions to $\dot{f}_\ast$,
and unsteady dynamics due to magnetospheric instabilities.
\cite{Romanova2004,DAngelo2010}
Some of the relevant issues are canvassed in Ref.\ \cite{Haskell2015}.
As a second illustrative example,
suppose the HMM detects a steady tone with minimal spin wandering
from a radio pulsar,
that displays strong timing noise at radio wavelengths.
Such an observation would arguably suggest,
that the gravitational wave signal is emitted
by the weakly coupled superfluid interior of the star as opposed to the crust
(which is locked magnetically to the radio pulses).
Furthermore,
if $f_\ast$ from the HMM approximately equals the time-averaged radio pulse frequency, 
it arguably represents partial evidence for pinning of the superfluid.
\cite{Jones2010,Melatos2015}
These and other possibilities will clarify themselves,
once detections are made routinely.

\section{Acknowledgements}
We would like to thank Paul Lasky, Chris Messenger, Keith Riles, Karl Wette, 
Letizia Sammut, John Whelan, Grant Meadors and the LIGO Scientific Collaboration 
Continuous Wave Working Group for detailed comments and informative discussions.
We especially thank Karl Wette for alerting us to the existence of the
phase extraction tool {\em XLALEstimatePulsarAmplitudeParams} in the LAL suite
and Grant Meadors for pointing us to the phase-sensitive formulation
of the ${\cal B}$-statistic in Ref.\ \cite{Dergachev2012}.
The synthetic data for Stage I of the Sco X-1 MDC were prepared primarily 
by Chris Messenger with the assistance of members of the MDC team 
\cite{Messenger2015}. 
We thank Chris Messenger and Paul Lasky for their assistance in handling 
the MDC data. 
We thank the anonymous referees for their constructive feedback.
P. Clearwater and L. Sun have been supported by Australian Postgraduate Awards. 
P. Clearwater was also a recipient of a scholarship from the
Commonwealth Scientific and Industrial Research Organisation,
Australia.
L. Sun has been a member of the LIGO Laboratory.
LIGO was constructed by the California Institute of Technology
and Massachusetts Institute of Technology
with funding from the National Science Foundation
and operates under cooperative agreement PHY-1764464.
Advanced LIGO was built under award PHY-0823459.
The research is supported by the Australian Research Council (ARC) 
Centre of Excellence for Gravitational Wave Discovery (OzGrav),
grant number CE170100004.

\appendix
\section{Viterbi algorithm
 \label{sec:vitappa}}
The Viterbi algorithm prunes the tree of possible hidden state sequences $Q$
by appealing to Bellman's Principle of Optimality:
if a subpath $\{ q^\ast(t_i),\cdots, q^\ast(t_j) \}$ is optimal,
then all of its subpaths are optimal as well.
\cite{Bellman1957}
Dynamic programming is exploited to implement the Principle of Optimality
in an efficient, recursive fashion.
\cite{Viterbi1967,Quinn2001,Suvorova2016}
Pseudocode describing the implementation is presented below in abridged form
for ease of reference.

At time $t_k$ ($1\leq k \leq N_T$), 
let the vector $\bm{\delta} (t_k)$ store the $N_Q$ maximum probabilities
\begin{equation}
 \delta_{q_i}(t_k) = 
 \mathop{\max} \limits_{q_j} \Pr[q(t_k) = q_i | q(t_{k-1}) = q_j; O^{(k)}]~,
\label{eq:vitappa1}
\end{equation}
with $1\leq i \leq N_Q$,
and let the vector $\bm{\Phi}(t_k)$ store the hidden states at $t_{k-1}$ 
leading to the corresponding maximum probabilities in $\bm{\delta}(t_k)$,
viz.
\begin{equation}
\Phi_{q_i}(t_k) = 
 \mathop{\arg \max} \limits_{q_j} \Pr[q(t_k) = q_i | q(t_{k-1}) = q_j; O^{(k)}]~,
\label{eq:vitappa2}
\end{equation}
with $O^{(k)} = \{ o(t_0),\dots,o(t_k) \}$ and
\begin{equation}
 \Pr [q(t_k) = q_i | q(t_{k-1}) = q_j; O^{(k)}] 
 =
 L_{o(t_k)q_i} A_{q_i q_j}\delta_{q_j}(t_{k-1})~.
\label{eq:vitappa3}
\end{equation}
The components of $\bm{\delta} (t_k)$ and $\bm{\Phi} (t_k)$ are filled
by running forward through the $N_T$ observations,
then the optimal path $Q^\ast(O)$ is reconstructed by backtracking.

$\emph{1. Initialization:}$
\begin{equation}
\delta_{q_i}(t_0) = L_{o(t_0)q_i} \Pi _{q_i},
\end{equation}
for $1 \leq i \leq N_Q$. 

$\emph{2. Recursion:}$
\begin{eqnarray}
\delta_{q_i}(t_k) &=& L_{o(t_k)q_i} \mathop{\max} \limits_{1 \leq j \leq N_Q} [A_{q_i q_j}\delta_{q_j}(t_{k-1})], \\
\Phi_{q_i}(t_k) &=& \mathop{\arg \max} \limits_{1 \leq j \leq N_Q} [A_{q_i q_j}\delta_{q_j}(t_{k-1})],
\end{eqnarray}
for $1 \leq i \leq N_Q$ and $1 \leq k \leq N_T$.

$\emph{3. Termination:}$
\begin{eqnarray}
\max \Pr(Q|O) &=& \mathop{\max} \limits_{q_j} \delta_{q_j}(t_{N_T}) \\
q^*(t_{N_T}) &=& \mathop{\arg \max} \limits_{q_j} \delta_{q_j}(t_{N_T})
\end{eqnarray}
for $1 \leq j \leq N_Q$.

$\emph{4. Optimal path backtracking:}$
\begin{equation}
q^*(t_k) = \Phi_{q^*(t_{k+1})}(t_{k+1})
\end{equation}
for $0 \leq k \leq N_T -1$.

\section{Drift time-scale
 \label{sec:vitappaa}}
A practical recipe for choosing the drift time-scale $T_{\rm drift}=t_{n+1}-t_n$ 
(see Section \ref{sec:vit2a}) when tracking $f_\ast(t)$
is described in Refs \cite{Suvorova2016} and \cite{Suvorova2017}.
In this appendix we generalize the recipe for the purpose of tracking 
$f_\ast(t)$ and $\Phi_\ast(t)$ in Version III of the HMM.

The choice of $T_{\rm drift}$ is governed by the packaging of input data
when computing the emission probability $L_{o_j q_i}$, which comes with
implicit assumptions about the signal properties in the interval 
$t_{n-1} \leq t \leq t_{n}$.
Importantly we require $L_{o_j q_i}$ to peak as sharply as possible
in the neighborhood of the truly occupied hidden state $q(t_n)$,
with $L_{o_j q_i} \approx \delta[q_i - q(t_n)]$ ideally,
in order to maximize $\Pr[Q^\ast(O) | O]$.
Typically $L_{o_j q_i}$ is computed from frequency-domain data
covering the whole interval $t_{n-1} \leq t \leq t_{n}$,
and $q(t)$ does not contain frequency-drift variables like $\dot{f}_\ast(t)$.
Therefore the matched filter that computes $L_{o_j q_i}$
(e.g.\ the ${\cal F}$- or ${\cal B}$-statistic)
assumes that $f_\ast(t)$ stays within a single, discrete bin
during every HMM time-step.
For this assumption to hold,
one must choose $T_{\rm drift}$ to satisfy
\begin{equation}
 \left|
  \int_t^{t+T_{\rm drift}}
  dt' \, \dot{f}_\ast(t')
 \right|
 < 
 \Delta f_{\rm drift}
\label{eq:vit6}
\end{equation}
for all $t$,
where $\Delta f_{\rm drift}$ is the separation
between adjacent frequency bins 
(which are assumed to be uniformly spaced in this paper, 
i.e.\ $\Delta f_{\rm drift}$ is independent of $q_i$).
A different method of computing $L_{o_j q_i}$,
e.g.\ from time-domain data, may impose a different constraint on $T_{\rm drift}$.

It is tempting to extend the above argument to $\Phi_\ast(t)$ 
and insist that it should stay within a single bin too
(of width $\Delta\Phi_{\rm drift}=\pi/16$ in this paper),
\footnote{
The analyst enjoys considerable freedom in setting $\Delta\Phi_{\rm drift}$,
as long as the peaks in the transition probability in Figure \ref{fig:vit1b}
are resolved.
In contrast, $\Delta f_{\rm drift}=(2T_{\rm drift})^{-1}$ is determined
by $T_{\rm drift}$.
See Section \ref{sec:vit2c} for details.
\label{foot:vit3}
}
but this is unnecessary.
Frequency-domain matched filters like the ${\cal F}$- and ${\cal B}$-statistic
do not assume that $\Phi_\ast(t)$ is constant 
for $t_{n-1} \leq t \leq t_{n}$;
they are well-behaved functions of $\Phi_\ast(t_{n-1})$ 
at the start of the HMM time-step.
Confining $\Phi_\ast(t)$ to a single phase bin would shorten
$T_{\rm drift}$ by a factor $\approx \pi/\Delta \Phi_{\rm drift}$,
widen every frequency bin by the same factor (Nyquist theorem),
and reduce proportionally the signal-to-noise ratio per frequency bin.
\citep{Jaranowski1998,Dergachev2010}

Naturally one does not know $\dot{f}_\ast(t')$ in (\ref{eq:vit6}) in advance,
so there is some trial and error involved in choosing $T_{\rm drift}$
through (\ref{eq:vit6}).
In this paper we focus on gravitational wave searches for isolated and accreting
neutron stars, whose rotatational irregularities have been studied extensively
in radio 
\cite{Cordes1985,Price2012}
and X-ray
\cite{Baykal1993,Bildsten1997}
timing experiments,
which yield autocorrelation time-scales of days to months.
These electromagnetic measurements therefore offer a starting point
to estimate $T_{\rm drift}$ for other objects in the same class,
where $\dot{f}_\ast(t')$ is not measured.
\cite{Mukherjee2018}
For reasons of convenience described in Section \ref{sec:vit2},
we elect to work with Fourier-transformed data in this paper,
\cite{Suvorova2016,Suvorova2017}
which come packaged in calibrated, conditioned
(anti-alias filtering, data drop-out),
short-time Fourier transforms (SFTs)
of duration $T_{\rm SFT} = 30\,{\rm min}$.
\cite{Mendell2002}
Hence one has $T_{\rm SFT} \leq T_{\rm drift}$ as a practical matter,
a constraint which would be absent in a time-domain analysis.
Given the wide range of measured auto-correlation time-scales,
one can envisage a hierarchical search strategy,
in which a search is repeated for several $T_{\rm drift}$ values
in the range $T_{\rm SFT} \leq T_{\rm drift} \leq T_{\rm obs}$,
where $T_{\rm obs} \sim 1 \, {\rm yr}$ is the total observation time.

\section{Phase-wrapped Ornstein-Uhlenbeck process
 \label{sec:vitappb}}
In this appendix we solve the Fokker-Planck equation corresponding to
the stochastic differential equations (\ref{eq:vit9}) and (\ref{eq:vit10})
to obtain the probability density function (PDF) $p(t,f_\ast,\Phi_\ast)$
and hence the HMM transition probabilities over the interval
$t_n \leq t \leq t_{n+1}$
given the initial state $q(t_n)=[f_\ast(t_n),\Phi_\ast(t_n)]$
or the final state $q(t_{n+1})=[f_\ast(t_{n+1}),\Phi_\ast(t_{n+1})]$.
The discussion follows Appendix A in Ref.\ \cite{Suvorova2018}.
Equations (\ref{eq:vit9}) and (\ref{eq:vit10})
are equivalent to traditional, spatial Brownian motion,
with $f_\ast$ and $\Phi_\ast$ playing the roles of velocity
and displacement respectively,
except that $\Phi_\ast$ is $2\pi$-periodic.

If the hidden state $q(t_n)$ occupied at the start of the HMM step
$t_n \leq t \leq t_{n+1}$ is known with certainty,
the PDF of the final state at $t=t_{n+1}$ is given by the solution
$p^{\rm F}(t,f_\ast,\Phi_\ast)$ of the forward Fokker-Planck equation
\cite{Gardiner1994}
\begin{equation}
 \frac{\partial p^{\rm F}}{\partial t}
 =
 \gamma p^{\rm F}
 + \gamma f_\ast \frac{\partial p^{\rm F}}{\partial f_\ast}
 - f_\ast \frac{\partial p^{\rm F}}{\partial \Phi_\ast}
 + \frac{\sigma^2}{2}
  \frac{\partial^2 p^{\rm F}}{\partial f_\ast^2}~,
\label{eq:vitappb1}
\end{equation}
evaluated at $t=t_{n+1}$ given
$p^{\rm F}(t_n, f_\ast, \Phi_\ast) =
 \delta[f_\ast - f_\ast(t_n) ]
 \delta[\Phi_\ast - \Phi_\ast(t_n) ]$.
If the final state $q(t_{n+1})$ is known with certainty,
the PDF of the initial state is given by the solution
$p^{\rm B}(t,f_\ast,\Phi_\ast)$ of the backward Fokker-Planck equation,
\begin{equation}
 \frac{\partial p^{\rm B}}{\partial t}
 =
 \gamma f_\ast \frac{\partial p^{\rm B}}{\partial f_\ast}
 - f_\ast \frac{\partial p^{\rm B}}{\partial \Phi_\ast}
 - \frac{\sigma^2}{2}
  \frac{\partial^2 p^{\rm B}}{\partial f_\ast^2}~,
\label{eq:vitappb2}
\end{equation}
evaluated at $t=t_{n}$ given
$p^{\rm F}(t_{n+1}, f_\ast, \Phi_\ast) =
 \delta[f_\ast - f_\ast(t_{n+1}) ]
 \delta[\Phi_\ast - \Phi_\ast(t_{n+1}) ]$.
Equation (\ref{eq:vitappb2}) is the adjoint of (\ref{eq:vitappb1}).
Upon multiplying (\ref{eq:vitappb1}) by the integrating factor
$\exp(-\gamma t)$,
we find 
\begin{equation}
 p^{\rm B}(t,f_\ast,\Phi_\ast)
 \propto 
 \exp(-\gamma t) p^{\rm F}(t,f_\ast,\Phi_\ast; \sigma^2 \mapsto -\sigma^2)~,
\label{eq:vitappb3}
\end{equation}
where
$\sigma^2 \mapsto -\sigma^2$ denotes replacing $\sigma^2$ by $-\sigma^2$
in $p^{\rm F}$.

Upon Fourier analysing $p^{\rm F}$, as in Ref.\ \cite{Suvorova2018},
we find that the characteristic function
\begin{eqnarray}
 \tilde{p}^{\rm F}(t,\kappa,m)
 & = &
 \int_0^{2\pi} d\Phi_\ast \int_{-\infty}^\infty df_\ast \,
 \exp(-im\Phi_\ast - i\kappa f_\ast) 
 \nonumber \\
 & & \times
 p^{\rm F}(t,f_\ast,\Phi_\ast)
\label{eq:vitappb4}
\end{eqnarray}
satisfies
\begin{equation}
 \frac{\partial \tilde{p}^{\rm F}}{\partial t}
 =
 (-\gamma \kappa + m) 
  \frac{\partial \tilde{p}^{\rm F}}{\partial \kappa}
 -
 \frac{\sigma^2 \kappa^2 \tilde{p}^{\rm F}}{2}~,
\label{eq:vitappb5}
\end{equation}
subject to the initial condition
\begin{equation}
 \tilde{p}^{\rm F}(t_n,\kappa,m)
 =
 \exp[-im\Phi_\ast(t_n)-i\kappa f_\ast(t_n)]~.
\label{eq:vitappb6}
\end{equation}
Equations (\ref{eq:vitappb5}) and (\ref{eq:vitappb6})
are solved by the method of characteristics to give
\begin{eqnarray}
 \tilde{p}^{\rm F}(t_{n+1},\kappa,m)
 & = &
 \exp[-im\Phi_\ast(t_n)-i\rho f_\ast(t_n)]
 \nonumber \\
 & & \times
 \exp\left[
  \frac{\sigma^2}{4\gamma}
  \left( \rho - \frac{m}{\gamma} \right)
  \left( \rho + \frac{3m}{\gamma} \right)
 \right]
 \nonumber \\
 & & \times
 \exp\left\{
  -\frac{\sigma^2}{2}
  \left[
   \frac{m^2 \tau}{\gamma^2}
   + \frac{2m}{\gamma^2} \left( \kappa - \frac{m}{\gamma} \right)
  \right]
 \right\}
 \nonumber \\
 & & \times
 \exp\left[
  -\frac{\sigma^2}{4\gamma}
  \left( \rho - \frac{m}{\gamma} \right)^2
  \exp( 2\gamma\tau )
 \right]~,
\label{eq:vitappb7}
\end{eqnarray}
with $\tau=t_{n+1}-t_n$ and
\begin{equation}
 \rho 
 = 
 \frac{m}{\gamma}
 + \left( \kappa - \frac{m}{\gamma} \right) \exp(-\gamma\tau)
\label{eq:vitappb8}
\end{equation}
and hence
\begin{eqnarray}
 \tilde{p}^{\rm F}(t_{n+1},f_\ast,\Phi_\ast)
 & = &
 (2\pi)^{-2} \sum_{m=-\infty}^\infty \exp(im\Phi_\ast)
 \nonumber \\
 & & \times
 \int_{-\infty}^\infty d\kappa \, 
 \exp(i\kappa f_\ast)
 \tilde{p}^{\rm F}(t,\kappa,m)~.
\label{eq:vitappb9}
\end{eqnarray}
By completing the square in the argument of the exponential
in (\ref{eq:vitappb7}), one finds that (\ref{eq:vitappb9})
can be written as a wrapped Gaussian.
\cite{Suvorova2018}

The solution (\ref{eq:vitappb3}) 
to the backward Fokker-Planck equation (\ref{eq:vitappb2})
provides an efficient route to calculating the maximum probabilities
at each HMM step,
which are stored in the vector $\bm{\delta} (t_k)$ in the Viterbi implementation
described in Appendix \ref{sec:vitappa}.
Equation (\ref{eq:vitappb3}), just like (\ref{eq:vitappb9}), 
can be expressed as a wrapped Gaussian, viz.
\begin{eqnarray}
 p^{\rm B}(t_n,{\bf q})
 & = &
 (2\pi)^{-1} ({\rm det}{\bf \Sigma} )^{-1/2}
 \nonumber \\
 & & \times
 \sum_{m=-\infty}^\infty 
 \exp[-({\bf q}-{\bf Q}_m) {\bf \Sigma}^{-1} ({\bf q} - {\bf Q}_m)^{\rm T}]~,
\label{eq:vitappb10}
\end{eqnarray}
with ${\bf q}=(f_\ast,\Phi_\ast)$ and matrix elements
\begin{eqnarray}
 ({\bf Q}_m)_1
 & = &
 f_\ast(t_{n+1}) \exp(-\gamma\tau )~,
\label{eq:vitappb11}
 \\
 ({\bf Q}_m)_2
 & = &
 \Phi_\ast(t_{n+1})
 + \frac{f_\ast(t_{n+1})}{\gamma} [ 1- \exp(-\gamma\tau )] 
 \nonumber \\
 & & 
 - 2\pi m~,
\label{eq:vitappb12}
 \\
 {\bf\Sigma}_{11}
 & = &
 \frac{\sigma^2}{2\gamma}
 [ 1- \exp(-2 \gamma\tau )]~,
\label{eq:vitappb13}
 \\
 {\bf\Sigma}_{12}
 =
 {\bf\Sigma}_{21}
 & = &
 \frac{\sigma^2}{2 \gamma^2}
 [ 1- \exp(-\gamma\tau )]^2~,
\label{eq:vitappb14}
 \\
 {\bf\Sigma}_{22}
 & = &
 \frac{\sigma^2}{2 \gamma^3}
 \{ 1 + 2\gamma\tau - [2-\exp(-\gamma\tau )]^2 \}~.
\label{eq:vitappb15}
\end{eqnarray}
We can then read off the moments $\langle f_\ast \rangle$,
$\langle \Phi_\ast \rangle$,
$\langle f_\ast^2 \rangle - \langle f_\ast \rangle^2$,
$\langle f_\ast \Phi_\ast \rangle 
 - \langle f_\ast \rangle \langle \Phi_\ast \rangle$,
and
$\langle \Phi_\ast^2 \rangle - \langle \Phi_\ast \rangle^2$
of $p^{\rm B}$ by inspection from
(\ref{eq:vitappb11})--(\ref{eq:vitappb15}) respectively.
\cite{Suvorova2018}

\section{Maximum likelihood alternatives to the ${\cal B}$-statistic
 \label{sec:vitappc}}
In this appendix, we review briefly the maximum likelihood formulas
for $L_{o(t_n) q_i}$ used in Versions I and II of the HMM,
which do not depend on rotational phase.
\cite{Suvorova2016,Suvorova2017}
We then present for completeness a natural, phase-dependent generalization of
these maximum likelihood formulas.
Empirical testing indicates, that the generalized formula yields
no discernible improvement in performance over Versions I and II of the HMM, 
unlike the ${\cal B}$-statistic presented in Section \ref{sec:vit4c}.

In Version I of the HMM,
\cite{Suvorova2016}
for an isolated source ($a_0=0$) with zero phase (cf.\ spin) wandering
($\Phi_{\rm w}=0$),
the log likelihood is just the $\mathcal{F}$-statistic,
$G(f_0)={\cal F}(f_0)$, viz.
\begin{equation}
 \mathcal{F}(f_0)
 =
 \frac{4{\bf F}(f_0) {\bf H}^{-1} {\bf F}(f_0)^\dagger}{T_{\rm obs} S_h(f_0)}~,
\label{eq:vitappc1}
\end{equation}
where a dagger denotes the Hermitian transpose,
with
\begin{equation}
 {\bf F}(f_0) = [F_{1a}(f_0),F_{1b}(f_0)]~,
\label{eq:vitappc2}
\end{equation}
\begin{equation}
 {\bf H} 
 = 
 \left(
  \begin{array}{cc}
  A & C \\
  C & B 
  \end{array}
 \right)~,
\label{eq:vitappc3} 
\end{equation}
$A=(a \| a)$, $B=(b \| b)$, and $C=(a \| b )$.
In the general case $A_{2i}\neq 0$, 
equations (\ref{eq:vitappc1})--(\ref{eq:vitappc3}) contain
additional, analogous terms involving $F_{2a}$ and $F_{2b}$,
obtained from $F_{1a}$ and $F_{1b}$ by replacing $f_0$ with $2f_0$.

For a binary source ($a_0 \neq 0$) with zero phase wandering
($\Phi_{\rm w} = 0$),
the log likelihood in Version I of the HMM is approximated by the Bessel-weighted
$\mathcal{F}$-statistic,
\begin{equation}
 G(f_0)
 =
 \sum_{s=-M'}^{M'}
 [ J_s(2\pi f_0 a_0) ]^2
 \mathcal{F}(f_0-s/P)~,
\label{eq:vitappc4}
\end{equation}
with $M'={\rm ceil}(2\pi f_0 a_0)$.
Equation (\ref{eq:vitappc4}) adds together the power in orbital sidebands 
incoherently; 
it takes no account of the relative Fourier phases of the sidebands.
This omission is corrected in Version II of the HMM,
\cite{Suvorova2017}
where $F_{1a}$ and $F_{1b}$ are replaced by $J_{1a}$ and $J_{1b}$,
defined by (\ref{eq:vit31}) and (\ref{eq:vit32}) respectively,
in order to include orbital phase information.
The log likelihood is calculated similarly to the binary-modulated
$\mathcal{F}$-statistic and yields the $\mathcal{J}$-statistic,
$G(f_0)=\mathcal{J}(f_0)$, with
\begin{equation}
 \mathcal{J}(f_0)
 =
 \frac{4{\bf J}(f_0) {\bf H}^{-1} {\bf J}(f_0)^\dagger}{T_{\rm obs} S_h(f_0)}
\label{eq:vitappc7}
\end{equation}
and
\begin{equation}
 {\bf J}(f_0) = [J_{1a}(f_0),J_{1b}(f_0)]~.
\label{eq:vitappc8}
\end{equation}
Equation (\ref{eq:vitappc7})
concentrates all the signal power in the
orbital sidebands into one $f_0$ bin, unlike (\ref{eq:vitappc4}),
as verified in Figure 1 in Ref.\ \cite{Suvorova2017}.
It is therefore as sensitive for binary sources,
as (\ref{eq:vitappc1}) is for isolated sources,
i.e.\ (\ref{eq:vitappc1}) and (\ref{eq:vitappc7}) can detect the same $h_0$ value.
\cite{Suvorova2017}

\begin{figure*}
\centering
\scalebox{0.65}{\includegraphics{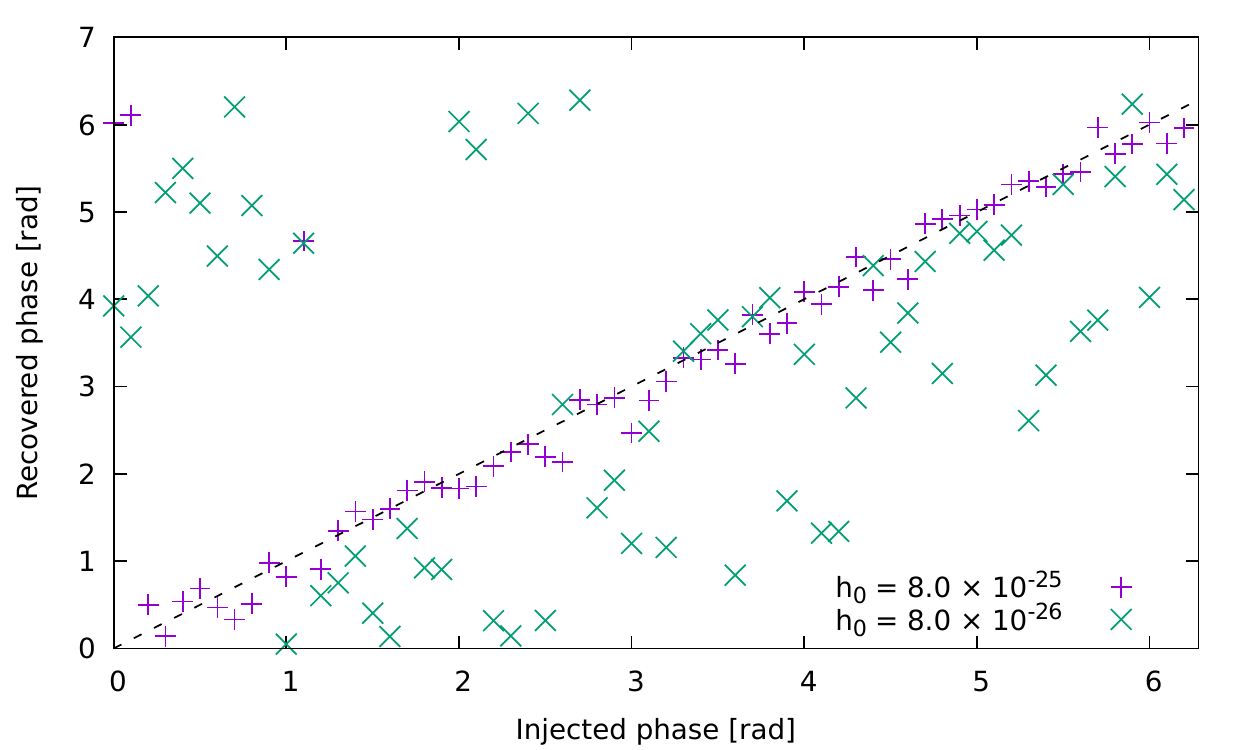}}
\caption{
Maximum likelihood phase tracking.
Estimated (vertical axis) versus injected (horizontal axis) phase
for $h_0 = 8.0\times 10^{-25}$ (purple points; 63 trials)
and $h_0 = 8.0\times 10^{-26}$ (green points; 63 trials),
using the maximum likelhood estimate returned by
the LAL function {\em XLALEstimatePulsarAmplitudeParams}.
}
\label{fig:vitappc1}
\end{figure*}

When the HMM tracks $\Phi_\ast(t)$ as well as $f_\ast(t)$,
it is tempting to generalize $G(f_0)$ to $G(f_0,\Phi_0)$,
where $\Phi_0$ is the trial phase,
by analogy with (\ref{eq:vitappc7}).
Firstly, one may try to incorporate the phase into the amplitudes $A_{1i}$,
as in Ref.\ \cite{PrixWhelan2007},
e.g.\ $A_{11}= A_+ \cos 2\psi \cos \Phi_{\rm w}
 - A_\times \sin 2\psi \sin \Phi_{\rm w}$.
Unfortunately, maximizing the likelihood $\Lambda'$
with respect to $A_{1i}$ returns estimators $\hat{A}_{1i}$,
which are rotated versions of the phase-independent estimators,
e.g.\ $\hat{A}_{11}$ becomes 
$\hat{A}_{11}\cos\Phi_{\rm w} + \hat{A}_{13}\sin\Phi_{\rm w}$.
The resulting ${\cal F}$-statistic is independent of phase, 
as shown in Appendix A in Ref.\ \cite{Suvorova2017} 
in the context of orbital phase.
Instead, one may try to factorize the ${\cal F}$-statistic
into a quadratic form constructed from complex amplitudes, 
multiply the complex amplitudes by the cosine of the phase,
and reassemble the quadratic form to obtain a real likelihood.
\footnote{
In non-gravitational-wave applications
where the signal is an unmodulated sinusoid
with a single polarization mode,
and the antenna beam-pattern does not vary diurnally,
this procedure yields the exact, maximum likelihood estimator.
\cite{Suvorova2018}
}
In this spirit, we define
\begin{equation}
 G(f_0,\Phi_0)
 =
 \frac{4{\bf R}(f_0,\Phi_0) {\bf H}^{-1} 
  {\bf R}(f_0,\Phi_0)^\dagger}{T_{\rm obs} S_h(f_0)}~,
\label{eq:vitappc11}
\end{equation}
with
\begin{equation}
 {\bf R}(f_0,\Phi_0) = [R_{1a}(f_0,\Phi_0),R_{1b}(f_0,\Phi_0)]~.
\label{eq:vitappc12}
\end{equation}
Numerical experiments reveal that (\ref{eq:vitappc11}) produces
no improvement in sensitivity compared to Version I of the HMM.
Essentially this is because noise in the phase estimate
defeats the HMM's ability to reject paths with inconsistent phase.
This can be seen by plotting the output of the function
{\em XLALEstimatePulsarAmplitudeParams} in the LAL suite,
which returns maximum likelihood estimates of the source parameters 
(including phase) given $F_{1a}$ and $F_{1b}$,
against the injected phase.
\footnote{
At the time of writing, 
{\em XLALEstimatePulsarAmplitudeParams} incorrectly adds $\pi$
to the phase. The error is corrected here.
}
Figure \ref{fig:vitappc1} demonstrates that the estimated and injected phases
are strongly correlated for $h_0 = 8.0\times 10^{-25}$.
However,
the correlation weakens appreciably for $h_0 = 8.0 \times 10^{-26}$
and even more so near the detection limit for Version III of the HMM
($h_0 = 1.3\times 10^{-26}$),
where the points scatter randomly (not plotted).
The Pearson correlation coefficient,
computed versus $h_0$ in Table \ref{tab:vitappc1},
exhibits the same behavior.

\begin{table}
	\vspace{0.5cm}
	\centering
	\setlength{\tabcolsep}{8pt}
	\begin{tabular}{ll}
		\hline
		\hline
		$h_0$ ($10^{-26}$) & Coefficient \\
		\hline
		80 & 0.978 \\
		8.0 & 0.464 \\
		1.7 & 0.156 \\
		1.3 & 0.059 \\
		\hline
		\hline
	\end{tabular}
	\caption{
Estimated versus injected phase:
Pearson correlation coefficient as a function of signal strength
for the maximum likelihood estimator
{\em XLALEstimatePulsarAmplitudeParams}
with $10^3$ realizations.
}
	\label{tab:vitappc1}
\end{table}

Note that the HMM tracks the phase difference between HMM steps;
the absolute phase enters through the prior and is not tracked explicitly.
This differs subtly from a fully coherent $\mathcal{F}$-statistic search
(without spin wandering), where $\mathcal{F}$ is evaluated as a function of
$\Phi_{\rm w}(t_0)$ as well as $f_0^{(k)}$, $\alpha$, and $\delta$.
\cite{Jaranowski1998}

\section{Validation tests
 \label{sec:vitappe}}
In this appendix, we present for completeness and reproducibility
the results of several validation tests applied to Version III of the HMM.
The tests relate to the PDF of the ${\cal B}$-statistic after a single HMM step,
the PDF of the block score after multiple HMM steps,
the detection probability as a function of $N_T$
for $T_{\rm drift}$ or $T_{\rm obs}$ fixed,
the effect of the block definition on the detector's performance,
and the conservation of signal power by the detection statistic.
The tests will help to guide future refinements of the HMM.

\subsection{PDF of the detection statistic
 \label{sec:vitappea}}
\begin{figure*}
	\centering
	\subfigure[]
	{
		\label{fig:vit4a}
		\scalebox{0.65}{\includegraphics{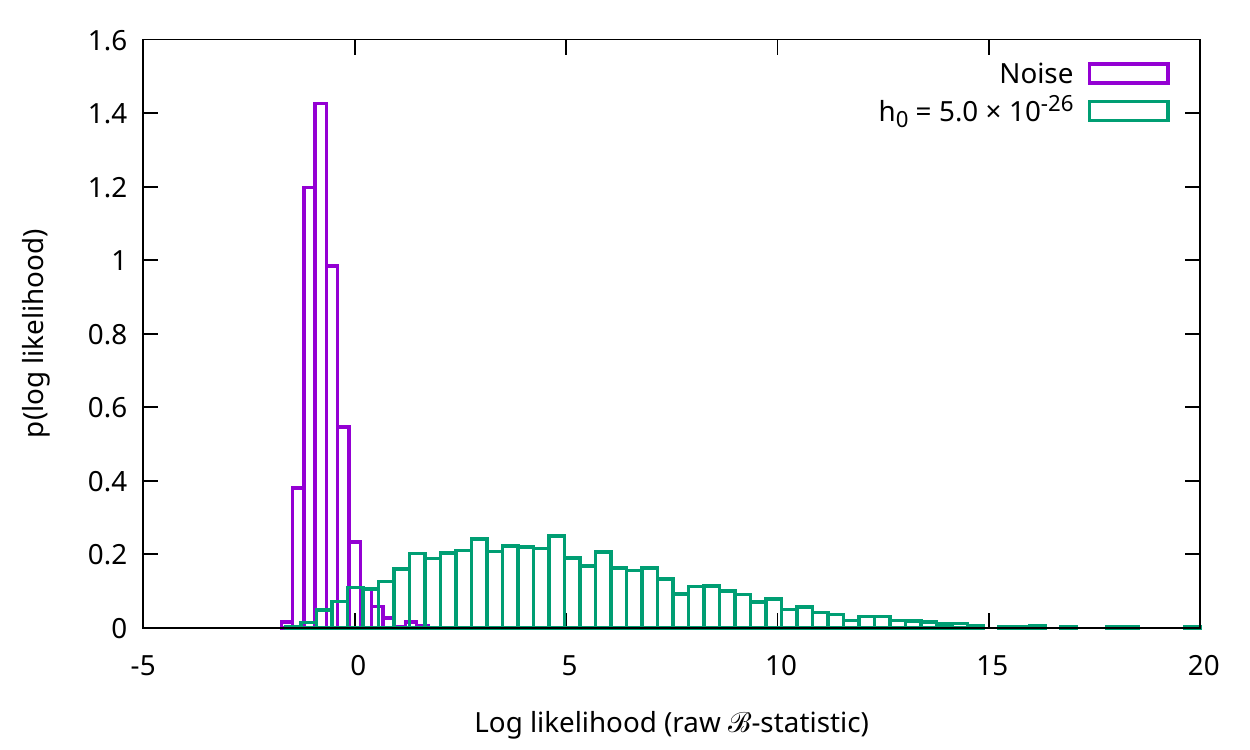}}
	}
	\subfigure[]
	{
		\label{fig:vit4b}
		\scalebox{0.65}{\includegraphics{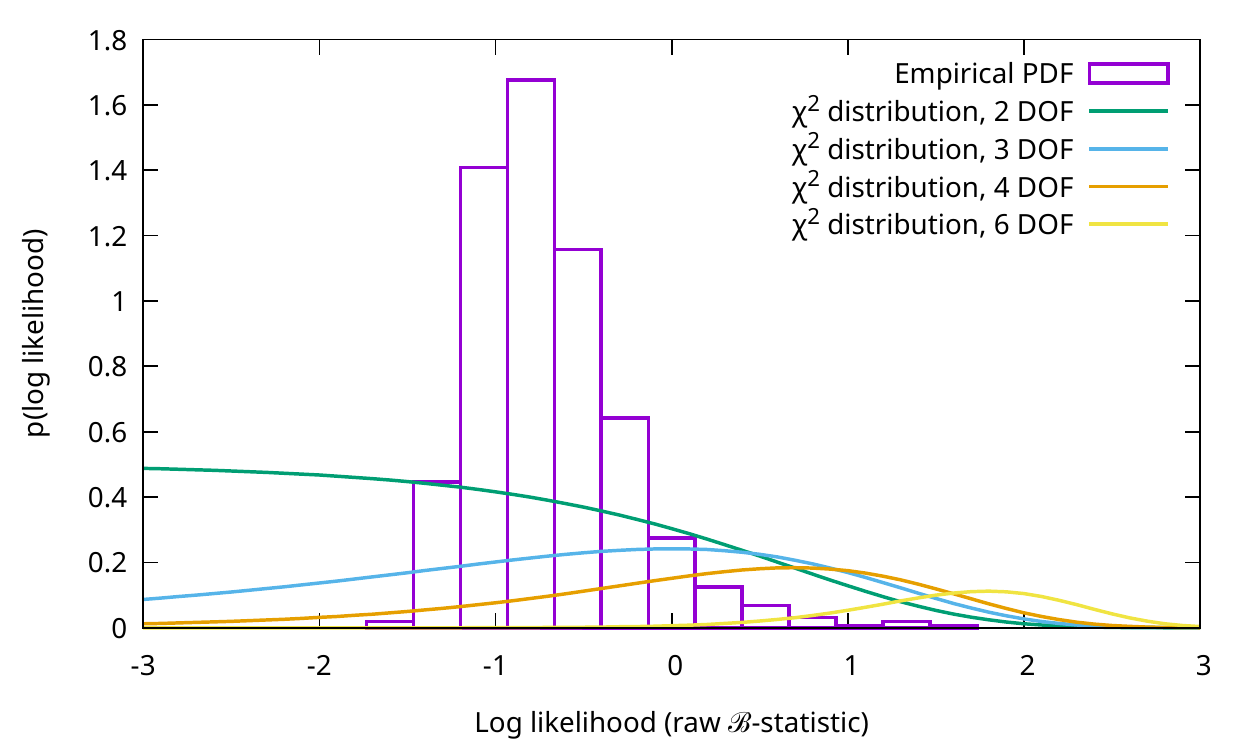}}
	}
	\subfigure[]
	{
		\label{fig:vit4c}
		\scalebox{0.65}{\includegraphics{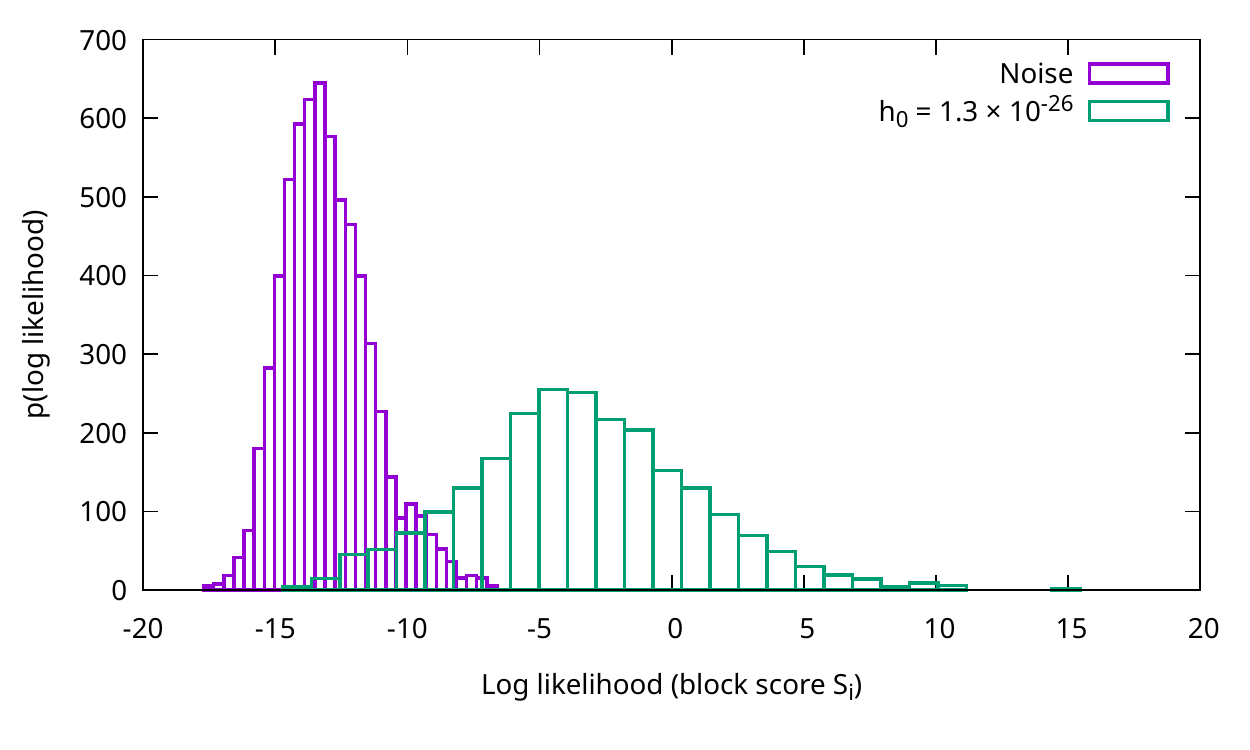}}
	}
	\caption{
Normalized PDF of the detection statistic for pure noise (purple histograms)
and a detected injection with the source parameters in Table \ref{tab:vit1}
(green histograms).
(a) Logarithm of the ${\cal B}$-statistic, $\ln {\cal B}(f_\ast,\Phi_\ast)$,
computed for a single HMM step
in the bin $(f_\ast,\Phi_\ast)$ containing the injection (where present),
with $h_0=0$ (purple histogram) and $h_0=5\times 10^{-26}$ (green histogram).
(b) Noise-only histogram from (a) rebinned over the domain $[-3,3]$
and overlaid with normalized, central, chi-squared distributions with
2, 3, 4, and 6 degrees of freedom (solid curves; color scheme in legend),
in order to test for congruence with the functional form of the
${\cal F}$-statistic PDF.
(c) Block score $S_i$ defined by (\ref{eq:vit45})
for the block containing the injection (where present),
with $h_0 = 0$ (purple histogram), $h_0 = 1.3\times 10^{-26}$ (green histogram),
and $N_T=37$.
Realizations: $2.5\times 10^3$ per histogram.
}
	\label{fig:vit4}
\end{figure*}

Figure \ref{fig:vit4}(a) displays the PDF of $\ln {\cal B}$
computed for a single HMM step
in pure noise ($h_0=0$; purple histogram) and for a relatively strong injection 
($h_0 = 5\times 10^{-26}$; green histogram).
The injection shifts the mode of the PDF to the right, as expected.
Figure \ref{fig:vit4}(b) investigates in more detail the functional form
of the noise-only PDF.
All the histograms and curves in Figure \ref{fig:vit4}(b) are normalized,
and the results are independent of $f_0$ and $\Phi_0$.
\footnote{
There is a weak dependence on the width of the running median window
applied to the power spectral density,
as for the ${\cal F}$-statistic.
\cite{SCO-X1-2015,Abbott2017ViterbiO1}
}
It is clear by inspection that the noise-only ${\cal B}$-statistic does not obey
a central chi-squared distribution with four degrees of freedom
(unlike the ${\cal F}$-statistic)
nor with two to six degrees of freedom.
The two statistics correspond to slightly different choices
of amplitude priors within a Bayesian framework but are otherwise the same,
with $|\ln {\cal B}-{\cal F}| \lesssim 0.05 {\cal F}$ 
for a wide range of signal and noise parameters.
\cite{Prix2009,Dergachev2012,Whelan2014,Dhurandhar2017}
However, by marginalizing over $\psi$, $\cos\iota$, and $h_0$
in (\ref{eq:vit38}),
one implicitly enforces constraints between $A_+$ and $A_\times$
and hence the four amplitudes $A_{1i}$ in (\ref{eq:vit20}),
so that the statistic is no longer the sum of four independent squares.

Detection with the HMM is performed using the block score $S$
defined in (\ref{eq:vit45}) in Section \ref{sec:vit5c}.
Figure \ref{fig:vit4}(c) displays histograms of $S$
after $N_T=37$ steps of the HMM
for pure noise ($h_0=0$; purple histogram)
and an injection below the single-step detection threshold
($h_0 = 1.3\times 10^{-26}$; green histogram).
The peaks of the noise-only and noise-plus-injection histograms 
are clearly separated,
demonstrating the discriminating power of the HMM.
The PDFs of $S$ are narrower than for $\ln {\cal B}$
and have thinner right-hand tails,
because the nonlinear maximization step in the Viterbi algorithm produces
an extreme value distribution similar to the Gumbel law.
\cite{Suvorova2017}
The maximum is taken over all Viterbi paths terminating in
a given frequency-phase bin,
so paths terminating in neighboring bins are correlated
because they share common subpaths.
There is no analytic expression for the PDF of 
$\ln \Pr[Q^\ast(O)|O]$ in the literature
to the best of our knowledge.
\cite{Suvorova2017}
We therefore rely on the empirical PDF in Figure \ref{fig:vit4}(c)
to set $S_{\rm th}(f)$ given $P_{\rm a}$.

\subsection{Detection probability versus $N_T$
 \label{sec:vitappeb}}
\begin{figure*}
	\centering
	\subfigure[]
	{
		\label{fig:vit6a}
		\scalebox{0.65}{\includegraphics{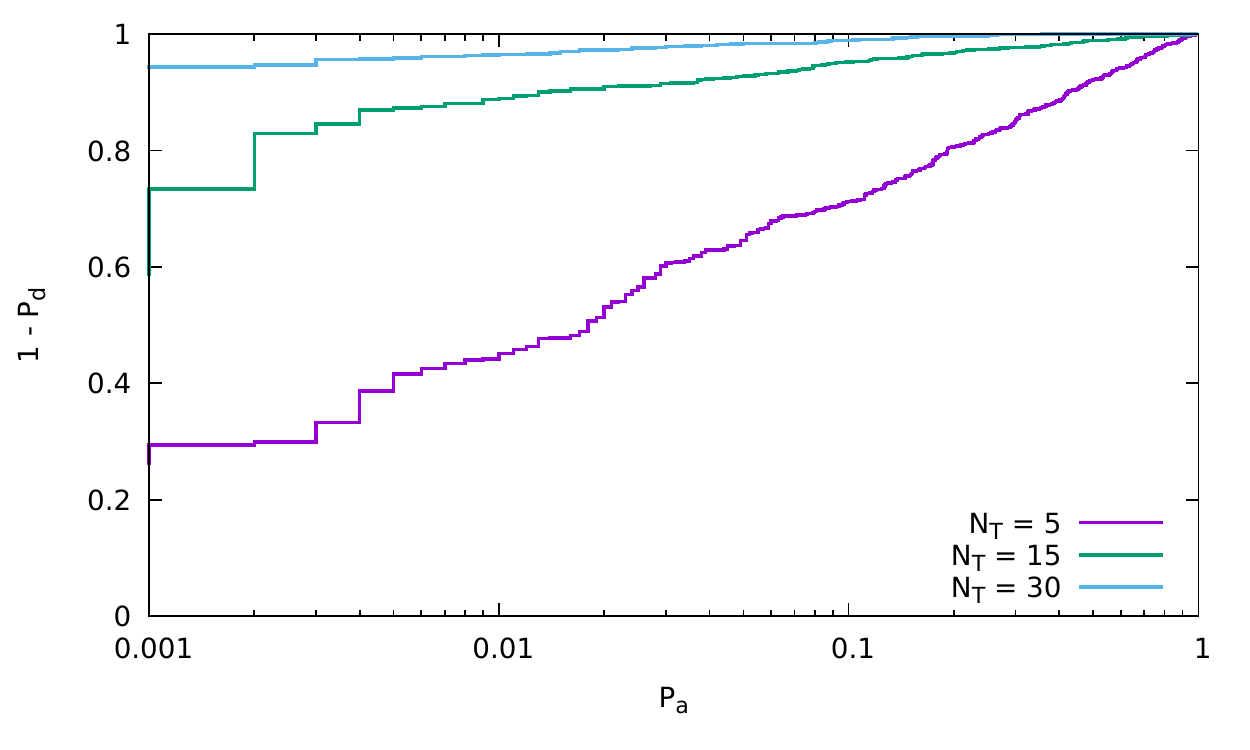}}
	}
	\subfigure[]
	{
		\label{fig:vit6b}
		\scalebox{0.65}{\includegraphics{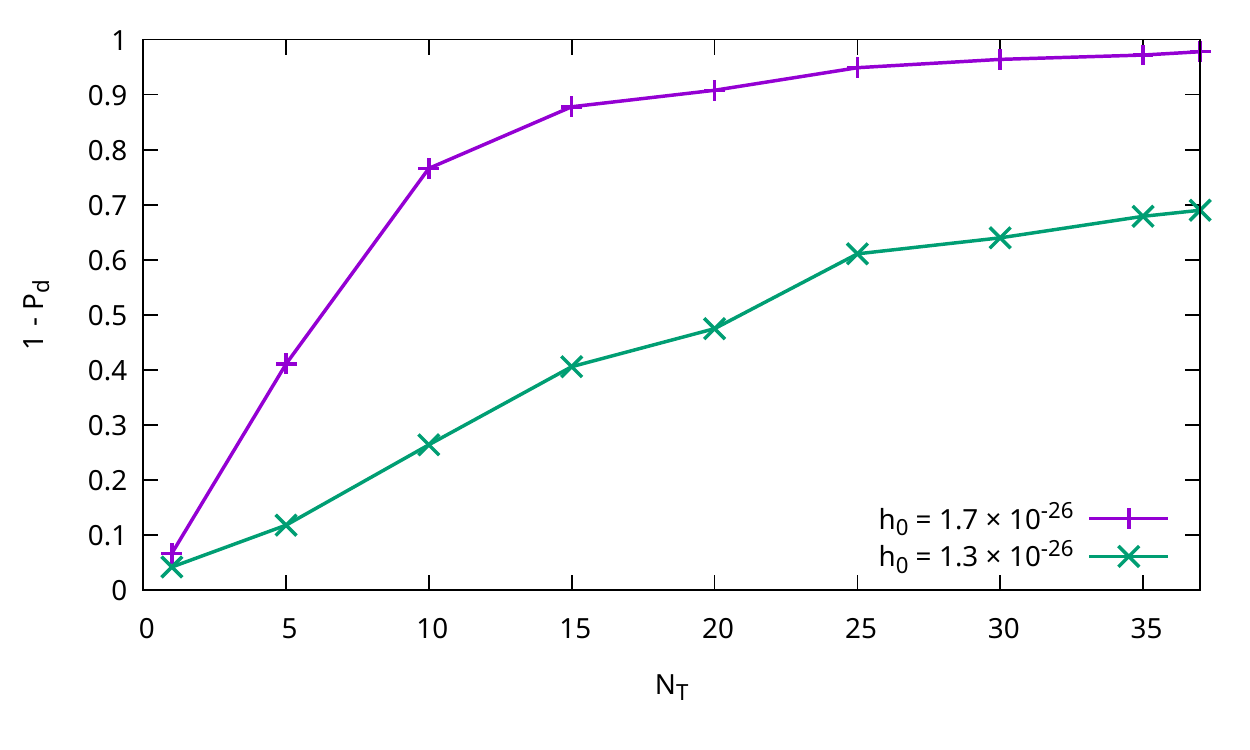}}
	}
	\subfigure[]
	{
		\label{fig:vit6c}
		\scalebox{0.65}{\includegraphics{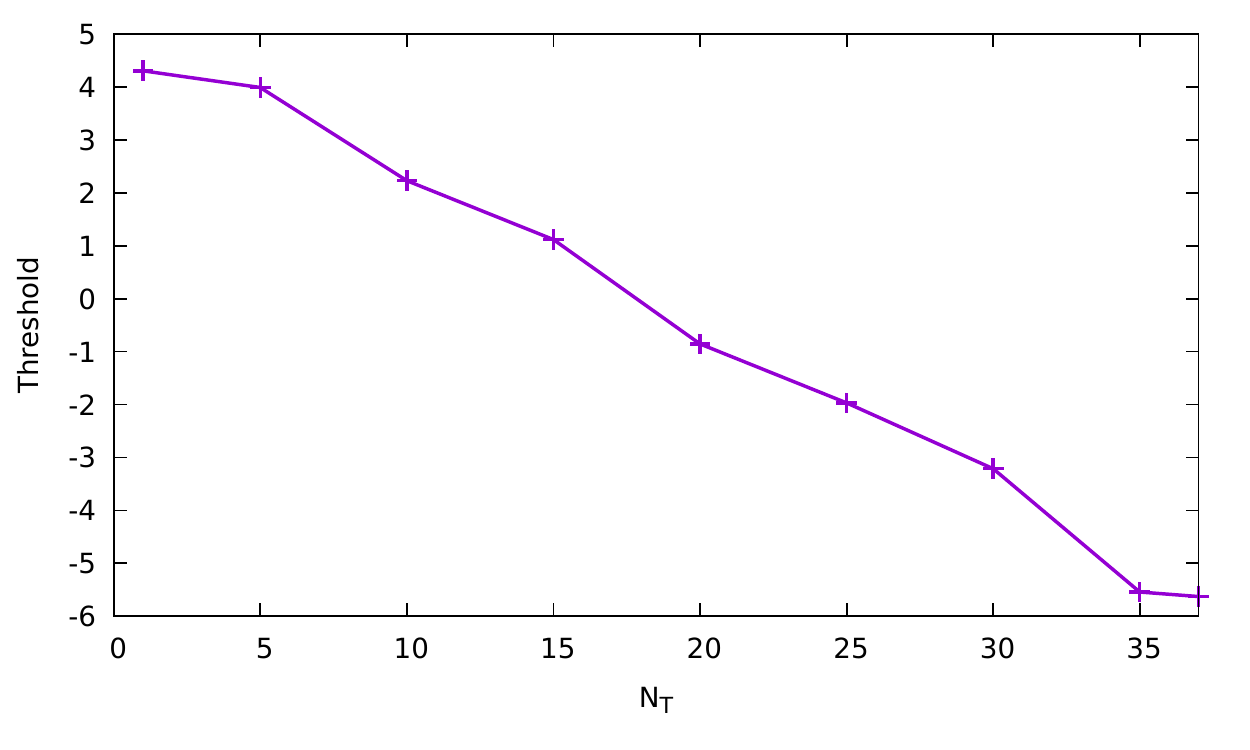}}
	}
	\caption{
Detector performance as a function of $N_T$
for $T_{\rm drift}=10\,{\rm d}$ fixed, 
$T_{\rm obs}=N_T T_{\rm drift} \propto N_T$ variable,
and the source parameters in Table \ref{tab:vit1}.
(a) ROC curves for $h_0 = 1.7\times 10^{-26}$ and
$N_T=5$ (purple curve), 15 (green curve), 30 (blue curve).
(b) Detection probability $1-P_{\rm d}$ versus $N_T$ for
$h_0 = 1.3\times 10^{-26}$ (green curve), $1.7 \times 10^{-26}$ (purple curve),
and $P_{\rm a}=10^{-2}$ per block.
(c) Block score threshold $S_{\rm th}$ [see (\ref{eq:vit45})]
versus $N_T$ for false alarm probability $P_{\rm a}=10^{-2}$ per block;
the number of bins per block, $\Pr(Q|O)$, and hence $S_{\rm th}$ depend on $N_T$.
All curves are calculated for Version III of the HMM.
Control parameters:
$\gamma=1.0\times 10^{-16}\,{\rm s^{-1}}$, 
$\sigma=3.7\times 10^{-10}\,{\rm s^{-3/2}}$.
Realizations: $10^3$ per curve.
}
	\label{fig:vit6}
\end{figure*}

Another important question is how the performance of the HMM
scales with $N_T$.
We formulate the question with respect to two practical scenarios:
(i) $T_{\rm drift}$ is fixed, and $T_{\rm obs} \propto N_T$ varies;
and (ii) $T_{\rm obs}$ is fixed, and $T_{\rm drift} \propto N_T^{-1}$ varies.
Figure \ref{fig:vit6} presents data for scenario (i).
As expected, the sensitivity of the HMM increases,
as $N_T$ and hence $T_{\rm obs}$ increase.
\cite{Suvorova2018}
We observe in Figure \ref{fig:vit6}(b) 
that the detection probability rises with $N_T$ at fixed $P_{\rm a}=10^{-2}$.
The same trend occurs in Figure \ref{fig:vit6}(a)
for $10^{-3}\leq P_{\rm a} \leq 1$.
Figure \ref{fig:vit6}(b) corresponds to a vertical cut 
at constant $P_{\rm a}= 10^{-2}$
through the family of ROC curves in Figure \ref{fig:vit6}(a).
One subtlety is that $S_{\rm th}$ depends on $N_T$ through
two countervailing factors.
The number of frequency bins per block is proportional to $N_T$,
so $S_{\rm th}$ should increase with $N_T$, ceteris paribus,
to keep $P_{\rm a}$ per block fixed;
but the product $\Pr(Q|O)$ in (\ref{eq:vit1}) decreases with $N_T$,
as more factors $L_{o(t_n)q(t_n)} A_{q(t_n)q(t_{n-1})}\leq 1$ are appended,
implying that $S_{\rm th}$ should decrease with $N_T$ for fixed $P_{\rm a}$.
The latter effect outweighs the former,
as is evident in Figure \ref{fig:vit6}(c);
the threshold decreases from $S_{\rm th}\approx 4.0$ for $N_T=5$
to $S_{\rm th}\approx -5.5$ for $N_T=35$.
In a genuine, astrophysical search one would typically set
$P_{\rm a} = 10^{-2}$ for the whole search band ($B\sim 1\,{\rm kHz}$),
or for sub-bands with $\Delta f_{\rm sub} \sim 1\,{\rm Hz}$
(to facilitate data handling),
and hence have $P_{\rm a} \ll 10^{-2}$ per block,
with $N_T \Delta f_{\rm drift} \ll \Delta f_{\rm sub} \leq B$.
The scalings with $N_T$ are the same in this regime,
but the ROC curves are time-consuming to generate by Monte Carlo simulations.
\footnote{
Occasionally situations may arise,
where it is desirable to hold the number of bins per block fixed
while varying $N_T$, e.g.\ when comparing results from two
data sets of different durations. 
We defer the analysis of such situations to future work.
}

\begin{figure*}
	\centering
	\subfigure[]
	{
		\label{fig:vit7a}
		\scalebox{0.65}{\includegraphics{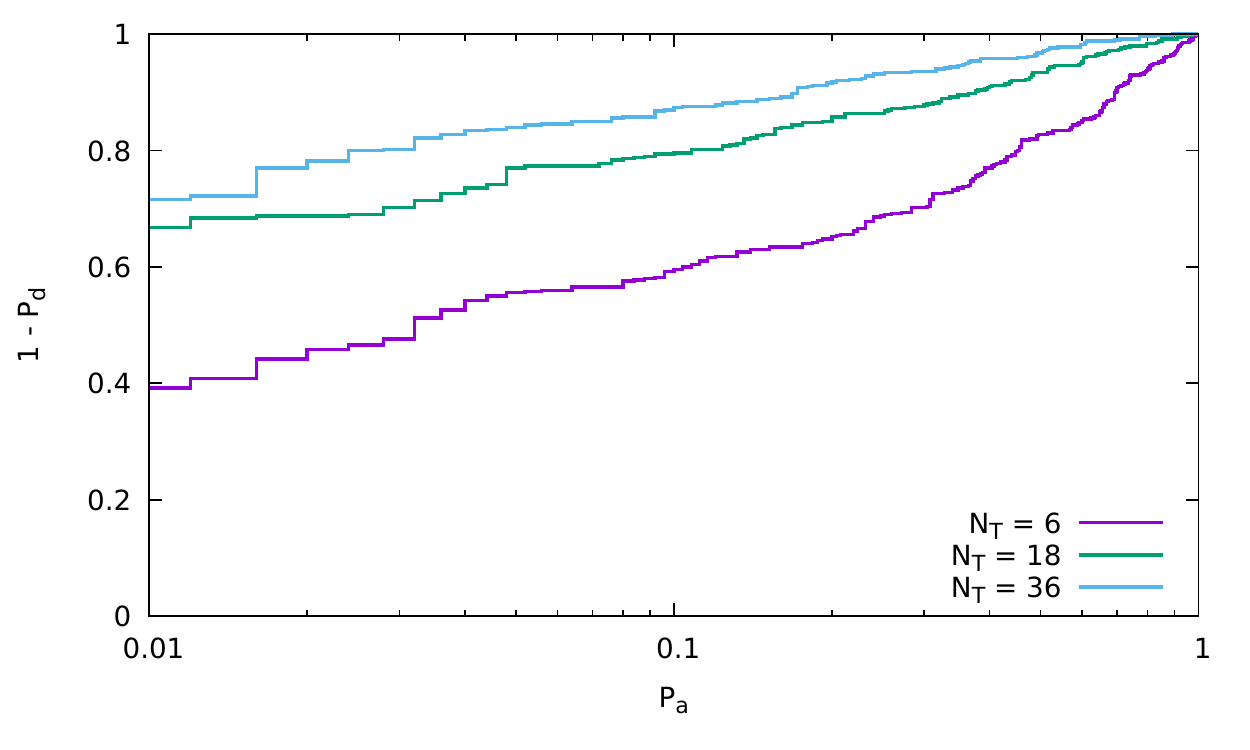}}
	}
	\subfigure[]
	{
		\label{fig:vit7b}
		\scalebox{0.65}{\includegraphics{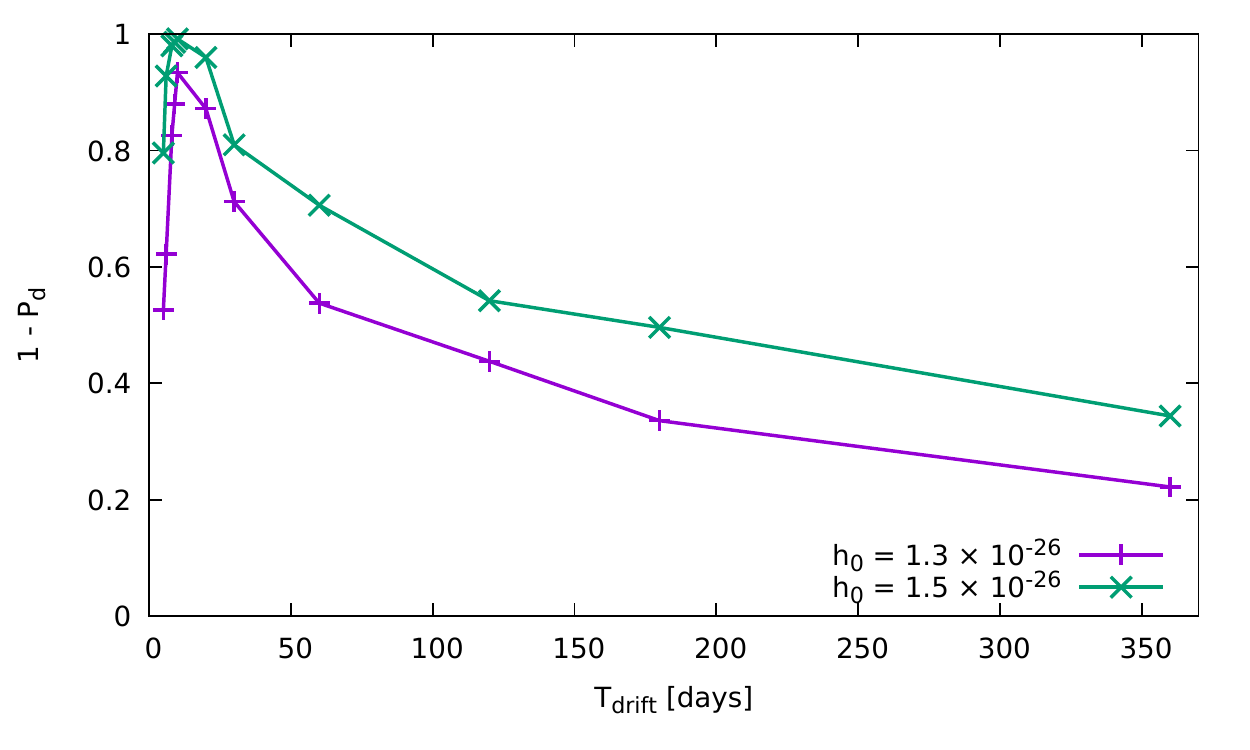}}
	}
	\subfigure[]
	{
		\label{fig:vit7c}
		\scalebox{0.65}{\includegraphics{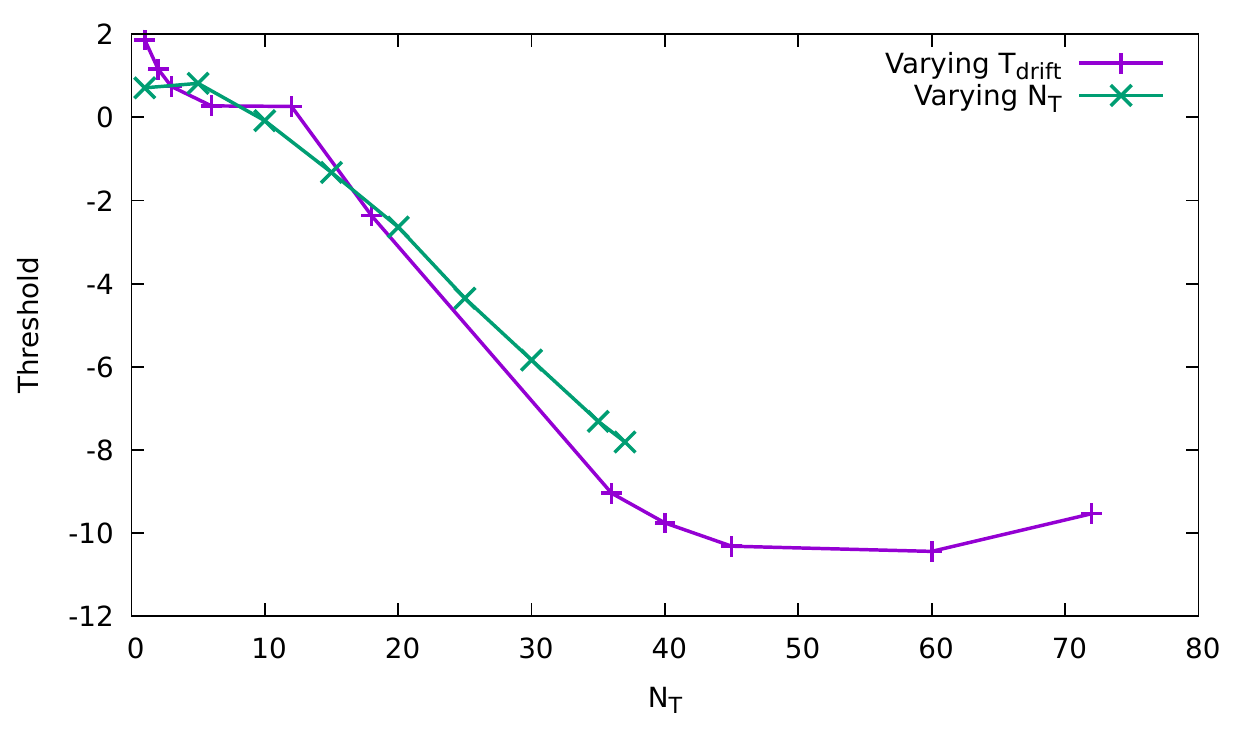}}
	}
	\caption{
Detector performance as a function of $N_T$
for $T_{\rm obs}=360\,{\rm d}$ fixed, 
$T_{\rm drift}=N_T^{-1} T_{\rm obs} \propto N_T^{-1}$ variable,
and the source parameters in Table \ref{tab:vit1}.
(a) ROC curves for $h_0 = 1.3\times 10^{-26}$ and
$N_T=6$ (purple curve), 18 (green curve), 36 (blue curve),
chosen to give an integer number of days per HMM step.
(b) Detection probability $1-P_{\rm d}$ versus $N_T$ for
$h_0 = 1.3\times 10^{-26}$ (purple curve), $1.5 \times 10^{-26}$ (green curve),
and $P_{\rm a}=10^{-2}$ per block.
(c) Block score threshold $S_{\rm th}$ [see (\ref{eq:vit45})]
versus $N_T$ for false alarm probability $P_{\rm a}=10^{-2}$ per block.
The number of bins per block and hence $S_{\rm th}$ scale with $N_T$,
with 
$T_{\rm drift}=N_T^{-1} T_{\rm obs} \propto N_T^{-1}$ variable
(purple curve) and $T_{\rm drift}=10\,{\rm d}={\rm constant}$
[green curve; copied from Figure \ref{fig:vit6}(c) for comparison].
All curves are calculated for Version III of the HMM.
Control parameters:
$\gamma=1.0\times 10^{-16}\,{\rm s^{-1}}$, 
$\sigma=3.7\times 10^{-10}\,{\rm s^{-3/2}}$.
Realizations: $10^3$ per curve.
}
	\label{fig:vit7}
\end{figure*}

Figure \ref{fig:vit7} presents data for scenario (ii) in the previous paragraph,
i.e.\ fixed $T_{\rm obs}$.
The trend with $N_T$ depends on whether $T_{\rm drift}\propto N_T^{-1}$
is less or greater than the characteristic time-scale over which the
signal frequency wanders.
\cite{Suvorova2018}
If $T_{\rm drift}$ is less than the wandering time-scale,
the detection probability decreases,
as $T_{\rm drift}$ decreases;
it is disadvantageous to shorten the coherent integration in a HMM segment,
when the frequency wanders by less than one bin during a segment.
We observe this behavior in Figure \ref{fig:vit7}(b)
to the left of the peak.
If $T_{\rm drift}$ is greater than the wandering time-scale,
the detection probability increases,
as $T_{\rm drift}$ decreases;
it is better to make the segments shorter,
as required by condition (\ref{eq:vit6}),
up to the point where the frequency wanders by roughly one bin during a segment.
We observe this behavior to the right of the peak in Figure \ref{fig:vit7}(b).
The behavior in Figure \ref{fig:vit7}(b) for $P_{\rm a}=10^{-2}$ per block
is consistent with the ROC curves in Figure \ref{fig:vit7}(a) over the range
$10^{-3}\leq P_{\rm a} \leq 1$.
The threshold decreases with $N_T$ in Figure \ref{fig:vit7}(c),
just like in Figure \ref{fig:vit6}(c),
because it is approximately independent of $T_{\rm drift}$.

\subsection{Block definition
 \label{sec:vitappec}}
\begin{figure*}
	\centering
	\subfigure[]
	{
		\label{fig:vit8a}
		\scalebox{0.65}{\includegraphics{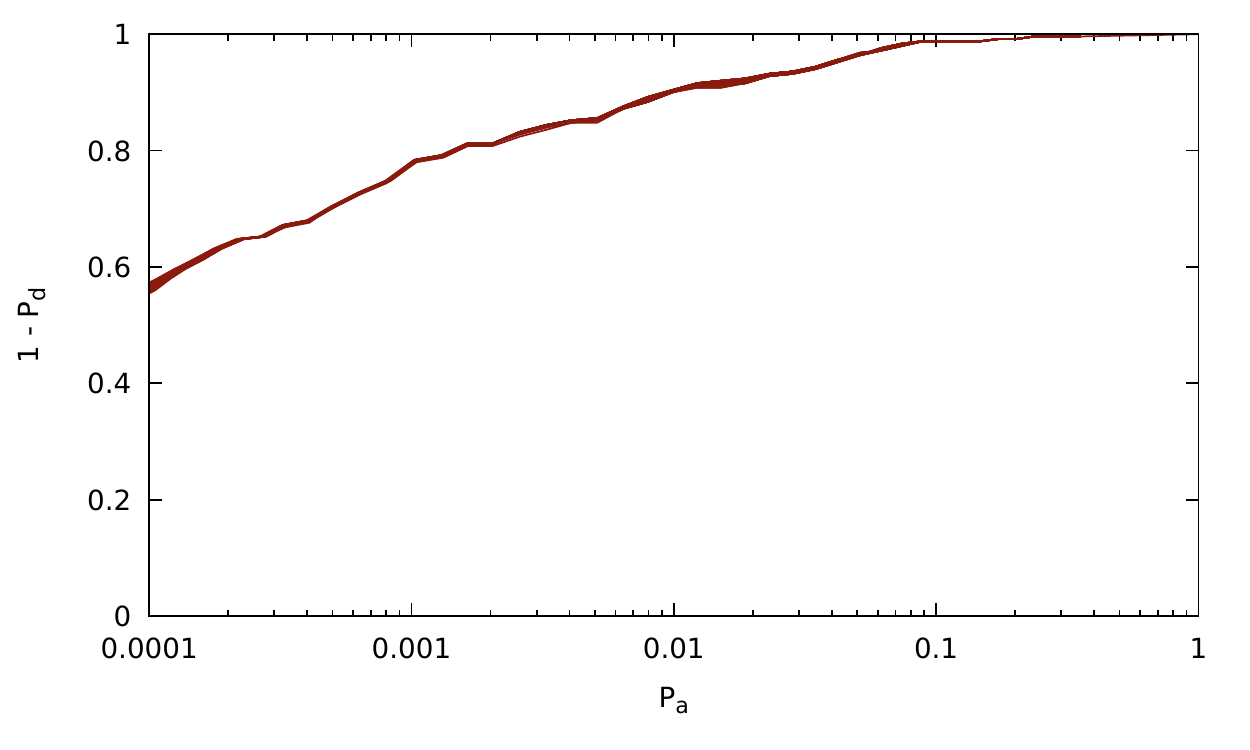}}
	}
	\subfigure[]
	{
		\label{fig:vit8b}
		\scalebox{0.65}{\includegraphics{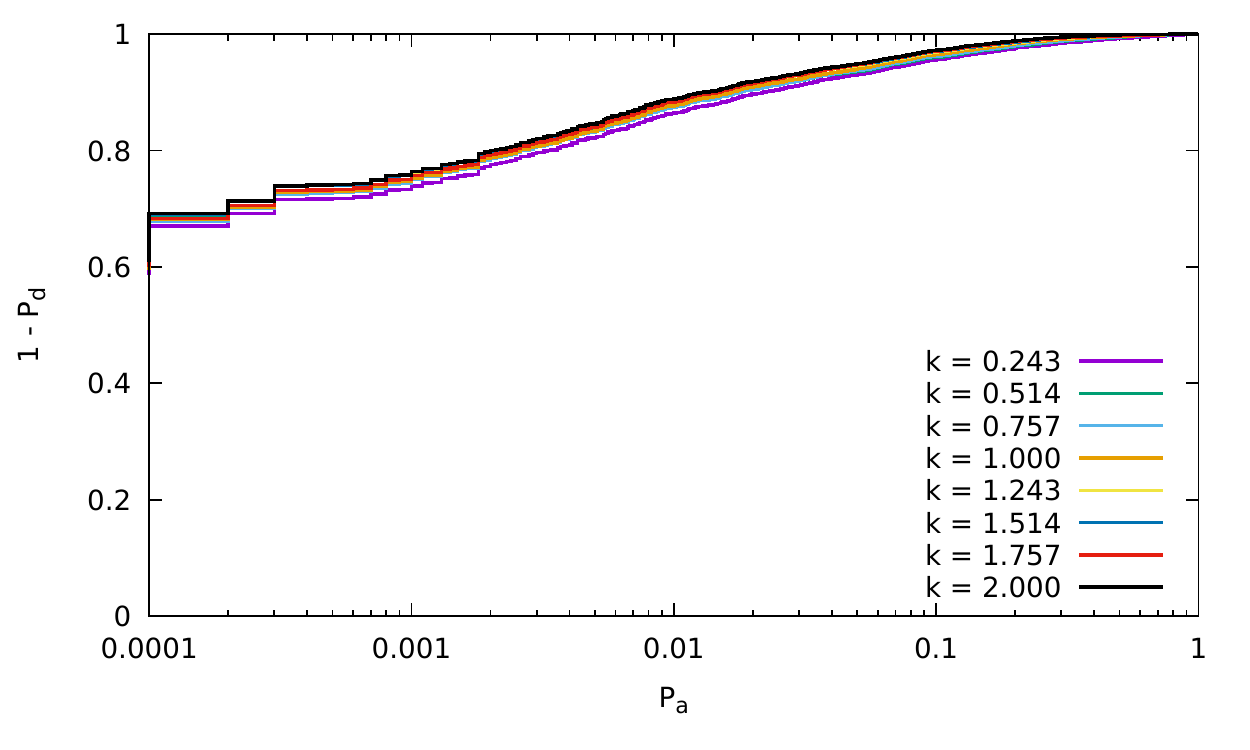}}
	}
	\caption{
Effects of block definition on performance.
(a) ROC curves for blocks of bandwidth $2N_T \Delta f_{\rm drift}$,
whose leftmost frequency bins are shifted
$1,2,\dots,N_T-1$ bins to the right of an arbitrary reference bin,
for $h_0=1.3\times 10^{-26}$, $T_{\rm drift}=10\,{\rm d}$, and $N_T=37$.
The 36 curves overlap closely and cannot be distingushed by eye.
(b) ROC curves for blocks of bandwidth $2kN_T \Delta f_{\rm drift}$,
with $k=0.243$, 0.514, 0.757, 1.00, 1.24, 1.51, 1.76, 2.00
(chosen to give an integer number of bins per block).
All curves are calculated for Version III of the HMM.
Source parameters: see Table \ref{tab:vit1}.
Control parameters:
$\gamma=1.0\times 10^{-16}\,{\rm s^{-1}}$, 
$\sigma=3.7\times 10^{-10}\,{\rm s^{-3/2}}$.
Realizations: $10^3$ per curve.
}
	\label{fig:vit8}
\end{figure*}

What happens when a candidate straddles the boundary between two blocks?
In this paper, we treat it as a special case, 
to be followed up through a veto procedure in a genuine astrophysical search.
Straddlers represent a modest fraction $\sim N_T^{-1/2}$ 
of all signals or false alarms.
\footnote{
Alternatively one can record straddlers on a candidate list and
consolidate candidates that share common sub-paths,
after all the data are analysed.
This complicates the statistical interpretation of the results,
because HMM paths with common sub-paths are correlated.
\cite{Suvorova2017}
}
Figure \ref{fig:vit8}(a) verifies that the absolute position of the 
block boundary does not affect the ROC curves appreciably.
It displays $N_T-1$ individual ROC curves
for $N_T-1$ different block boundaries,
in which the leftmost frequency bin is shifted right by
$1,2,\dots, N_T-1$ bins relative to an arbitrary, reference bin.
The curves overlap closely and are barely distinguishable by eye.

Likewise we find that the performance of the HMM depends weakly
on the bandwidth of each block.
It is unlikely for a path to drift by $\approx N_T$ bins
after $N_T \gg 1$ HMM steps,
even when the tails in $A_{q_j q_i}$ with $|j-i| > 1$ are preserved,
as in Appendix \ref{sec:vitappb}
(cf.\ truncated $A_{q_j q_i}$ with $|j-i|\leq 1$ in Ref.\ \cite{Suvorova2016}).
Figure \ref{fig:vit8}(b) verifies this property by plotting multiple ROC curves
for block widths $2kN_T\Delta f_{\rm drift}$ with $0.2\leq k \leq 2$.
Again the curves overlap closely.
We use $k=1$ henceforth in this paper.

\subsection{Conservation of signal power
 \label{sec:vitapped}}
In Version II of the HMM, based on the ${\cal J}$-statistic,
$J_{1a}(f_0)$ and $J_{1b}(f_0)$ marshal the Doppler-shifted signal power
into one frequency bin by coherently summing orbital sidebands
weighted by $J_s(2\pi f_0 a_0) e^{-is\phi_{\rm a}}$.
It turns out that the same holds true empirically for the ${\cal B}$-statistic,
although there exists no formal mathematical proof at the time of writing;
it may not be possible to derive the ${\cal B}$-statistic for a binary source 
exactly as a Jacobi-Anger expansion of the ${\cal B}$-statistic for an isolated source,
by analogy with the ${\cal J}$-statistic.
This appendix verifies numerically that 
minimal power is lost or dispersed into neighboring frequency bins,
when the ${\cal B}$-statistic is evaluated using
$J_{1a}(f_0)$ and $J_{1b}(f_0)$.

\begin{figure*}
	\centering
	\subfigure[]
	{
		\label{fig:vit12a}
		\scalebox{0.65}{\includegraphics{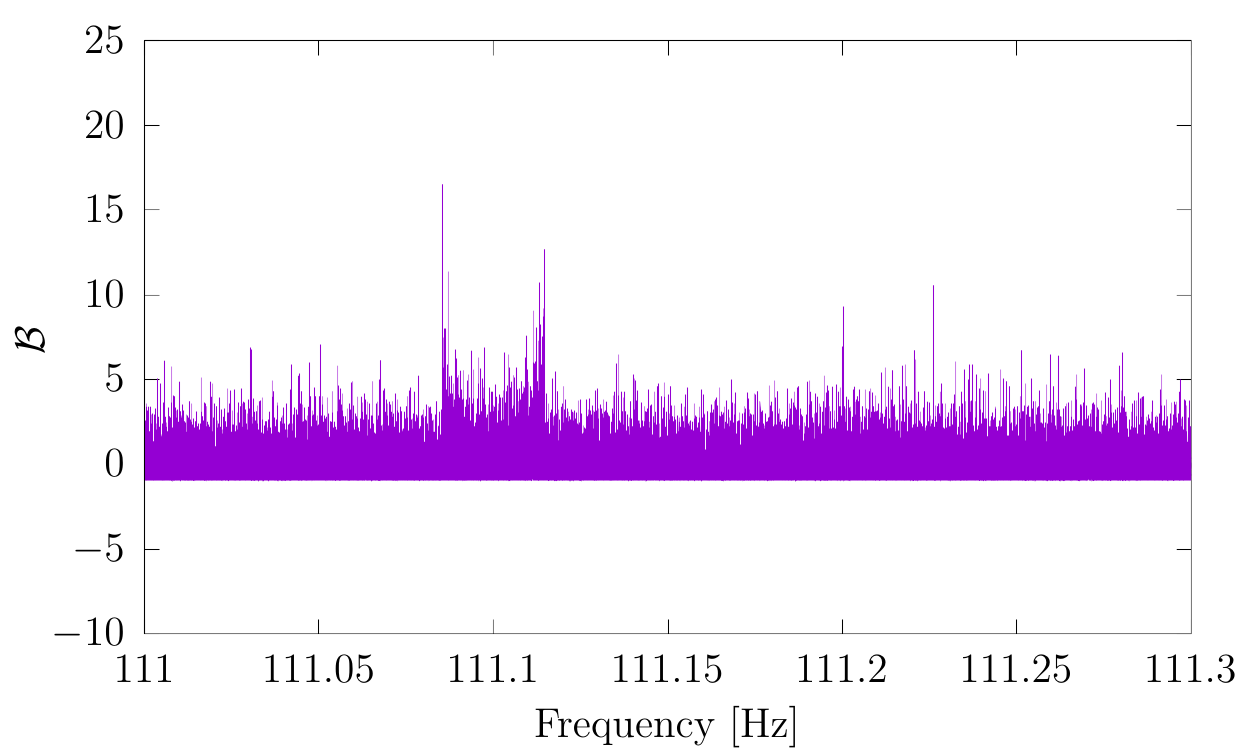}}
	}
	\subfigure[]
	{
		\label{fig:vit12b}
		\scalebox{0.65}{\includegraphics{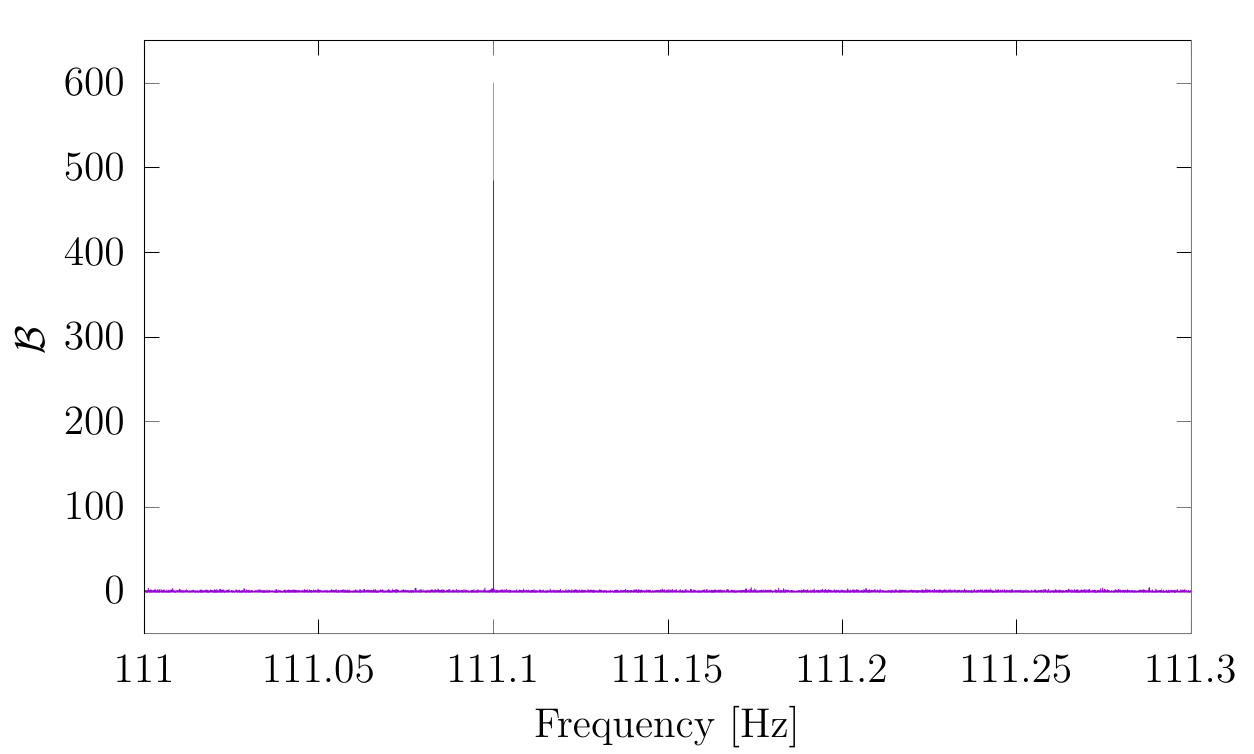}}
	}
	\caption{
Conservation of signal power.
${\cal B}$-statistic at the correct phase bin versus frequency (in Hz)
for a strong binary injection with $h_0=8\times 10^{-25}$ 
and source parameters drawn from Tables \ref{tab:vit1} and \ref{tab:vit2}.
(a) ${\cal B}$ evaluated with $F_{1a}$ and $F_{1b}$ 
in (\ref{eq:vit38})--(\ref{eq:vit44}).
(b) ${\cal B}$ evaluated with $J_{1a}$ and $J_{1b}$ (Jacobi-Anger version)
in (\ref{eq:vit38})--(\ref{eq:vit44}).
Note the different vertical scales in (a) and (b).
}
	\label{fig:vit12}
\end{figure*}

Figure \ref{fig:vit12}(a) graphs ${\cal B}(f_0,\Phi_\ast)$ versus $f_0$ 
(evaluated for $\Phi_\ast$ in the injected bin)
for a strong binary signal
using $F_{1a}$ and $F_{1b}$ to evaluate ${\cal B}$.
As the orbital motion is not accounted for,
${\cal B}$ displays a comb of orbital sidebands at $f_\ast+s/P$,
which fill the band $111.09 \leq f_0 / (1\,{\rm Hz}) \leq 111.11$.
The comb exhibits the classic two-horned envelope familiar from
the Sideband algorithm,
\cite{Sideband-Messenger2007,Sammut2014}
because the source spends more time moving
perpendicular to the plane of the sky 
(when the orbital Doppler shift is a maximum)
than moving perpendicular to the line of sight
(when the Doppler shift is zero).
Figure \ref{fig:vit12}(b) shows the same thing as Figure \ref{fig:vit12}(a)
but with $F_{1a}$ and $F_{1b}$ replaced by $J_{1a}$ and $J_{1b}$
when computing ${\cal B}$.
The sidebands now merge into one peak,
which is $\approx 40$ times higher than the tallest peak
in the comb in Figure \ref{fig:vit12}(a)
(note the different scales).
Identical behavior is seen in Figure 1 in Ref.\ \cite{Suvorova2017}
for the ${\cal J}$-statistic instead of the ${\cal B}$-statistic.

\section{Representative phase paths recovered by the HMM for a source in a binary
 \label{sec:vitappd}}
In this appendix, 
we examine for completeness the optimal phase paths $\Phi_\ast(t)$
recovered by Version III of the HMM 
for the representative examples of binary sources
studied in Section \ref{sec:vit6a}.

\begin{figure*}
	\centering
	\subfigure[]
	{
		\label{fig:vitappd2a}
		\scalebox{0.65}{\includegraphics{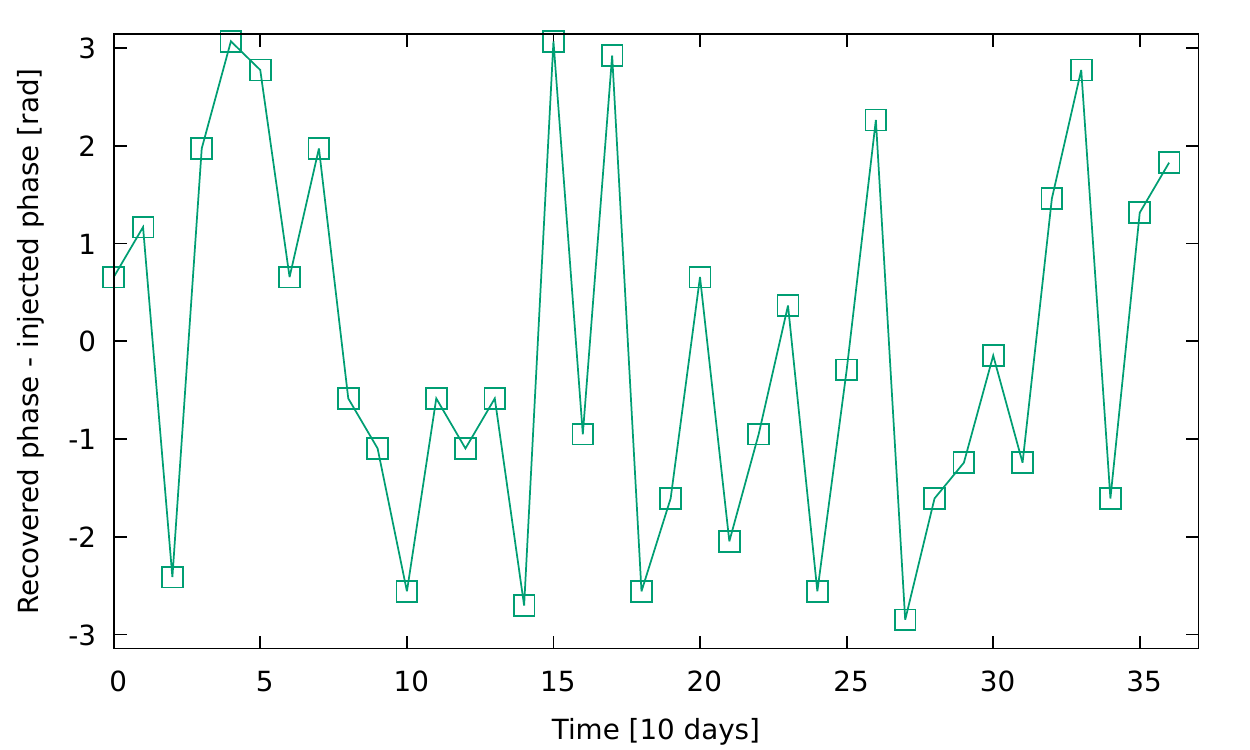}}
	}
	\subfigure[]
	{
		\label{fig:vitappd2b}
		\scalebox{0.65}{\includegraphics{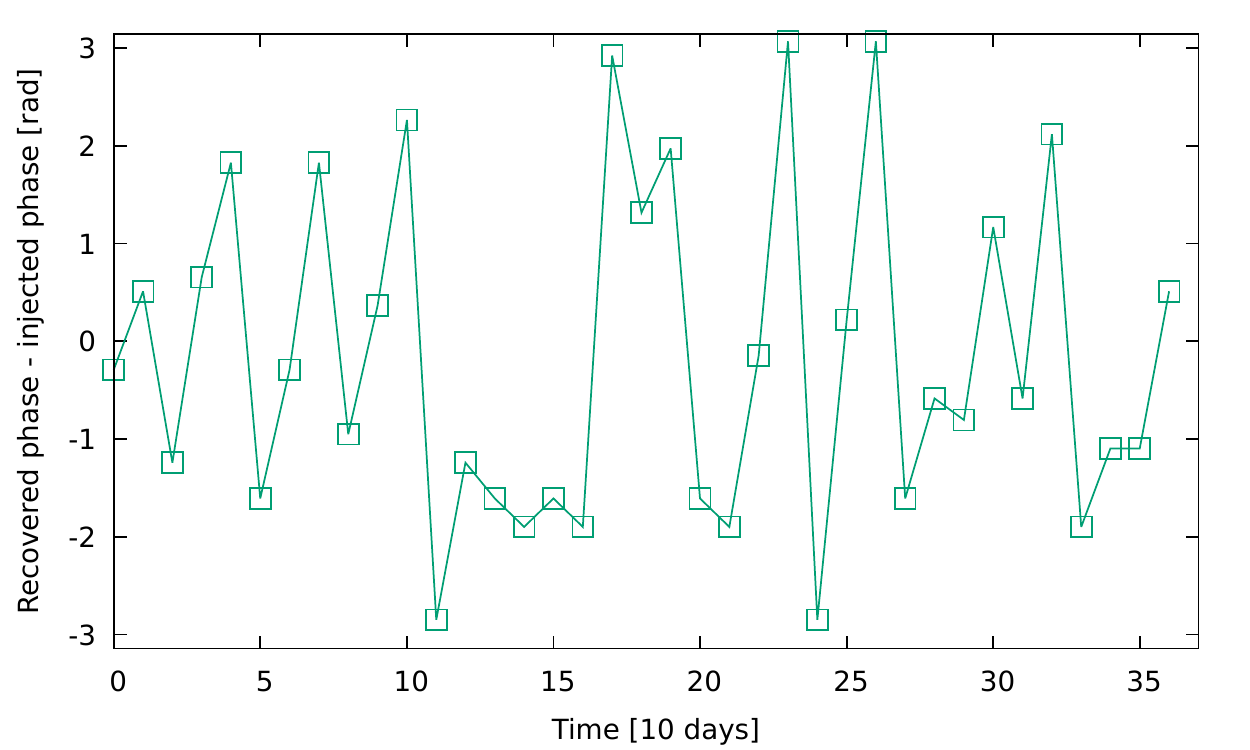}}
	}
	\subfigure[]
	{
		\label{fig:vitappd2c}
		\scalebox{0.65}{\includegraphics{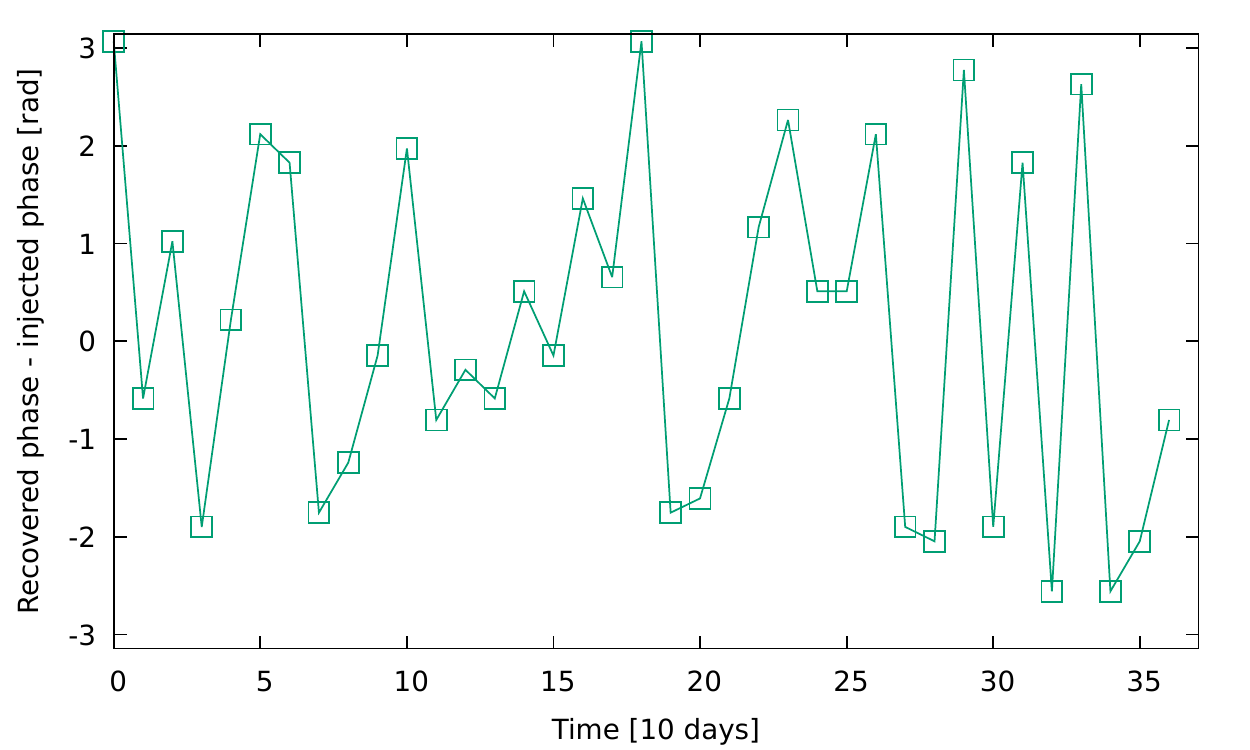}}
	}
	\caption{
Phase tracking in a representative source in a binary.
Layout as for Figure \ref{fig:vitappd1} but for the three sources
in Figure \ref{fig:vit10}.
}
	\label{fig:vitappd2}
\end{figure*}

Figure \ref{fig:vitappd2} displays the absolute error between 
the injected and recovered phase for the
three synthetic binary sources tracked in Figure \ref{fig:vit10}.
The interpretation is the same as in Section \ref{sec:vit5b}.
The phase error jumps around, 
even after unwinding the phase wrapping,
because the ${\cal B}$-statistic spreads the signal power
over multiple phase bins.
On balance, though, the imperfect phase tracking delivers
improved sensitivity,
as evidenced by comparing Figures \ref{fig:vit10}(a) and \ref{fig:vit10}(b)
and the ROC curves in Section \ref{sec:vit6b}.

\bibliography{viterbi_bib}

\end{document}